\documentclass[fleqn,usenatbib,useAMS]{mnras}

\usepackage{graphicx}	
\usepackage{amsmath}	
\usepackage{amssymb}	
\usepackage{multicol}        
\usepackage{pdflscape}	
\usepackage{mathptmx}
\usepackage{txfonts}
\usepackage{tikz}

\usepackage[T1]{fontenc}
\usepackage{ae,aecompl}

\usepackage{newtxtext}
\usepackage{multirow}
\usepackage{multicol}
\usepackage{subcaption}
\usepackage{url}
\usepackage{gensymb}
\usepackage{adjustbox}
\usepackage{booktabs} \newcommand{\ra}[1]{\renewcommand{\arraystretch}{#1}}
\usepackage[font=scriptsize]{caption} 
\newcommand{\HII}{{H}{\sc {ii}}}
\newcommand{\mm}{$\mu$m}

\usepackage{mathptmx}
\usepackage{txfonts}

\usepackage[T1]{fontenc}

\DeclareRobustCommand{\VAN}[3]{#2}
\let\VANthebibliography\thebibliography
\def\thebibliography{\DeclareRobustCommand{\VAN}[3]{##3}\VANthebibliography}

\usepackage{graphicx}	
\usepackage{amsmath}	
\usepackage{amssymb}	
\usepackage{amstext}
\usepackage{amsopn}

\usepackage{amssymb}	

\title[Machine Learning Approaches for classifying SFGs and AGN]{Machine Learning Approaches for Classifying Star-Forming Galaxies and Active Galactic Nuclei from MIGHTEE-Detected Radio Sources in the COSMOS Field}

\author[W. Silima et al.]{
Walter Silima,$^{1,2}$  
Fangxia An,$^{3,4,1}$\thanks{E-mail: anfangxia@ynao.ac.cn; fangxiaan@gmail.com}
Mattia Vaccari,$^{2,1,5}$
Eslam A. Hussein,$^{1}$ 
S. Randriamampandry$^{6}$
\\
$^{1}$Inter-University Institute for Data Intensive Astronomy (IDIA), Department of Physics and Astronomy, University of the Western Cape, 7535 Bellville, Cape Town,\\ South Africa\\
$^{2}$Inter-University Institute for Data Intensive Astronomy (IDIA), Department of Astronomy, University of Cape Town, 7701 Rondebosch, Cape Town, South Africa\\
$^{3}$Yunnan Observatories, Chinese Academy of Sciences, Kunming 650216, People’s Republic of China\\
$^{4}$Purple Mountain Observatory, Chinese Academy of Sciences, 10 Yuanhua Road, Qixia District, Nanjing 210023, People's Republic of China\\
$^{5}$INAF - Istituto di Radioastronomia, via Gobetti 101, 40129 Bologna, Italy\\
$^{6}$A\&A, Department of Physics, Faculty of Sciences, University of Antananarivo, P.O. Box 906, Antananarivo 101, Madagascar\\
}

\date{Accepted 2025 September 26. Received 2025 September 25; in original form 2025 April 10}

\pubyear{2025}

\begin{document}
\label{firstpage} 
\pagerange{\pageref{firstpage}--\pageref{lastpage}}
\maketitle

\begin{abstract}
    Radio synchrotron emission originates from both massive star formation and black hole accretion, two processes that drive galaxy evolution. Efficient classification of sources dominated by either process is therefore essential for fully exploiting deep, wide-field extragalactic radio continuum surveys. In this study, we implement, optimize, and compare five widely used supervised machine-learning (ML) algorithms to classify radio sources detected in the MeerKAT International GHz Tiered Extragalactic Exploration (MIGHTEE)–COSMOS survey as star-forming galaxies (SFGs) and active galactic nuclei (AGN). Training and test sets are constructed from conventionally classified MIGHTEE-COSMOS sources, and 18 physical parameters of the MIGHTEE-detected sources are evaluated as input features. As anticipated, our feature analyses rank the five parameters used in conventional classification as the most effective: the infrared-radio correlation parameter ($q_\mathrm{IR}$), the optical compactness morphology parameter (class$\_$star), stellar mass, and two combined mid-infrared colors. By optimizing the ML models with these selected features and testing classifiers across various feature combinations, we find that model performance generally improves as additional features are incorporated. 
Overall, all five algorithms yield an $F1$-score (the harmonic mean of precision and recall) $>90\%$ even when trained on only 20\% of the dataset. Among them, the distance-based $k$-nearest neighbors classifier demonstrates the highest accuracy and stability, establishing it as a robust and effective method for classifying SFGs and AGN in upcoming large radio continuum surveys.
\end{abstract}

\begin{keywords}
methods: observational -- software: machine learning -- galaxies: evolution -- galaxies: formation -- radio continuum: galaxies
\end{keywords}



\section{Introduction}
The radio continuum emission from galaxies is powered by star formation (SF) and black hole accretion, the two dominant physical processes that drive galaxy evolution. SF-related radio emission originates from supernova-accelerated cosmic ray (CR) electrons gyrating within galactic magnetic fields, producing non-thermal synchrotron radiation, and from Coulomb scattering between free ions and electrons in \HII\ regions, resulting in thermal free-free emission. Both synchrotron and free-free emissions remain unaffected by dust obscuration, which is critical for obtaining an unobstructed view of SF in galaxies \citep[see][for a review]{Condon92}. The synchrotron emission from relativistic jets and outflows powered by black hole (BH) accretion dominates the radio emission of luminous radio sources, namely radio galaxies \citep{Sadler89, Miley08}. The feedback processes associated with BH accretion play a crucial role in regulating galaxy growth, with jets and outflows potentially expelling star-forming gas from galactic bulges and quenching star formation in galaxies. Consequently, distinguishing between SF-dominated and active galactic nuclei (AGN)-dominated radio emission is crucial for utilizing the radio continuum in exploring cosmic evolution. 

Newly constructed and upgraded radio interferometric arrays in the past two decades, such as the Australian Square Kilometre Array Pathfinder \citep[ASKAP,][]{Hotan21}, Murchison Widefield Array \citep[MWA,][]{Lonsdale09}, MeerKAT \citep{Jonas16}, the Low Frequency Array \citep[LOFAR,][]{van13}, and upgraded Giant Metrewave Radio Telescope \citep[uGMRT,][]{Swarup91}, have led to a new generation of large extragalactic radio continuum surveys \citep[e.g.,][]{Ocran20, Ishwara20, heywood2022mightee, Best23, Hale24}. Some of these deep surveys achieve an angular resolution of $\la$ 5$\arcsec$ \citep[e.g.,][]{Smolcic17, Jimenez24}, while the majority of current extragalactic radio continuum surveys operate at resolutions of $\sim$6-10$\arcsec$, with sensitivities reaching the $\mu$Jy-level. Combined with survey areas covering dozens of square degrees, these surveys have led to an exponential increase in the number of radio sources detected over the past two decades \citep{Norris17, Best23, Hale24}. This rapid expansion in data volume demands the development of efficient and automated techniques to classify the sources detected from these surveys as SF-dominated or AGN-dominated before further investigating their physical nature.

Machine learning (ML) is firmly established in astronomy and has been widely used in various research areas, such as galaxy (morphology) classification \citep[e.g.,][]{BALL_2010, An18}, discovery/prediction of astrophysical activities \citep[e.g.,][]{Florios18, Mahabal19}, estimation of photometric redshifts \citep[e.g.,][]{Li23}, noise analysis in gravitational wave detection \citep[e.g.,][]{Biswas13, George18}, and for many other applications \citep[see][for a review]{Fluke20}. Automated classification has recently been adopted in SFG-AGN separation but with only one particular ML algorithm \citep{Karsten23}. 

In this work, we implement and optimize five widely used supervised ML algorithms, namely Logistic Regression \citep[LR,][]{menard2010logistic}, Support Vector Machine \citep[SVM,][]{cristianini2000introduction}, K-Nearest Neighbour \citep[\textit{k}NN,][]{peterson2009k}, Random Forest \citep[RF,][]{breiman2001random}, and Extreme Gradient Boosting, commonly known as XGBoost \citep[XGB,][]{chen2016xgboost}, to classify SF-dominated or AGN-dominated radio sources from the MeerKAT International GHz Tiered Extragalactic Exploration (MIGHTEE) survey \citep{jarvis2017meerkat, heywood2022mightee,Hale24}. Since these ML models are based on distinct algorithmic approaches, we also aim to assess their relative effectiveness in classifying star-forming galaxies (SFGs) and AGN from radio continuum surveys. Sources detected from the MIGHTEE-COSMOS early science data have been classified as SF-dominated or accretion-dominated following the traditional SFG/AGN classification diagnostic \citep{whittam2022mightee}. We use these classifications to construct the training set and optimize the different ML algorithms.

The successful adoption of ML will efficiently provide accurate SFGs/AGN samples, which is essential for scientific studies based on recently completed or ongoing high-sensitivity and wide-field extragalactic radio continuum surveys, and eventually, the surveys conducted by the Square Kilometre Array \citep[SKA,][]{Dewdney09}, next-generation $Karl\,G.\,Jansky$ Very Large Array \citep[ngVLA,][]{Murphy18}, and the Five-hundred-meter Aperture Spherical Radio Telescope (FAST) Core Array \citep{Jiang24}.

We describe the MIGHTEE-COSMOS data as well as the ancillary data used in this work in Section $\S$\ref{s:data}. The data analyses and feature selection of ML are described in Section $\S$\ref{s:analyses}. We show the results of our ML application in Section $\S$\ref{s:results}. 
Our results are discussed and summarized in Sections $\S$\ref{s:discussion} and $\S$\ref{s:conclusion} respectively. Throughout this paper, we adopt the AB magnitude system \citep{Oke74} and assume a flat $\Lambda$CDM cosmological model with the Hubble constant $H_0$\,= 67.27\,km\,s$^{-1}$\,Mpc$^{-1}$, matter density parameter $\Omega_{\rm m}$\,=0.32, and cosmological constant $\Omega_{\Lambda}$=\,0.68 \citep{Planck16}.

\section{MIGHTEE-COSMOS Data}\label{s:data}
The MIGHTEE survey is one of the MeerKAT large survey projects, conducted by an international collaboration of researchers. 
MIGHTEE targets four extensively studied extragalactic fields: the Cosmological Evolution Survey (COSMOS) field, the Extended Chandra Deep Field-South (E-CDFS), the European Large Area Infrared Survey South 1 (ELAIS-S1) field, and the XMM-Newton Large Scale Structure (XMM-LSS) field, covering a total of 20 square degrees with $\mu$Jy-level sensitivity \citep{jarvis2017meerkat}. The survey includes deep GHz radio continuum \citep{heywood2022mightee, Hale24}, spectral line \citep{Maddox21}, and polarization \citep{Taylor24} observations, aimed at exploring cosmic evolution.

This work utilizes MIGHTEE early science radio continuum data in the COSMOS field, as released and fully described by \cite{heywood2022mightee}. The COSMOS field was observed for a total of 17.45 hours on source between 2018 and 2019 using MeerKAT's L-band receivers (856--1712\,MHz), with a single pointing centered at RA=10$^{\rm h}$00$^{\rm m}$28.6$^{\rm s}$, Dec =\,+02$^{\rm d}$12$^{\rm m}$21$^{\rm s}$. The MIGHTEE-COSMOS early science radio data were processed with Briggs' robust weighting values of 0.0 and -1.2 \citep{1995AAS...18711202B}. The former yielded more sensitive imaging data with a thermal noise of 1.7\,$\mu$Jy\,beam$^{-1}$ and a circular synthesized beam size of 8.6$\arcsec\times$8.6$\arcsec$. It is important to note that the high-sensitivity data are limited by classical confusion at the center, increasing the mean noise to 4-5\,$\mu$Jy\,beam$^{-1}$ \citep{heywood2022mightee}. While the full coverage of the MIGHTEE-COSMOS early science data spans 1.6\,deg$^{2}$, we restrict the analyses within the central 0.86\,deg$^{2}$, where the radio data are deepest and multi-wavelength cross-matching has been completed for the MIGHTEE sources \citep{Whittam24}.

\subsection{MIGHTEE-COSMOS Multi-wavelength catalogue}
\label{sec: catalogue}
As described in \citep{Whittam24}, there are 6102 radio components with peak brightnesses that exceed the local background noise by 5$\sigma_{\rm local}$ within the central 0.86\,deg$^{2}$ of the MIGHTEE-COSMOS field. \cite{whittam2022mightee,Whittam24} identified the host galaxy for 5223 out of 6102 radio-detected sources by visual cross-matching the MIGHTEE sources with $K_{\rm s}$-band-detected sources from the fourth data release (DR4) of the UltraVISTA survey \citep{bowler2020lack, Adams21}. Details of the visual cross-matching are presented in \cite{Whittam24}.

Using the position of the host galaxies, \cite{whittam2022mightee,Whittam24} also identified multi-wavelength counterparts for the 5223 radio sources detected in the MIGHTEE survey. Here, we briefly summarize the identified multi-wavelength counterparts of MIGHTEE sources as reported by \cite{whittam2022mightee,Whittam24}. Of the 5223 MIGHTEE sources with UltraVISTA $K_{\rm s}$-band counterparts, 572 (11\%) were detected in X-ray observations. The optical, near-infrared (NIR) counterparts of MIGHTEE sources were identified using the optical and near-infrared broad-band photometric catalogue created by \cite{Adams21}, which include $YJHK_{\rm s}$-band data from the UltraVISTA DR4, as well as \textit{grizy}-bands data from Hyper Suprime-Cam Subaru Strategic Program \citep[HSC SSP;][]{Tanaka17}, deep $u^*$-band data from the Canada–France–Hawaii Telescope Legacy Survey \citep[CFHTLS;][]{Cuillandre12}, and mid-infrared (MIR) data from the Spitzer Infrared Array Camera (IRAC) at 3.6 and 4.5\,\mm. \cite{whittam2022mightee} also used the high-resolution \textit{Hubble Space Telescope (HST)} Advanced Camera for Surveys (ACS) $I$-band imaging data \citep{Scoville07} and found 4697 out of 5223 MIGHTEE sources have \textit{HST I}-band counterparts. Furthermore, to identify the MIR counterparts of MIGHTEE sources, \citet{whittam2022mightee} used data at 5.8 and 8.0\,\mm\ from the Spitzer Large-Area Survey with Hyper-Supprime-Cam \citep[SPLASH;][]{Steinhardt14}, accessed through the COSMOS2015 catalog \citep{laigle2016cosmos2015}. As a result, 4815 of the 5223 MIGHTEE sources have detections at 5.8 and 8.0\,\mm. Far-infrared (FIR) counterparts were identified using data from the Herschel Extragalactic Legacy Project \citep[HELP;][]{Vaccari2016Universe, Shirley21}. Among the 5223 MIGHTEE sources, 4540 were detected at 24, 100, and 160\,\mm\ using the Multiband Imaging Photometer \citep[MIPS;][]{Rieke04} on the Spitzer Space Telescope and the Photodetector Array Camera and Spectrometer \citep[PACS;][]{Poglitsch10}. Additionally, 4957 out of the 5223 sources were detected at 250, 350, and 500\,\mm\ using the Spectral and Photometric Imaging Receiver \citep[SPIRE;][]{Griffin10} on Herschel. Furthermore, \cite{whittam2022mightee} cross-matched the optical positions of MIGHTEE sources with very long baseline interferometry (VLBI) observed sources, finding that 255 of the 5223 sources have VLBI detections \citep{Herrera17}.

This work also uses the estimated redshift and stellar mass of MIGHTEE sources from the MIGHTEE-COSMOS multi-wavelength catalogue. As reported by \cite{Whittam24}, 2427 of the 5223 MIGHTEE sources have spectroscopic redshifts compiled from the literature. For the remaining 2796 sources, their photometric redshifts were determined by \cite{Hatfield22} using a hierarchical Bayesian approach that integrates two distinct methodologies, as detailed by \cite{Duncan18}. The stellar masses of MIGHTEE sources were estimated using AGNFITTER SED-fitting. Details of the estimation and comparisons of stellar mass estimates across different SED-fitting codes are presented in \cite{whittam2022mightee}.

\subsection{MIGHTEE-COSMOS Conventional Classification}
\label{sec: manual classification}
Utilizing the well-matched MIGHTEE-COSMOS multi-wavelength catalogue, \cite{whittam2022mightee} classified AGN and SFGs from the MIGHTEE-COSMOS survey by using five conventional classification techniques: radio excess, MIR colour-colour, optical morphology, X-ray luminosity, and the VLBI criteria \citep[Table 1,][]{whittam2022mightee}. 

Radio-excess AGN were identified as sources with significantly more radio emission than expected from star formation alone, determined by the infrared-radio correlation (IRRC) quantified as $q_\mathrm{IR}$ in \cite{whittam2022mightee}. The MIR colour-colour diagram defined by \cite{2012ApJ...748..142D} was used to identify sources exhibiting power-law emissions from the torus, classifying them as MIR AGN in \cite{whittam2022mightee}. Optical point-like AGN were identified using \textit{HST} ACS \textit{I}-band imaging, based on the principle that the emission from the nucleus outshines that of the host galaxy. Sources with a Source-Extractor (\href{https://www.astromatic.net/software/sextractor/}{SExtractor}) compactness parameter, class$\_$star, $\geq$ 0.9 were classified as optical point-like AGN in \cite{whittam2022mightee}. As some of the brightest AGN exhibit characteristic accretion-related X-ray emissions, X-ray AGN were identified by applying a rest-frame (0.5-10\,keV) X-ray luminosity threshold of $L_x \ge  10^{42}$\,erg\,$s^{-1}$ \citep{2004ApJS..155..271S}. Finally, VLBI AGN were classified as sources with a brightness temperature exceeding that of typical SFGs. We refer the reader to \cite{whittam2022mightee} for details about the conventional classification of MIGHTEE-COSMOS radio sources. 

\cite{whittam2022mightee} classified a source as an AGN if it satisfied any one (or more) of AGN criteria. Sources that did not meet any of the AGN criteria across all five diagnostic methods were classified as SFGs. However, due to the limited depth of the X-ray observations, only sources with $z < 0.5$ could be confidently classified as `not X-ray AGN' if they were undetected in the X-ray. For X-ray undetected sources with $z>0.5$, their potential X-ray luminosity might be above the classification threshold of $L_x \ge 10^{42}$\,erg\,$s^{-1}$, and therefore, are unable to fulfill the `not X-ray AGN' criteria. If these sources were classified as `not AGN' based on the other four diagnostics, \citet{whittam2022mightee} introduced an additional category, namely `probable SFG'. 

Figure~\href {fig:1}{1} shows the overall classification completeness and the completeness for each diagnostic method of MIGHTEE-COSMOS detected radio sources. Due to the low completeness and unpredictability of the X-ray and VLBI classifications, these two features are excluded from our ML analysis. Consequently, we merge the `probable SFG' category into the `SFG' class in this work. Table \ref{tab:mightee classes} summarizes the number of sources in each class used in this work.

\begin{figure}
    \centering
    \includegraphics[width=1\linewidth]{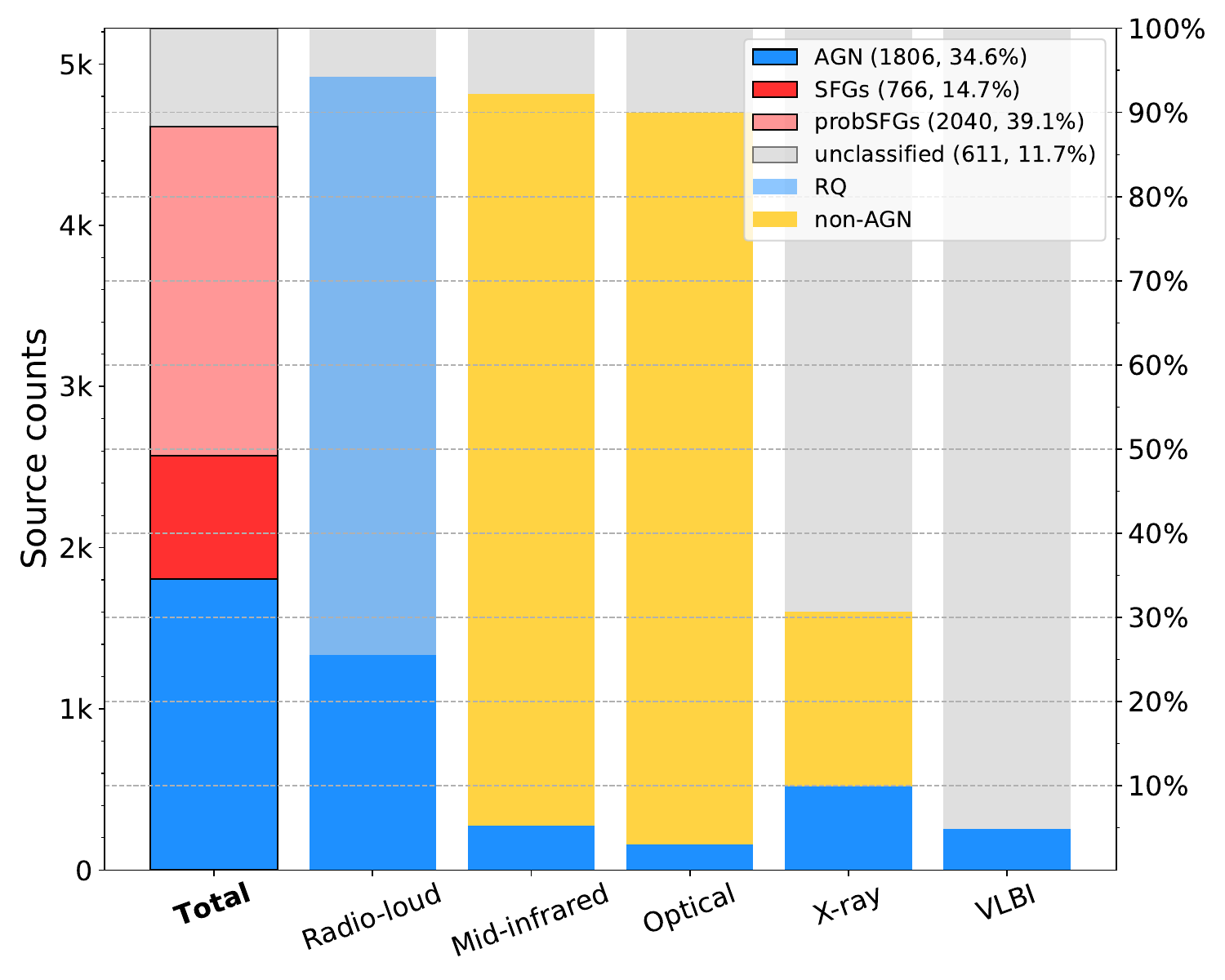}
    \label{fig:1}
    \caption{The bar plot illustrates the completeness of the overall classification (total) and the completeness for each diagnostic method of MIGHTEE-COSMOS detected radio sources. The categories of sources are colour-coded, with AGN shown in blue, SFGs in red, probable SFGs in light red, radio quiet (RQ) in light blue, non-AGN in yellow, and unclassified sources in grey.}
\end{figure}

\begin{table}
    \centering
    \caption{Number of MIGHTEE-COSMOS sources per class}
    \label{tab:mightee classes}
    \ra{1.3}
    \begin{tabular}{@{}rrrrcrrrcrrr@{}}\toprule
    
    Overall Class & Number of Sources \\\midrule
    
    AGN & 1806 \\ 
    SFG & 2806 \\
    Unclassified & 611\\\bottomrule
    \end{tabular}
\end{table}

\section{Analyses}\label{s:analyses}
A successful supervised classification relies on the quality and representativeness of the dataset used in model development, the careful selection and tuning of adjustable model parameters, and the robustness of the evaluation criteria (Section~$\S$\ref{metric}) employed to assess the model's performance. 
To ensure optimal results, we adhere to the standard workflow for supervised ML, which consists of the following steps: 
\begin{enumerate}
\item feature analysis and selection ( Sections~$\S$\ref{s:f-analyses}, $\S$\ref{s:3.3}, and $\S$\ref{s:selection}),
\item building a training set using the selected features (Section~$\S$\ref{s:SV_ML}), training of the ML model to create a classifier, and hyperparameter optimization (Section~$\S$\ref{subs:hyperparameter}), and
\item applying the classifier to predict the class labels of the test sample. However, in this study, we evaluate the performance of ML classification using only the validation dataset (Section~$\S$\ref{s:results}).
\end{enumerate}

\subsection{Evaluation Metrics}
\label{metric}
Evaluation metrics are used to evaluate the performance of ML models. In this work, we adopt the classification metrics \textit{Precision}, \textit{Recall} and \textit{$F1$-score}, with the latter being particularly effective for imbalanced datasets \citep{hossin2015review,unbalancef1}.

Table \ref{tab:matrix} outlines the confusion matrix for a binary classification scenario. If we consider the AGN class as positive and the SFG class as negative, a True Positive (TP) refers to the number of labeled AGN that are correctly classified by the ML models. Conversely, False Positives (FP) represent cases where SFGs are incorrectly identified as AGN. 
Similarly, False Negatives (FN) occur when AGN are misclassified as SFGs, while True Negative (TN) refers to the number of SFGs that are accurately classified by the ML models.

\begin{table}
    \centering
    \caption{Confusion matrix for binary classification. Actual represents the actual labeled class (in our case, the manually labeled class, i.e. AGN or SFG), while predicted represents the class predicted by the ML algorithms.} \label{tab:matrix}
    \begin{adjustbox}{width=0.45\textwidth}
        \begin{tabular}{ |c|c|c|c| }
    \cline{3-4}    
    \multicolumn{2}{c|}{} & \multicolumn{2}{c|}{Actual} \\ \cline{3-4}
    \multicolumn{2}{c|}{} & Positive & Negative\\ \cline{1-4}
    \multirow{2}{*}{ Predicted}  & Positive &\textcolor{blue}{True Positive (TP)} & \textcolor{red}{False Positive (FP)} \\ \cline{2-4}
    {} & Negative &\textcolor{red}{False Negative (FN)} & \textcolor{blue}{True Negative (TN)}\\ \cline{1-4}
        \end{tabular}
    \end{adjustbox}
\end{table}

The classification metrics are derived from the confusion matrices. \textit{Precision} quantifies the accuracy of positive predictions made by the ML models, defined as:
\begin{equation} \label{e:equation1}
\centering
Precision=\dfrac{\rm TP}{\rm TP+FP},
\end{equation}
whereas \textit{Recall} evaluates the model's ability to minimize false negatives, expressed as:
\begin{equation} \label{e:equation2}
    Recall = \frac{\rm TP} {\rm TP + FN}.
\end{equation}
Lastly, the \textit{$F1$-score}, which represents the harmonic mean of \textit{Precision} and \textit{Recall}, offers a comprehensive measure of the model's performance. It is defined as: 
\begin{equation} \label{e:equation3}
    F1  = \frac{\rm 2TP}{\rm (2TP + FP + FN)},
\end{equation}
and utilised as the main evaluation metric in this work (Section~$\S$\ref{s:results}).

\subsection{Feature Analysis}
\label{s:f-analyses}
As outlined in Table \ref{tab:mightee classes}, \cite{whittam2022mightee} classified a total of 1806 AGN and 2806 SFGs from the MIGHTEE-COSMOS survey, utilizing five conventional classification diagnostics. This labeled sample of 4612 sources is used to construct training and test datasets, enabling the evaluation of ML models based on various input features. As described in Appendix \ref{A:feature-completeness} and shown in Figure~\ref{fig:allperm}, in this study, we incorporate all available photometric data from the MIGHTEE-COSMOS multi-wavelength catalogue \citep{Whittam24}, alongside the conventional classification diagnostics used in \citet{whittam2022mightee}, to evaluate their efficiency in distinguishing between SFGs and AGN in the MIGHTEE-COSMOS survey.


\begin{figure}
    \begin{subfigure}{0.23\textwidth}
        \includegraphics[width=\linewidth]{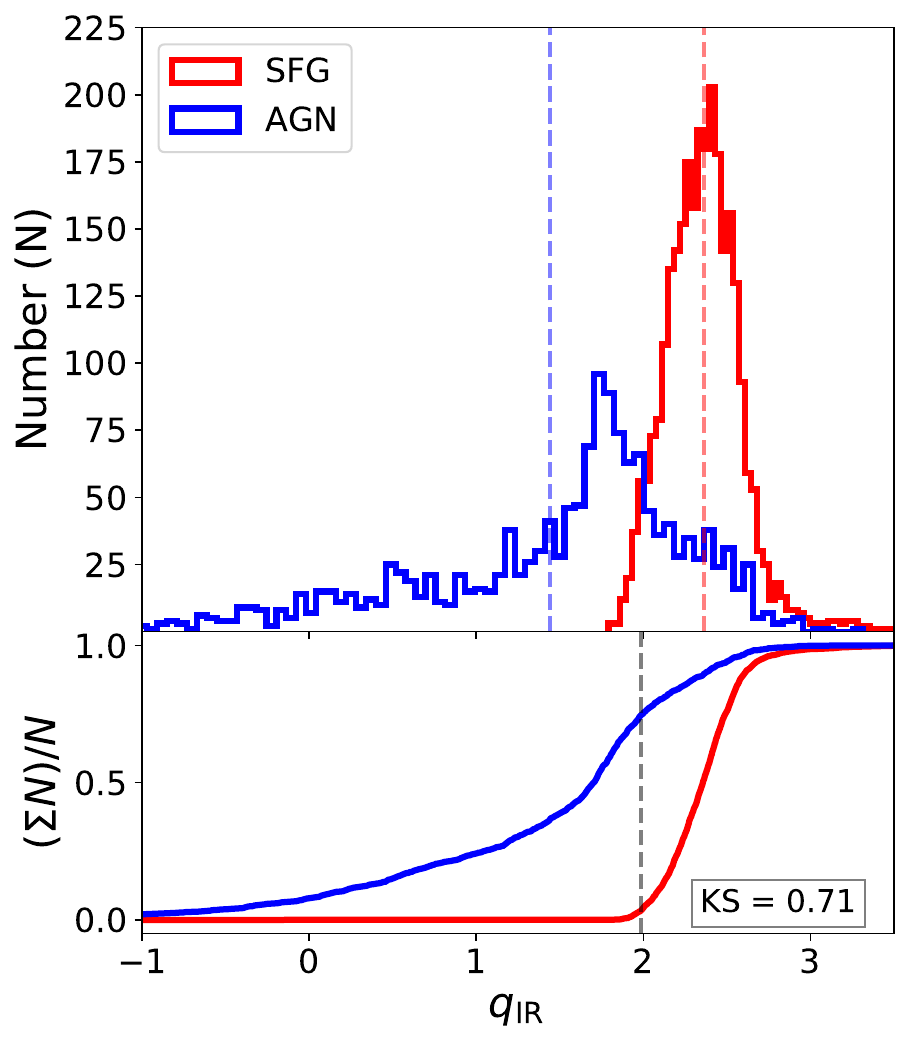}
        \label{fig:4.2a}
    \end{subfigure}\hfill
    \begin{subfigure}{0.24\textwidth}
        \includegraphics[width=\linewidth]{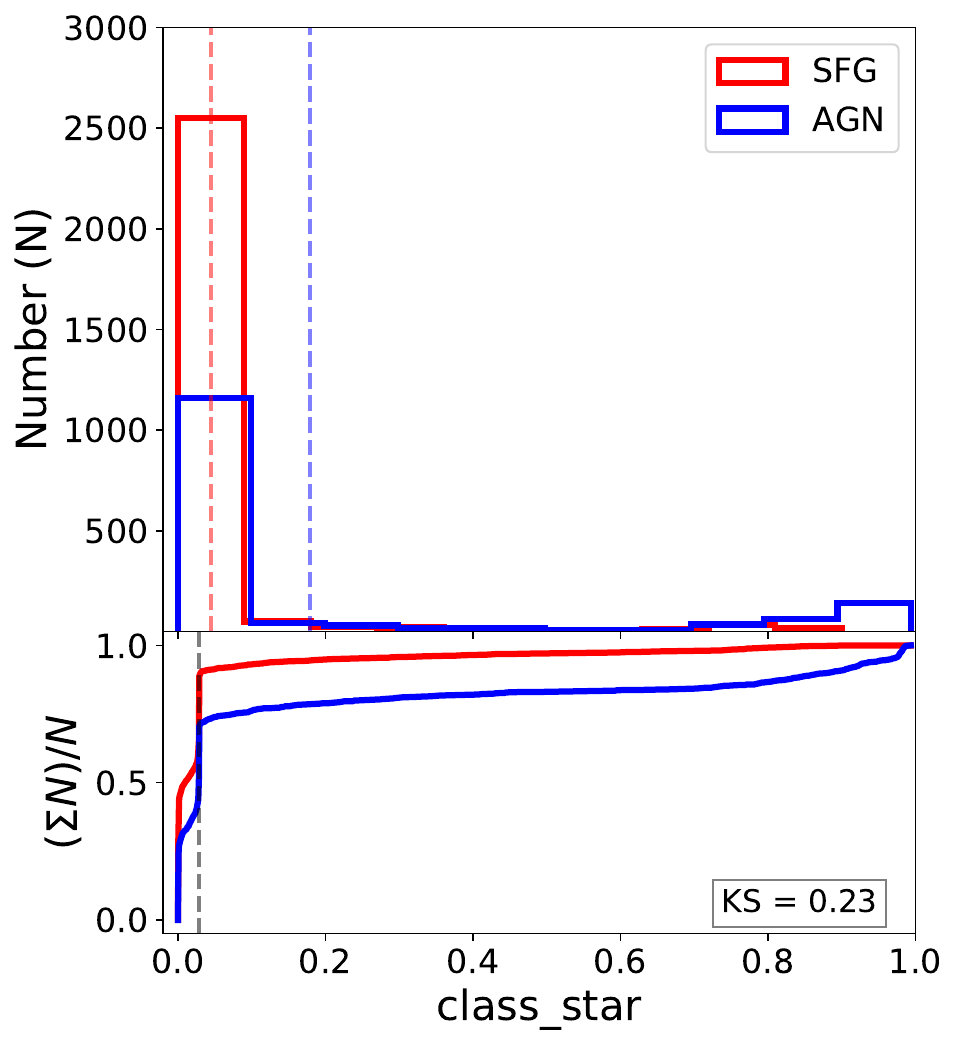}
        \label{fig:4.2b}
    \end{subfigure}
    \begin{subfigure}{0.23\textwidth}
        \includegraphics[width=\linewidth]{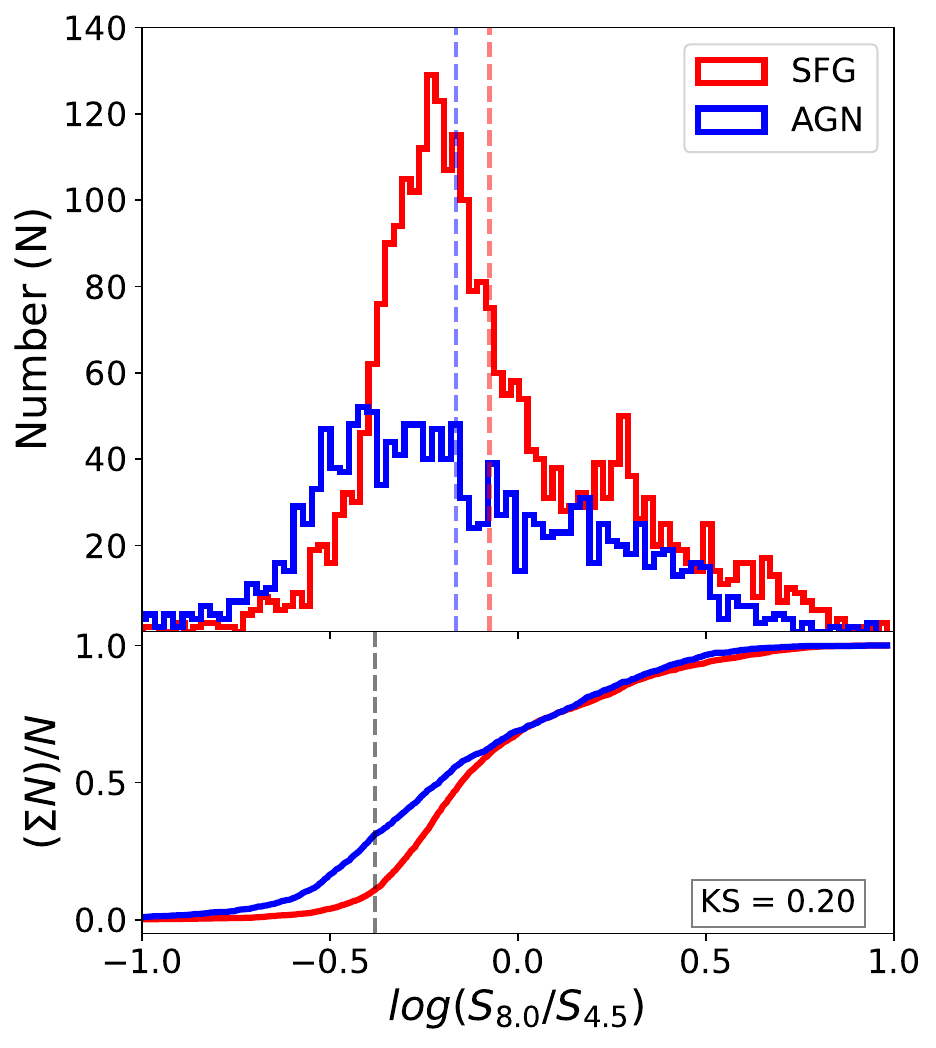}
        \label{fig:4.2c}
    \end{subfigure}\hfill
    \begin{subfigure}{0.23\textwidth}
        \includegraphics[width=\linewidth]{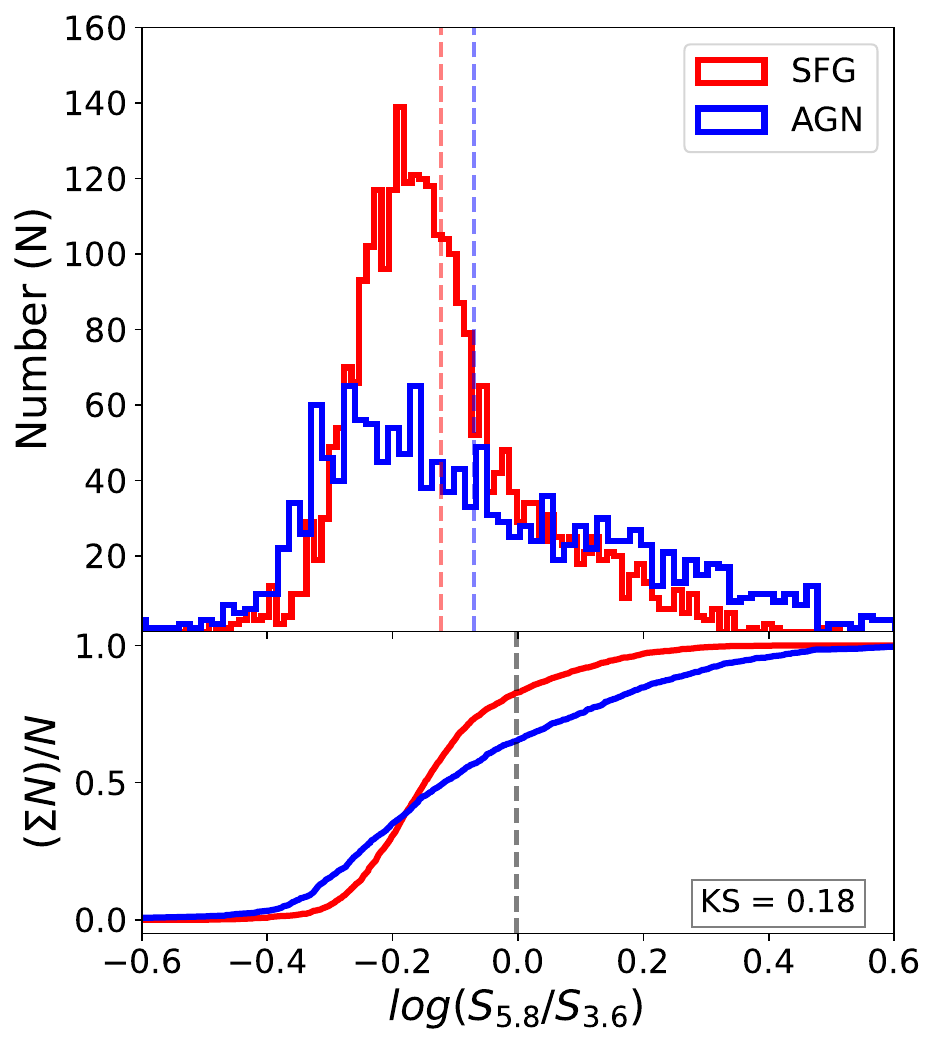}
        \label{fig:4.2d}
    \end{subfigure}
    \begin{subfigure}{0.23\textwidth}
        \includegraphics[width=\linewidth]{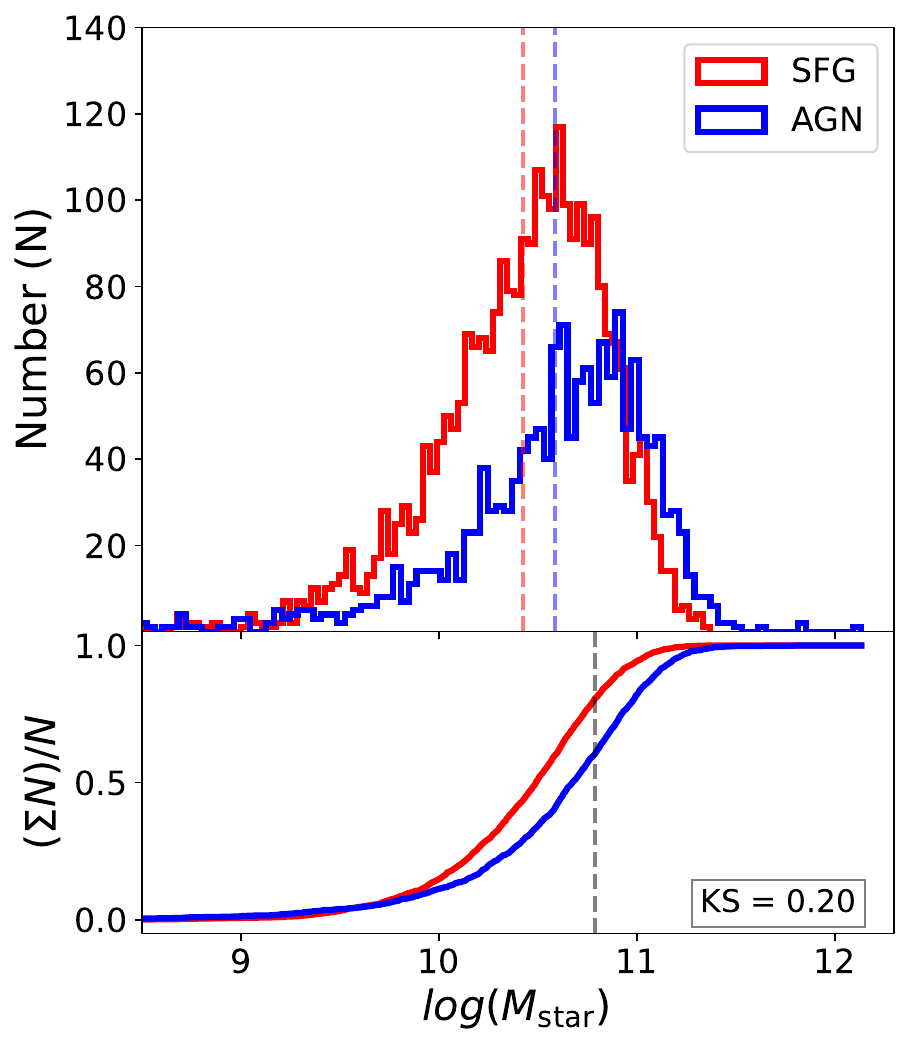}
        \label{fig:4.2e}
    \end{subfigure}\hfill
    \begin{subfigure}{0.23\textwidth}
        \includegraphics[width=\linewidth]{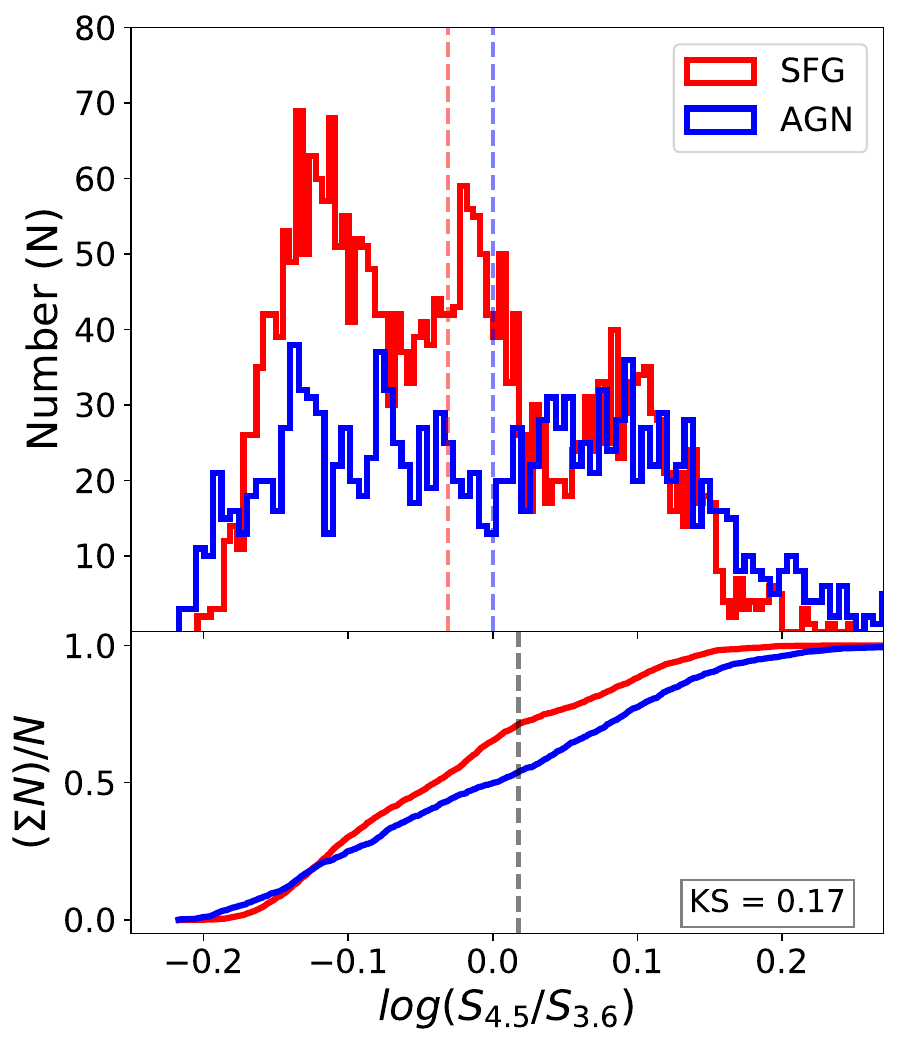}
        \label{fig:4.2f}
    \end{subfigure}

    \caption{Histograms (top) and Kolmogorov–Smirnov (K-S) test results (bottom) for AGN (blue) and SFGs (red), in the MIGHTEE-COSMOS catalog. Among the 18 parameters considered for selecting input features for ML, these six exhibit the highest significance based on the K-S statistic. The K-S value for each feature is displayed in the bottom-right corner of each panel. The features are sorted by the significance level of the K-S statistic (from left to right, top to bottom). In the top panels, vertical dashed lines indicate the mean of each distribution. In the bottom panels, the Y-axis shows the cumulative distribution function (CDF), with vertical dashed lines marking the point of maximum separation between the two distributions.}
    \label{fig:4.2}
\end{figure}

\subsubsection{One Dimensional Analysis}\label{s:3.2.1}
The performance of supervised ML models is highly dependent on the selection of features used for training. Effective feature selection reduces the dimensions of the data, enabling the model to perform more efficiently. However, identifying the most critical features is a non-trivial task and often requires advanced ML techniques. This section examines two methods for analyzing a low-dimensional feature space.

A straightforward method for identifying key features in binary classification is to examine the histograms of each feature for the two classes \citep{zhang2003data}, as illustrated in Figure \ref{fig:4.2}. Greater separation between the distributions of the two classes indicates that the feature is more effective in distinguishing between them. A quantitative measure of this separation is the difference in the \textit{means} of the two distributions (Figure~\ref{fig:4.2}). Additionally, in Figure \ref{fig:4.2}, we also present the results of Kolmogorov–Smirnov (K-S) tests, which assess the statistical differences between the two populations for each feature \citep{berger2014kolmogorov}.

To determine the input features for training ML models, we first incorporate all twelve-band optical to MIR photometries from the MIGHTEE-COSMOS multi-wavelength catalogue (Section~$\S$\ref{sec: catalogue}), including four HSC $griz$-band flux densities, four UltraVISTA $YJHK_{\rm s}$-band photometries, and four IRAC 3.6, 4.5, 5.8, and 8.0\,\mm-band flux densities. From these photometries, we derive 15 color indices: three MIR colors (log(S$_{8.0}$/S$_{4.5}$), log(S$_{5.8}$/S$_{3.6}$), log(S$_{4.5}$/S$_{3.6}$)) and 12 NIR and optical colors. A full description of these color indices is provided in Appendix \ref{A:feature-completeness}. 

Additionally, we include other measurements available in the MIGHTEE-COSMOS multi-wavelength catalogue, particularly the conventional classification diagnostics used by \cite{whittam2022mightee}, such as the IRRC parameter $q_\mathrm{IR}$, stellar mass log($M_{\rm star}$), and the optical compactness parameter class$\_$star. As discussed in Section \S\ref{sec: manual classification}, we exclude X-ray luminosity and VLBI detection from the input features due to their low completeness and the unpredictability.

In total, we include 18 physical parameters in our analysis. Figure~\ref{fig:4.2} highlights the six features with the greatest significance level based on the K-S statistic. The K-S value for each feature is displayed in the bottom-right corner of each panel in Figure~\ref{fig:4.2}, indicating that these six features exhibit the greatest differences in cumulative distribution functions (CDFs) between SFGs and AGN. As illustrated in Figure \ref{fig:4.2}, both the histogram and K-S test results demonstrate that $q_\mathrm{IR}$ is the most discriminative feature for differentiating SFGs from AGN among the MIGHTEE-detected radio sources, followed by the optical compactness parameter class$\_$star. Stellar mass (log($M_{\rm star}$)), along with three IRAC colours show slight variations in ranking between the two methods, yet consistently rank among the most effective features for classifying SFGs and AGN in the MIGHTEE dataset.

However, as shown in Figure~\ref{fig:4.2}, while clear differences exist between the distributions of SFGs and AGN across many features, substantial overlap occurs in individual features. Nonetheless, as demonstrated by our subsequent analyses and the results presented in Section~$\S$\ref{s:results}, the performance of all ML models improves by incorporating multiple features when classifying radio sources as SFGs or AGN.

 \begin{figure}
     \begin{subfigure}{0.24\textwidth}
         \includegraphics[width=\linewidth]{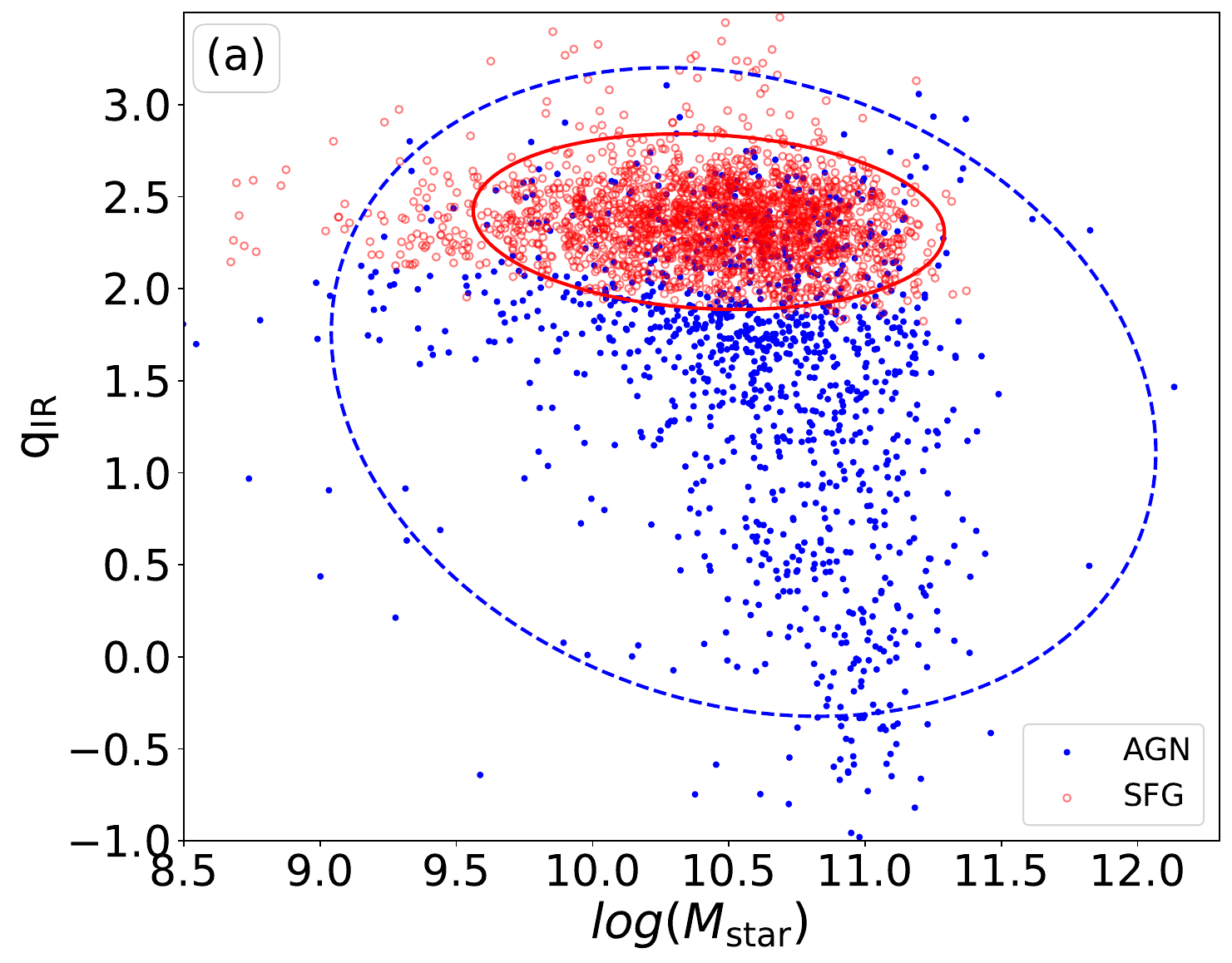}
         \phantomcaption\label{fig:4.3a}
     \end{subfigure}\hfill 
     \begin{subfigure}{0.24\textwidth}
         \includegraphics[width=\linewidth]{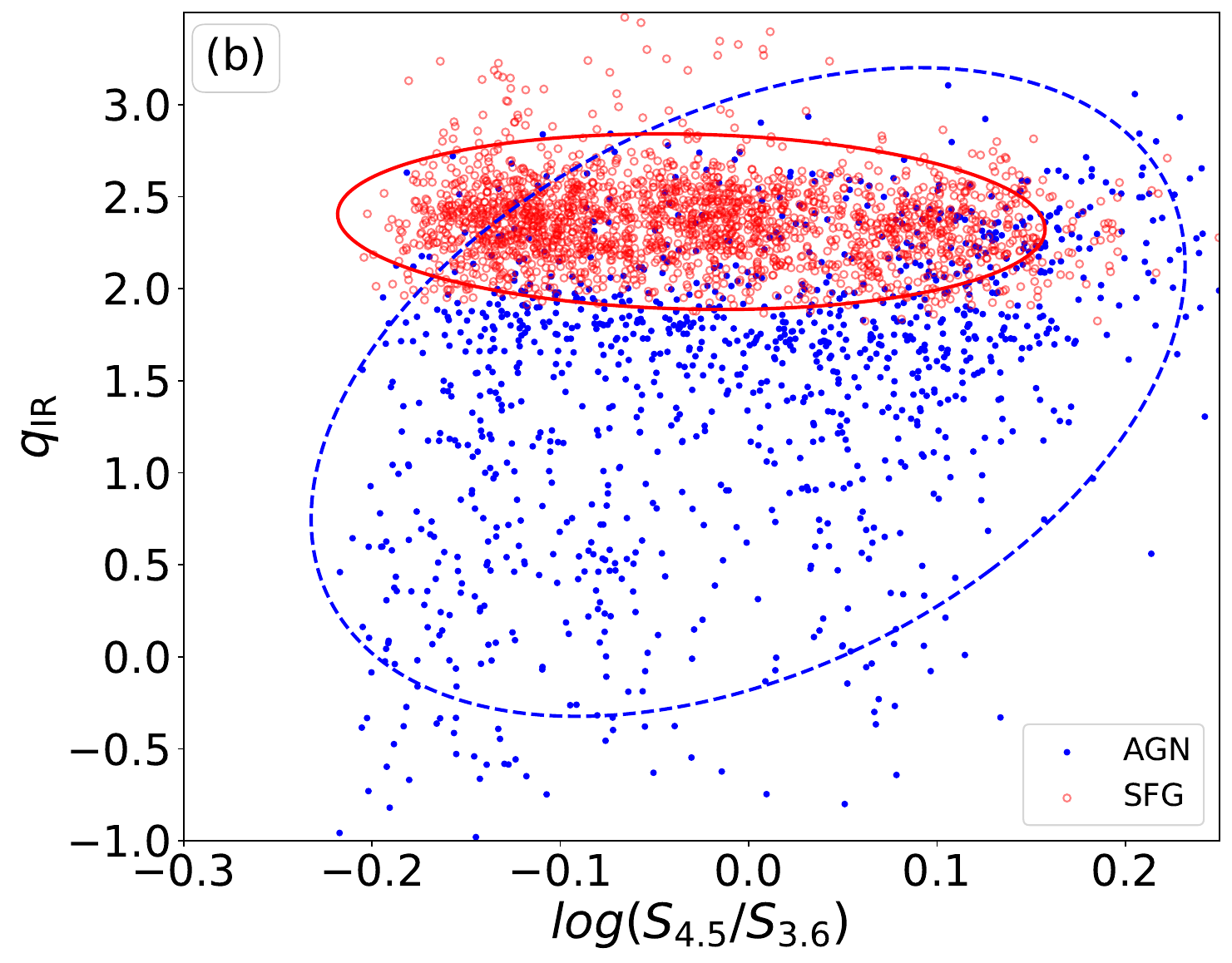}
         \phantomcaption\label{fig:4.3b}
     \end{subfigure}\hfill
     \begin{subfigure}{0.24\textwidth}
         \includegraphics[width=\linewidth]{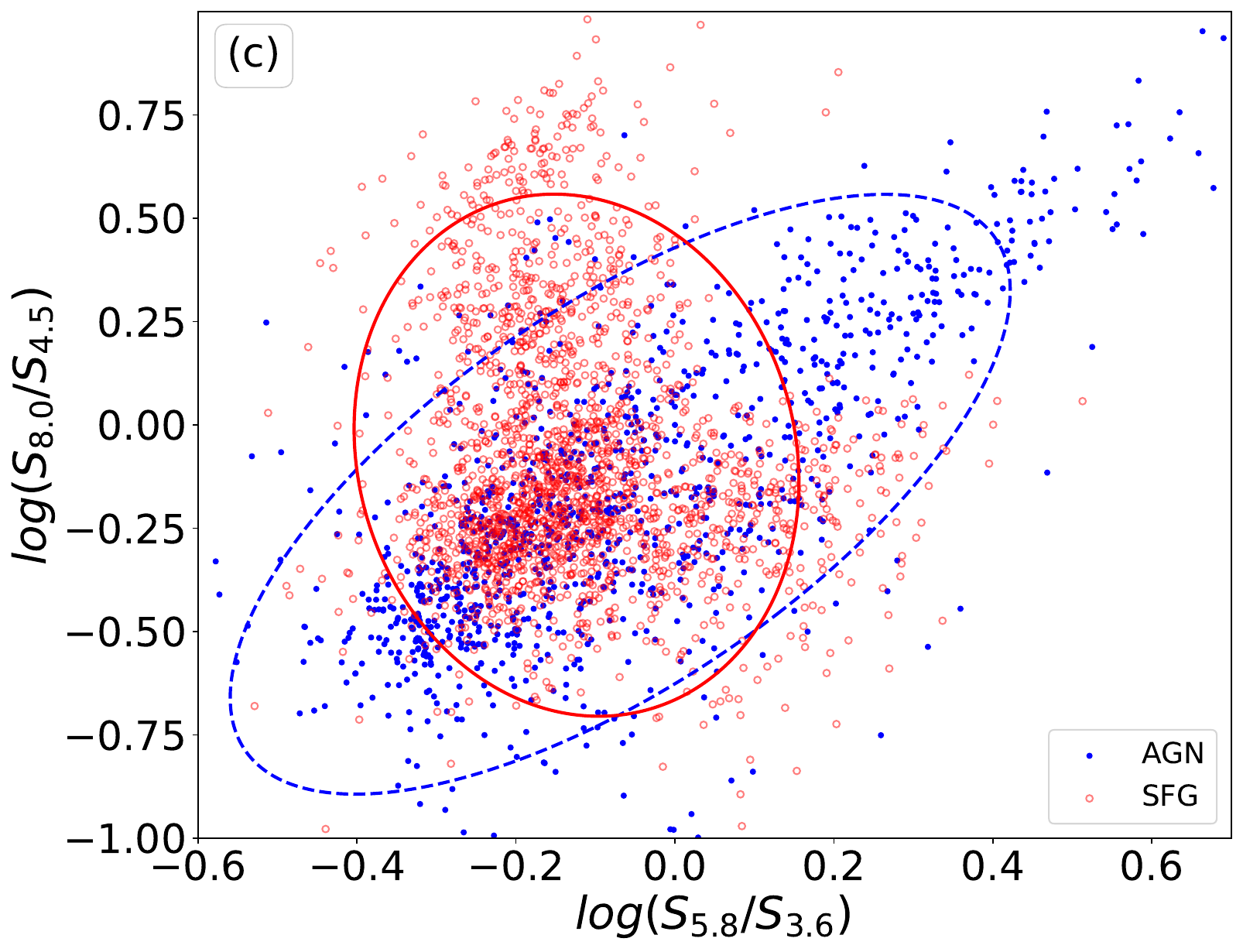}\quad
         \phantomcaption\label{fig:4.3d}
     \end{subfigure}\hfill
     \begin{subfigure}{0.24\textwidth}
    \includegraphics[width=\linewidth]{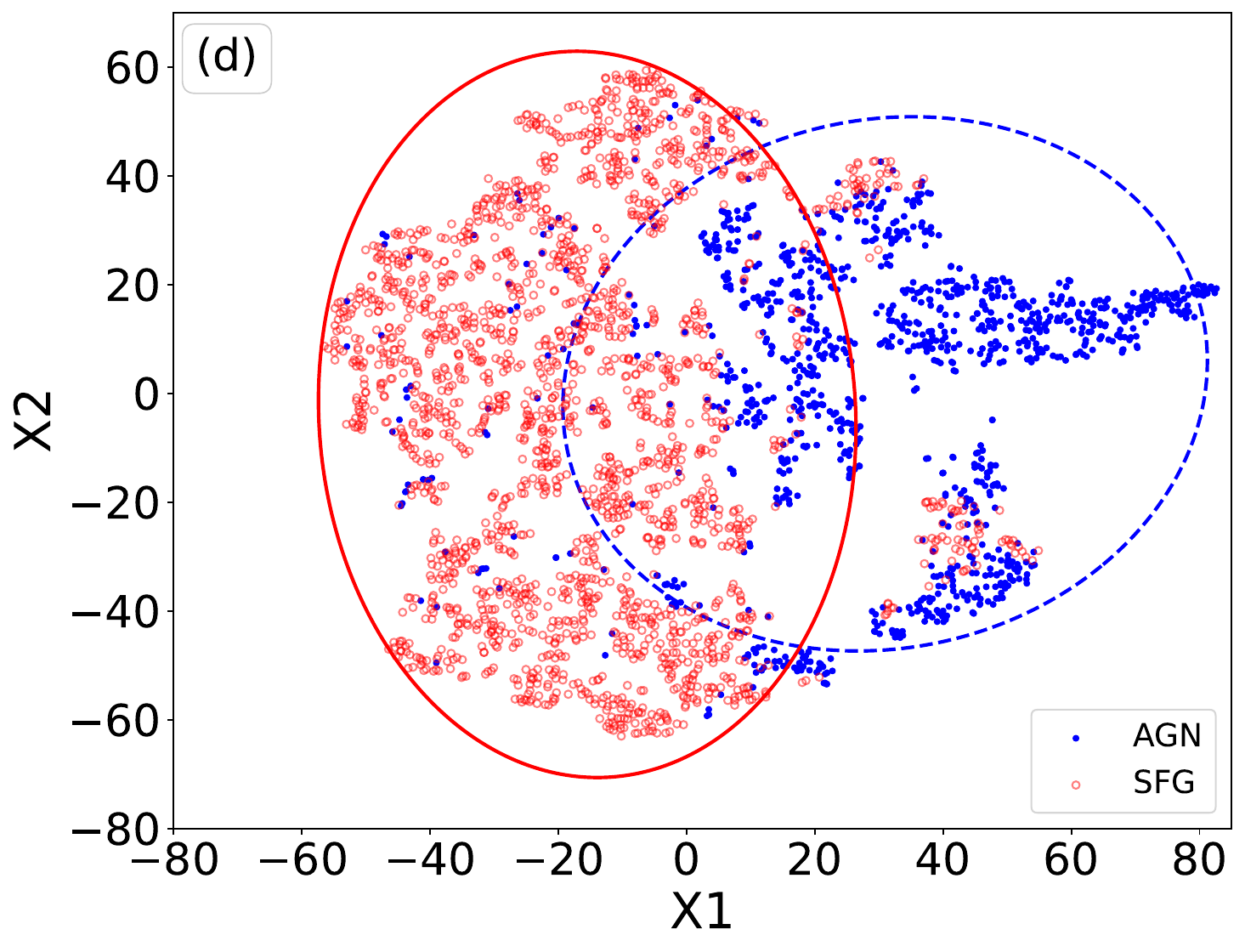}
         \phantomcaption\label{fig:4.3e}
      \end{subfigure}
     \caption{Feature correlation plots for pairs of features selected for classifying SFGs and AGN in the MIGHTEE-COSMOS radio source. Figure \ref{fig:4.3e} shows two-dimensional feature space generated by t-SNE. The open red circles represent SFGs, while the blue dots represent AGN. The solid red ellipses outline the 95\% confidence for SFGs, while the dashed blue ellipses represent the 95\% confidence for AGN. The orientation and shape of each ellipse represent the strength and direction of the correlation between the paired features and the corresponding galaxy classifications.} 
     \label{fig:4.3 corelation}
 \end{figure}

\subsubsection{Feature Correlation (Two Dimensional analysis)}\label{s:3.2.2}
Since conventional classifications of SFGs and AGN, including that of \cite{whittam2022mightee}, rely on combinations of multiple features, we generate all possible pairs of the six features shown in Figure~\ref{fig:4.2} to investigate the correlation between features and their impact on the performance of ML models in classifying SFGs and AGN from the MIGHTEE-COSMOS survey. A total of 15 correlation plots are created, with three highlighted in Figure~\ref{fig:4.3 corelation} and the remainder provided in Figure~\ref{fig:4.3App-corr} (Appendix \ref{B:correlation}).

Figure~\ref{fig:4.3a} presents the $q_\mathrm{IR}$ plots against stellar mass. \cite{whittam2022mightee} apply the mass- and redshift-dependent IRRC from \cite{delvecchio2021infrared} to identify radio-excess AGN. As a result, the combination of these two features tends to miss radio-quiet AGN, causing most SFGs to fall within the 95\% confidence ellipse of AGN (indicated by the blue dashed ellipse in Figure~\ref{fig:4.3a}). Although combining $q_\mathrm{IR}$ with log($S_{4.5}$/$S_{3.6}$) marginally enhances the classification of radio-quiet AGN, the confidence ellipses of the two populations remain significantly overlapped, as illustrated in Figure~\ref{fig:4.3b}. The  Figure~\ref{fig:4.3d} (also see Figure~\ref{fig:Appd}) illustrates the two populations plotted in the IRAC colour-colour feature space. Despite the substantial overlap between AGN and SFGs, the confidence ellipses exhibit different correlation directions: AGN display a positive correlation with the two IRAC colours, whereas SFGs show a negative correlation. This suggests a potential association between the IRAC colour-colour and the two classes. Thus, combining the two IRAC colours improves the performance of ML models in distinguishing between SFGs and AGN among the MIGHTEE-detected radio sources, which may reflect the established MIR colour-colour classification diagnostic \citep{2012ApJ...748..142D}.

The remaining feature correlation plots, presented in Figure~\ref{fig:4.3App-corr}, suggest that in certain cases, combining two features may enhance the performance of ML models in classifying SFGs and AGN from the MIGHTEE-detected radio sources, despite the substantial overlap between the confidence ellipses of the two populations. As demonstrated in Section $\S$\ref{s:results}, the combination of multiple feature improves the ML models' ability to distinguish SFGs from AGN, motivating the use of t-distributed Stochastic Neighbor Embedding (t-SNE) \citep{van2008visualizing} to reduce the six input features used in Section $\S$\ref{s:results} to a two-dimensional feature space. The resulting correlation plot, shown in Figure~\ref{fig:4.3e}, reveals a remarkable improvement in the separation between the two populations when compared with the initial feature pairs presented in the first three plots of Figure~\ref{fig:4.3 corelation} and the correlation plots in Figure~\ref{fig:4.3App-corr}.

 However, since t-SNE features are generated through an unsupervised process and do not lend themselves to straightforward physical interpretation \citep{balamurali2016t}, these features were not used to train the ML models in this study. In the subsequent sections, we further explore the significance of features selected from the MIGHTEE-COSMOS multi-wavelength catalogue through automated techniques to assess their contribution to model performance.

\subsection{Automated Feature Analysis} \label{s:3.3}
 In this section, we employ three automated methods that do not rely on the built-in feature importance algorithms of the ML model, making them independent of the ML model's internal mechanisms. These methods include permutation and RF feature importances (detailed in Section~\ref{subs:3.3.1}), sequential feature importance (Section~\ref{subs:3.4.2}), and the receiver operating characteristic (ROC) curves (Section~\ref{subs:3.4.3}), which are used to assess the significance of the selected features in classifying SFGs and AGN from radio sources. For comparison, in this section, we utilize the RF ML model, which has built-in algorithms to measure the importance of features, allowing us to evaluate its results alongside the permutation method (illustrated in Figure~\ref{fig:feature improtance}).

\subsubsection{Feature Importance}\label{subs:3.3.1}
Permutation feature importance is defined as the decrease in an ML model score (we use $F1$-score as the evaluation metric) when a single feature values are randomly shuffled. This shuffling disrupts the true relationship between the feature and the target variable, resulting in degraded model performance. The extent of the performance drop reflects the feature's importance, i.e., features causing greater drops when their relationship is disrupted are considered more significant. A detailed mathematical explanation of this method is provided in
\cite{molnar2025}.

\begin{figure}
\centering
    \begin{subfigure}{0.24\textwidth}
        \includegraphics[width=\linewidth]{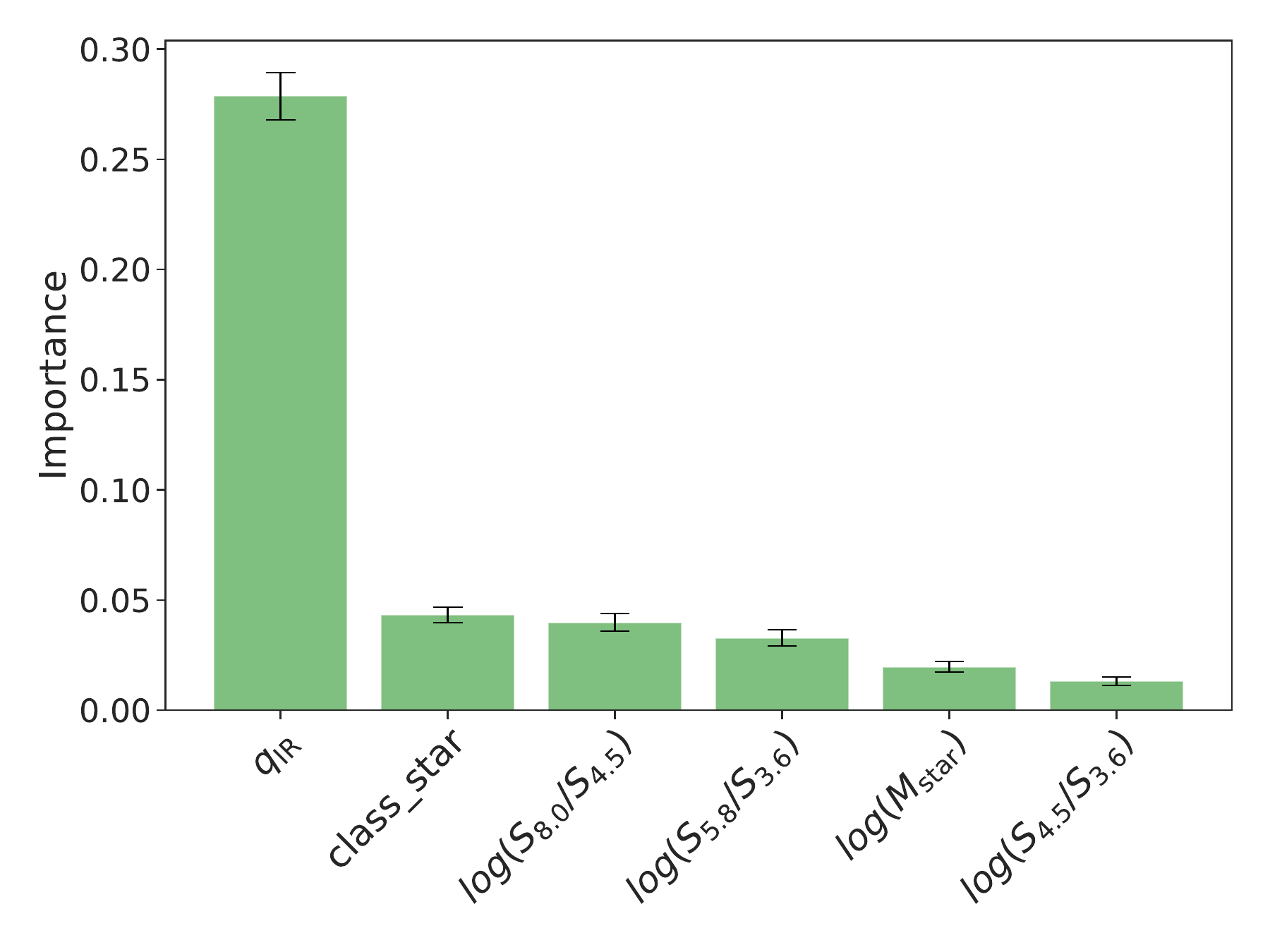}
        \caption{Permutation-based feature-importance.} \label{fig:5a} 
    \end{subfigure}\hfill 
    \begin{subfigure}{0.24\textwidth}
        \includegraphics[width=\linewidth]{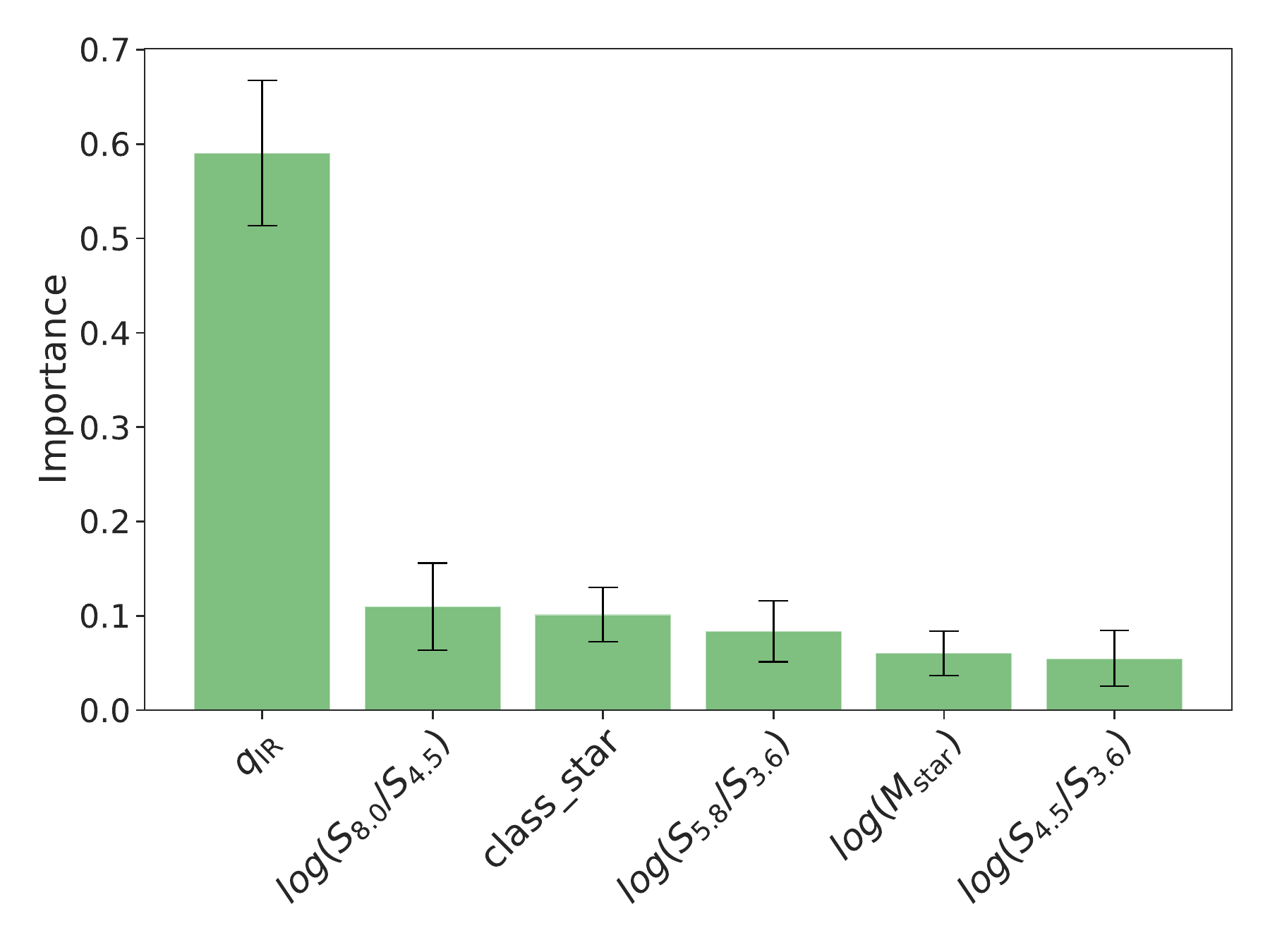}
        \caption{Random Forest model feature importance.}
        \label{fig:5b}
    \end{subfigure}
    \caption{Feature importance estimated by the Permutation ({\it left}) and RF ({\it right}) models. The importance in the Permutation model is derived from the mean scores based on 1000 permutations. For the RF model, importance is computed by measuring the reduction in impurity within a decision tree node when a specific feature is used to split the data. The evaluation metric used is the $F1$-score.}\label{fig:feature improtance}
\end{figure}

Figure~\ref{fig:5a} presents the results of permutation feature importance in distinguishing SFGs and AGN from the MIGHTEE-COSMOS survey. The importance score is defined as the mean performance score obtained over 1,000 permutations. In this section, we present the permutation feature importance for the six most effective features. Complete results for all 18 features selected from the MIGHTEE-COSMOS multi-wavelength catalogue are shown in Figure~\ref{fig:allperm} and are discussed in Section~$\S$\ref{s:discussion}. Notably, Figure~\ref{fig:5a} shows that permutation feature importance yields a ranking consistent with that of the one-dimensional analysis (Section~\S\ref{s:3.2.1}), further confirming these features' efficiency in classifying radio-detected sources as SFGs or AGN.

Figure~\ref{fig:5b} illustrates the feature importance determined by the RF model. The built-in RF importance is computed using two methods: \textit{Gini importance} (also known as mean decrease impurity (MDI)) and \textit{Mean Decrease Accuracy} (MDA). In this study, we employ MDI since MDA closely mirrors the Permutation feature importance method. The details of RF MDI can be found in \cite{li2019debiased}. In brief, this importance is calculated by evaluating the reduction in impurity (or randomness) within a decision tree node when a specific feature is used to split the data. The RF model also identifies $q_{\rm IR}$ as the most effective feature, though it switches the ranking of class$\_$star and log($S_{8.0}$/$S_{4.5}$). However, the difference in importance scores between these two features is negligible.

\subsubsection{Sequential Feature Importance}\label{subs:3.4.2}
In this subsection, we apply a sequential feature selection approach to determine and evaluate the importance of multiple features. This method reduces the initial set of N features to M features, where M$<$N. The selected M features are optimized and used as input for the ML models. For more details on the implementation of sequential feature selection, please refer to the official documentation at \href{https://scikit-learn.org/stable/modules/generated/sklearn.feature_selection.SequentialFeatureSelector.html}{scikit-learn: Sequential Feature Selection} \footnote{\url{https://scikit-learn.org/stable/modules/generated/sklearn.feature_selection.SequentialFeatureSelector.html}}.

In this study, the six most effective features identified in the previous subsections are used to run the sequential feature selection model five times, with M ranging from 1 to 5. As shown in Table \ref{knn.tab}, the model ranks $q_\mathrm{IR}$, class$\_$star, and log($S_{8.0}$/$S_{4.5}$) as the three most essential features, respectively. Compared to the results from the permutation and RF feature importance models, the feature selection model alters the ranking of log$(M_{\rm star})$ and log($S_{5.8}$/$S_{3.6}$). This is likely due to the sequential permutation method, which tends to consider only one feature when two or more features are highly correlated.

\begin{table}
    \centering
    \caption{Sequential feature importance results}
    \label{knn.tab}
    \ra{1.3}
    \begin{tabular}{@{}rrrrcrrrcrrr@{}}\toprule
    
    N$^{a}$ & M$^{b}$ & Features selected \\\midrule
    6 & 1 & $q_\mathrm{IR}$ \\
    6 & 2 & $q_\mathrm{IR}$ and class$\_$star \\
    6 & 3 & $q_\mathrm{IR}$, class$\_$star, and log$(S_{8.0}/S_{4.5})$ \\
    6 & 4 & $q_\mathrm{IR}$, class$\_$star, log$(S_{8.0}/S_{4.5})$, and log$(M_{\rm star})$ \\
    6 & 5 & $q_\mathrm{IR}$, class$\_$star, log$(S_{8.0}/S_{4.5})$, log$(M_{\rm star})$, and log$(S_{5.8}/S_{3.6})$ \\\bottomrule
    \end{tabular}
    $^{a}$ N represents the initial set of features, where $N=6$ in this study.\\
    $^{b}$ M is the reduced set of features, $M<N$.\\
\end{table}

\subsubsection{ROC Curve}\label{subs:3.4.3}
The ROC curve is a graphical tool used to assess the performance of ML classifiers across all classification thresholds. It plots the true positive rate (TPR or {\it Recall}, as defined in Equation~\ref{e:equation2}) against the false positive rate (FPR), which is defined as:
\begin{equation} \label{e:equation4}
    FPR = \frac{\rm FP} {\rm FP + TN}.
\end{equation}

\begin{figure}
    \centering
    \includegraphics[width=1\linewidth]{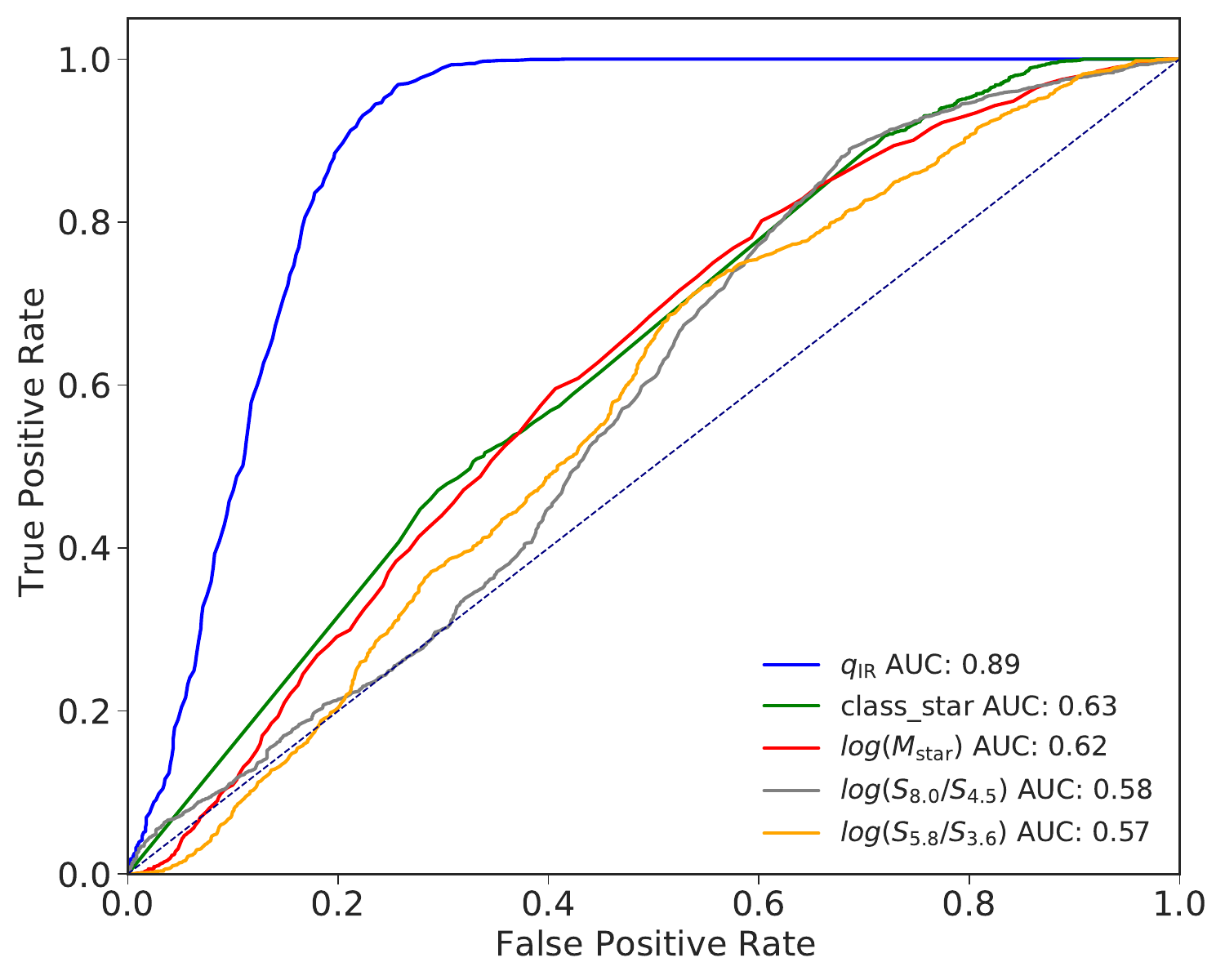}
    \caption{Receiver Operating Characteristic (ROC) curves for the five selected features, namely, the $q_\mathrm{IR}$ (blue), class$\_$star (green), log$(M_{\rm star})$ (red), log($S_{8.0}$/$S_{4.5}$) (grey), and log($S_{5.8}$/$S_{3.6}$) (yellow).}
    \label{fig:4.5}
\end{figure}

In this subsection, we also begin with a total of 18 features, comprising both the multi-wavelength photometric measurements from the MIGHTEE-COSMOS catalogue and the conventional classification diagnostics previously employed by \cite{whittam2022mightee}. The ROC curves, computed based on thresholds applied individually to each feature, are presented in Figure~\ref{fig:4.5}. The area under the curve (AUC) is used to evaluate the effectiveness of these features in distinguishing SFGs from AGN in the MIGHTEE-COSMOS survey. 

To mitigate potential bias arising from highly informative features, such as $q_{\rm IR}$, which may dominate the feature space and obscure the contribution of other variables, we also implement an iterative feature-ranking procedure. In each iteration, the most dominant feature, based on AUC performance, is removed from the feature set, and the process is repeated on the remaining features. This process continues until a complete ranking is established. The AUC values of the top-ranked features at each iteration are summarized in Table~\ref{tab:roc}.

\begin{table}
    \centering
    \caption{AUC values of the top-ranked features identified at each iteration}
    \label{tab:roc}
    \ra{1.3}
    \begin{tabular}{@{}cccrcrrrcrrr@{}}\toprule
    Iteration run & Number of total features & Top-ranked features & AUC \\\midrule
    0 & 18 & $q_\mathrm{IR}$ & 0.886 \\ 
    1 & 17 & class$\_$star & 0.630 \\
    2 & 16 & log$(M_{\rm star})$ & 0.619 \\
    3 & 15 & log($S_{8.0}$/$S_{4.5}$) & 0.580 \\
    4 & 14 & log($S_{5.8}$/$S_{3.6}$) & 0.574 \\
    5 & 13 & log($i$/$z$) & 0.570 \\
    ...&... &... &... \\
    9 & 9 & log($S_{4.5}$/$S_{3.6}$) & 0.542 \\
    ...&... &... &... \\
    17& 1 & $log(H/K_{s})$  & 0.500\\\bottomrule
    \end{tabular}
\end{table}

As expected, the $q_\mathrm{IR}$ achieves the maximum AUC, indicating it is the most significant feature for distinguishing SFGs and AGN from the radio sources,  followed by class$\_$star. Contrary to the findings in previous subsections, the ROC curves suggest that log$(M_{\rm star})$ is the third most important feature. The IRAC colour indices, log($S_{8.0}$/$S_{4.5}$) and log($S_{5.8}$/$S_{3.6}$), are identified as the fourth and fifth most informative features, respectively, although their individual discriminative power remains low, with AUC$\lesssim$0.6. In contrast, log($S_{4.5}$/$S_{3.6}$) fails to rank among the top six features, displaying an AUC score close to 0.5, indicating a performance akin to random classification.

\subsection{Feature Selection}
\label{s:selection}
Sections~$\S$\ref{s:f-analyses} and $\S$\ref{s:3.3} detail our feature analyses, which combine one-dimensional, two-dimensional, ML-independent, and ML-dependent analyses to identify the most effective features for classifying SFGs and AGN among the radio-detected sources in the MIGHTEE-COSMOS survey. Across all feature analysis methods, the $q_\mathrm{IR}$ parameter consistently emerges as the most significant feature. This is likely because the majority (74\%) of AGN in our sample are radio-excess AGN, which are conventionally distinguished from SFGs based on their $q_\mathrm{IR}$ values. The optical compactness parameter, class$\_$star, is consistently ranked among the top three. In addition, log$(M_{\rm star})$ and two IRAC colours, namely, log($S_{8.0}$/$S_{4.5}$) and log($S_{5.8}$/$S_{3.6}$), are generally among the five most important features across most analyses. By contrast, the remaining IRAC colour, log($S_{4.5}$/$S_{3.6}$), although occasionally ranked sixth, is shown by the ROC-based AUC metric, which is a model-independent measure of feature discriminative power, to perform comparably to random classification. 

Another crucial factor in feature selection is completeness, defined as the fraction of sources with measured values for the chosen features. As discussed in Section $\S$\ref{sec: manual classification}, while X-ray and VLBI detections are highly effective diagnostics for identifying AGN, their limited completeness results in approximately 70\% of MIGHTEE sources remaining unclassified if only these two features are used (as shown in Figure~\href{fig:1}{1}).

Balancing completeness and classification efficiency, we select five key features for the subsequent ML analyses: the IRRC parameter ($q_\mathrm{IR}$), optical compactness (class$\_$star), stellar mass (log$(M_{\rm star})$), and  two IRAC colours: log($S_{8.0}$/$S_{4.5}$) and log($S_{5.8}$/$S_{3.6}$). A detailed description of these features is provided in Section $\S$\ref{sec: catalogue}, where they are outlined as conventional diagnostics frequently employed in the literature to classify sources as SFGs or AGN. The effectiveness of these features in ML-based classification of SFGs and AGN among MIGHTEE-detected sources is thoroughly evaluated in Sections $\S$~\ref{s:f-analyses} and $\S$\ref{s:3.3}. 

\begin{figure}
    \centering
    \includegraphics[width=1\linewidth]{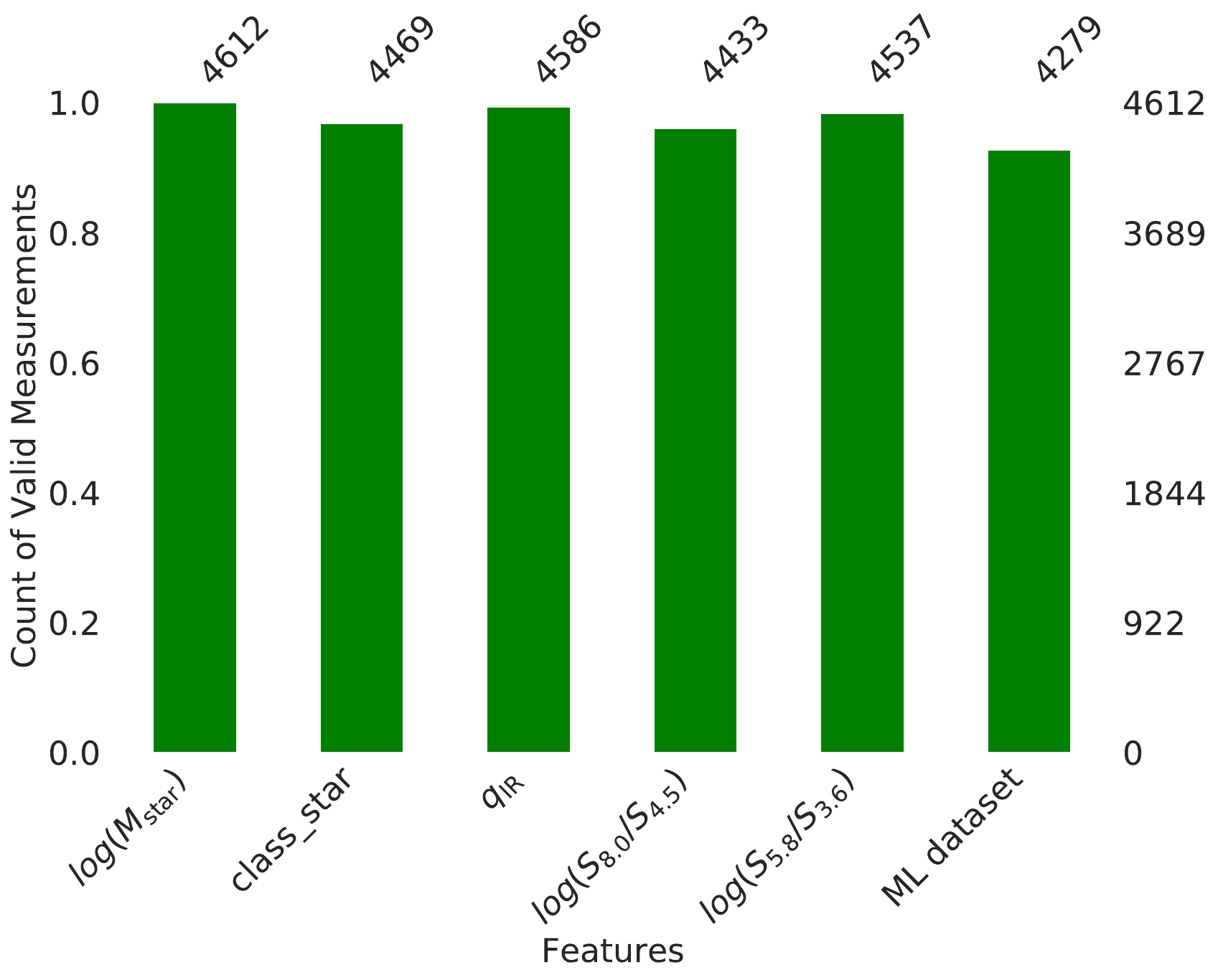}
    \caption{The completeness of the five features selected to train ML models from the MIGHTEE-COSMOS multi-wavelength catalogue. The left vertical axis indicates the completeness fraction for each feature, while the right vertical axis displays the corresponding numbers. The bottom horizontal axis lists the names of the selected features, and the top horizontal axis represents the total number of valid measurements for each feature. ML dataset bar represents the number of sources in the final sample used for ML.} \label{fig:completeness}
\end{figure}

As shown in Figure \ref{fig:completeness}, for the 4612 labeled sources in \cite{whittam2022mightee}, the completeness of these selected features $>96\%$ (4433/4612). Due to the unpredictable nature of some of these astronomical features, we include 4279 sources with valid measurements for all five features in our ML analyses. This approach is justified, as sources with missing measurements account for approximately 7\% of the entire dataset. As shown in Figure~\ref{fig:color_ml_results}, the inclusion of additional features does not improve the performance of ML classification but slightly reduces the completeness of the dataset available for ML analysis.

\subsection{Supervised ML Classification}\label{s:SV_ML}
As illustrated in the previous subsection, the dataset used for ML analyses consists of 4279 sources, with 1,526 classified as AGN and 2753 as SFGs according to \cite{whittam2022mightee}. This sample is used to construct the training and test datasets and to optimize the ML models.

For binary classification, two distinct approaches can be employed: the first involves a straightforward dichotomous distinction between the two classes, where class labels 0 or 1 are assigned to an unknown source. The second approach models the probability $P(y|X)$, providing both a class label and the probability of class membership for a given source. In this study, we implement five different supervised classification algorithms. SVM uses the first approach, while the remaining four, namely, LR, \textit{k}NN, RF, and XGB, adopt the second approach, estimating class probabilities.

The implementations of these classification algorithms are readily available through widely-used open-source libraries such as \href{https://scikit-learn.org/stable/}{scikit-learn}\footnote{scikit-learn: \url{https://scikit-learn.org/stable/}} and \href{https://xgboost.readthedocs.io/en/stable/}{\textit{XGBoost}}\footnote{XGBoost: \url{https://xgboost.readthedocs.io/en/stable/}}. Therefore, detailed descriptions of these algorithms are omitted here, and the focus of this work is on optimizing these models for classifying SFGs and AGN from the radio-detected sources.

\subsubsection{Hyperparameter Optimization}
\label{subs:hyperparameter}

Hyperparameters determine the structure and behavior of an ML model before training, and optimizing them remains a trial-and-error process. Two common approaches to identifying the optimal set of hyperparameters for maximizing model performance are grid search and random search parameter tuning. Grid search exhaustively explores all possible combinations of hyperparameters, systematically evaluating each one to identify the best performance configuration. In contrast, random search generates random combinations of hyperparameters and evaluates a subset of them. The classifier with the best accuracy from the random search is then considered optimal.

In this work, we perform $k$-fold cross-validation using the grid search technique to identify the optimal hyperparameters for each ML model. Cross-validation (CV) is a resampling method used to evaluate the generalization ability of predictive models and to prevent overfitting \citep{berrar2019cross}. The data is split into $k$ folds with each fold used for testing while the remaining $k$-1 folds are used for training. For this work, we use $k=3$, as illustrated in Figure \ref{fig:2grid}. We highlight some optimized hyperparameters for different ML models in  Appendix~\ref{app:hyperparameter}.

\begin{figure}
    \centering
    \includegraphics[width=1\linewidth]{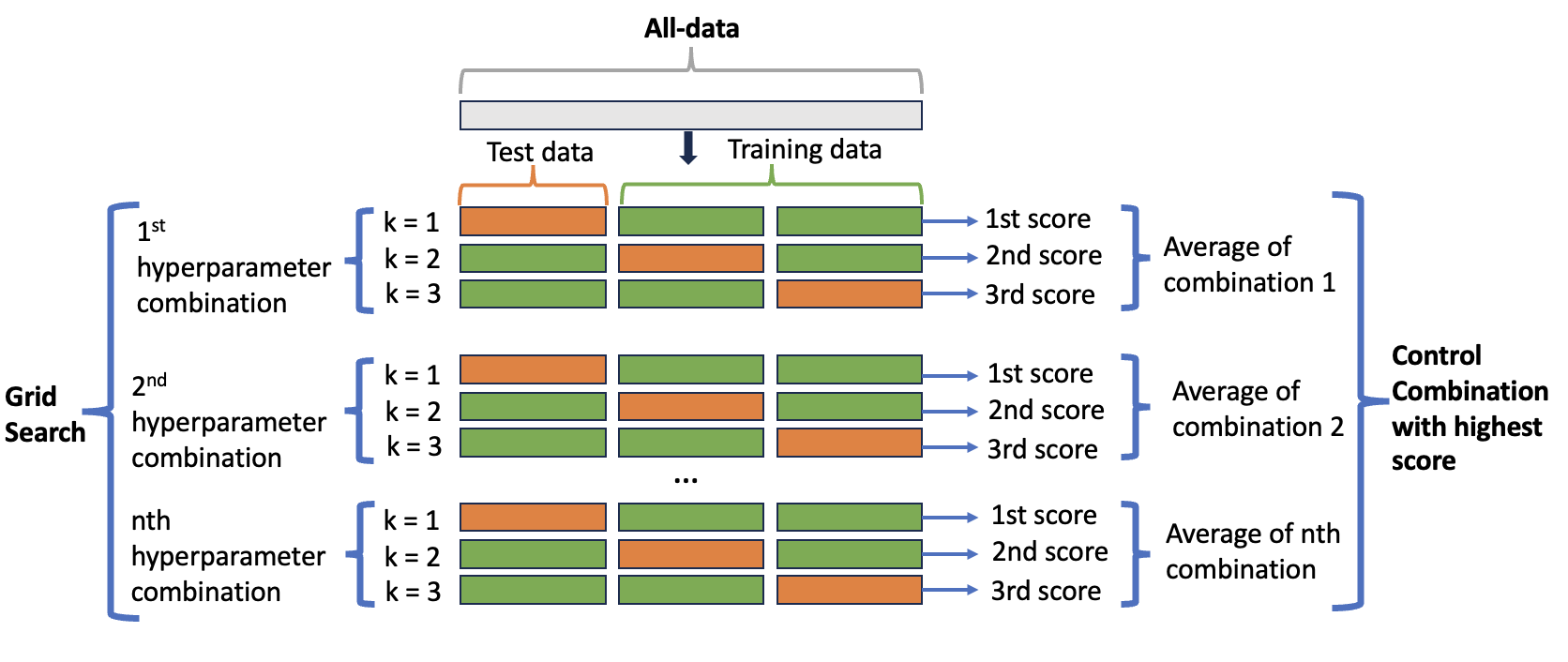}
    \caption{The three-fold cross-validation hyperparameter tuning using grid search technique. The data is split into three folds with each fold used for testing (highlighted in red) while the remaining two folds are used for training (green). Image inspired by \citep{shatnawi2022shear}.} 
    \label{fig:2grid}
\end{figure}

\section{Results}\label{s:results}
In this section, we present the results of ML approaches for classifying SFGs and AGN among radio sources detected by the MIGHTEE survey using selected input features and optimized ML models. As outlined previously, five widely used supervised ML models, namely, LR, SVM, \textit{k}NN, RF, and XGB, are employed for this classification task. Our feature analyses suggest that combining multiple features improves the performance of ML models in distinguishing between SFGs and AGN in radio-detected sources (Section \S\ref{s:3.2.2}). We, therefore, create  five distinct feature combinations (Table \ref{tab:fea}), guided by the ROC-based AUC metric, which offers a model-independent evaluation of feature importance.


\subsection{Cross-validation Results}
\label{s:crossvalidation results}
The performance of ML models in classifying SFGs and AGN from radio surveys is systematically evaluated through cross-validation techniques. We use random stratified sampling to partition our ML dataset, comprising 4279 MIGHTEE sources, into \textit{training} and \textit{validation} sets. Specifically, we implement several training-to-validation splits, i.e., [1:4], [2:3], [3:2], and [4:1], to assess model performance across a range of training data ratios.

These variations in the sizes of the training and validation datasets are motivated by the anticipated scale of future radio continuum surveys, such as those planned for SKA1 and the full SKA, which are projected to detect billions of radio sources. In contrast, current radio surveys have identified only tens of thousands of sources. As a result, the labeled data available for training ML models is likely to constitute merely a small fraction of the total dataset expected from future surveys. Therefore, it is imperative to evaluate the models’ ability to maintain robust performance in scenarios where training data is limited.

\begin{figure}
    \begin{subfigure}{0.46\textwidth}
        \includegraphics[width=1\linewidth]{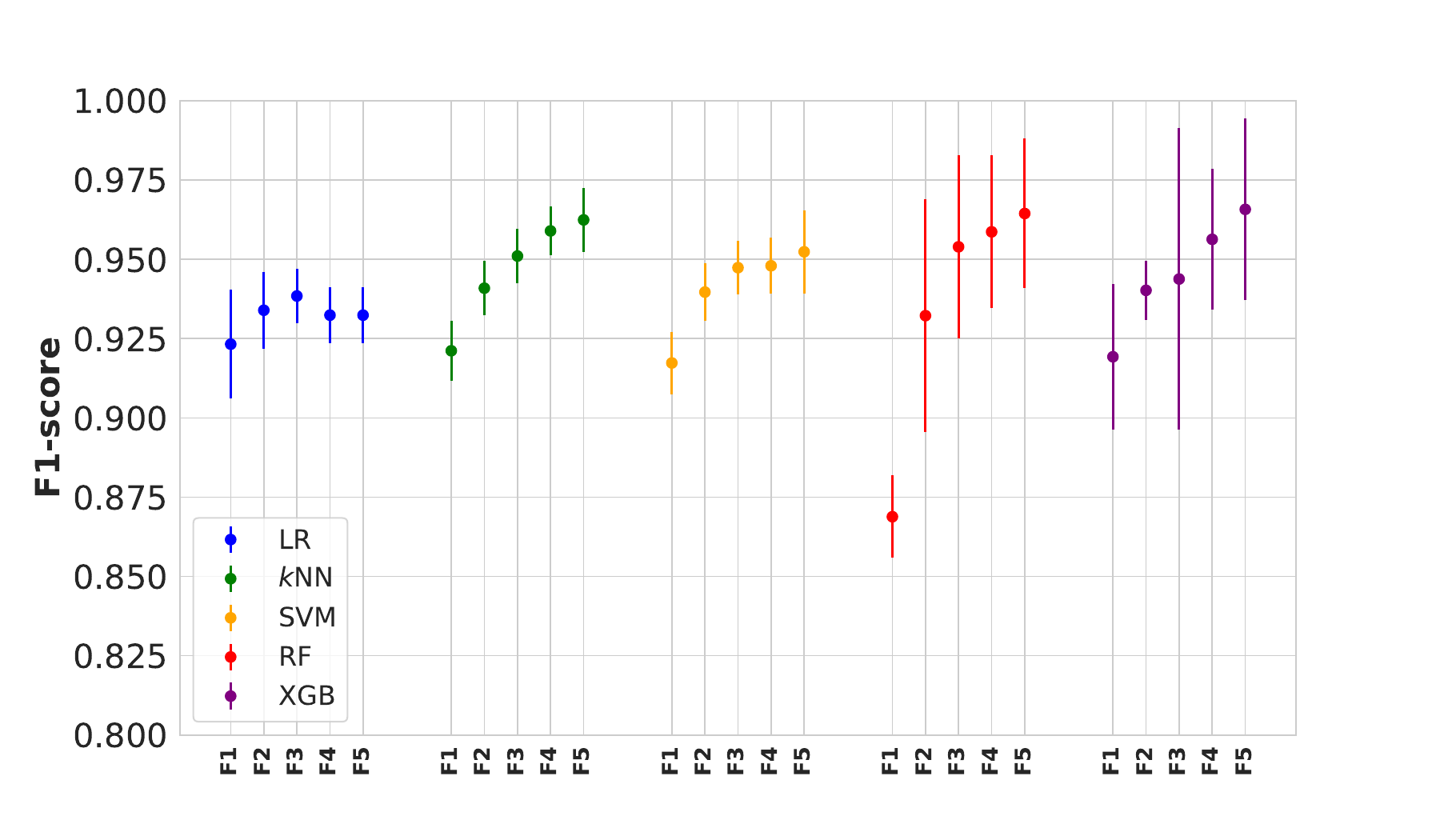} 
        \caption{Models trained on $80\%$ of the full ML dataset} \label{fig:4.8.1}
    \end{subfigure}\hfill
    \begin{subfigure}{0.46\textwidth}
        \includegraphics[width=1\linewidth]{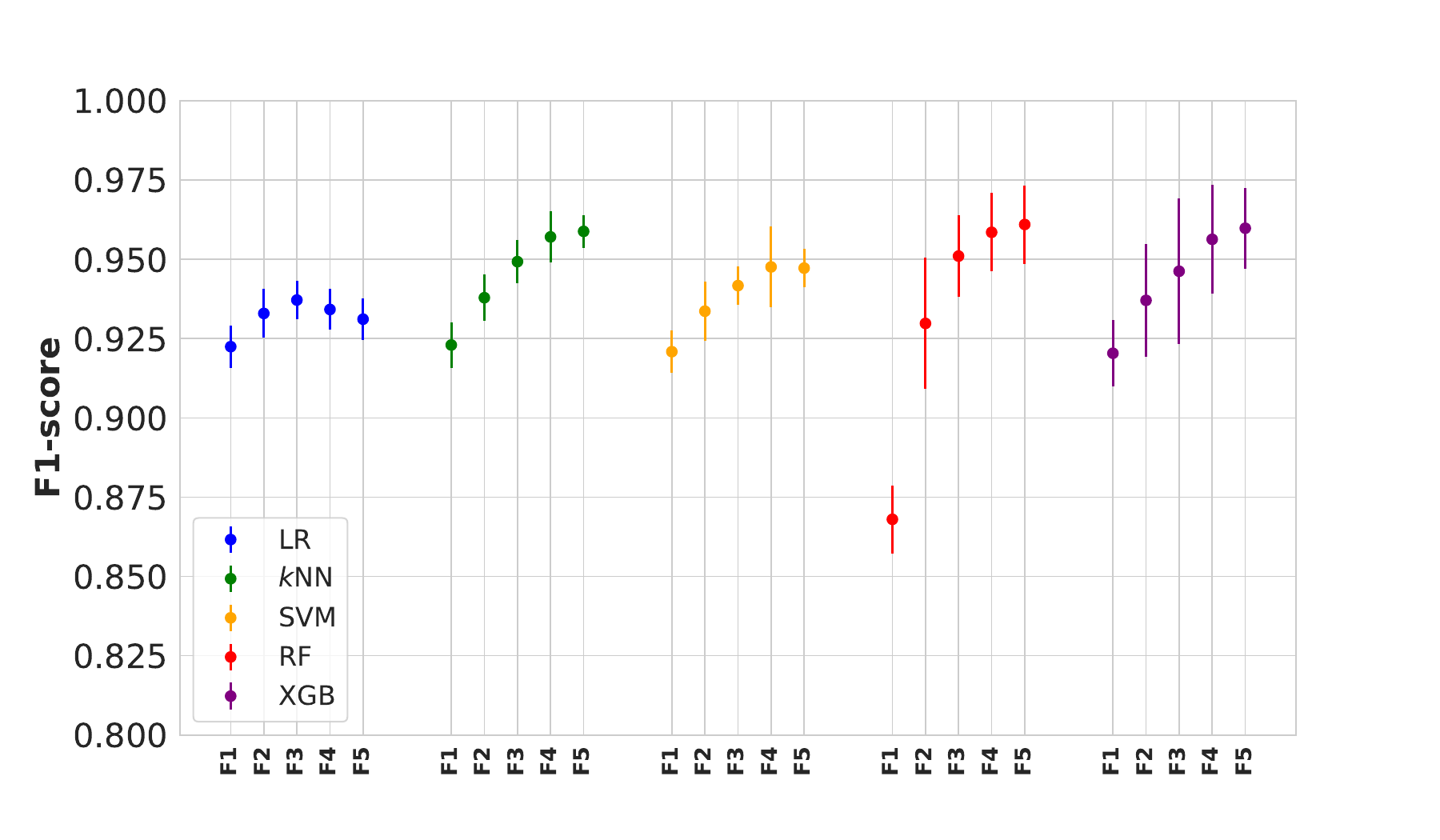} 
        \caption{Models trained on $60\%$ of the full ML dataset}\label{fig:4.8.2}
    \end{subfigure}
    \begin{subfigure}{0.46\textwidth}
        \includegraphics[width=1\linewidth]{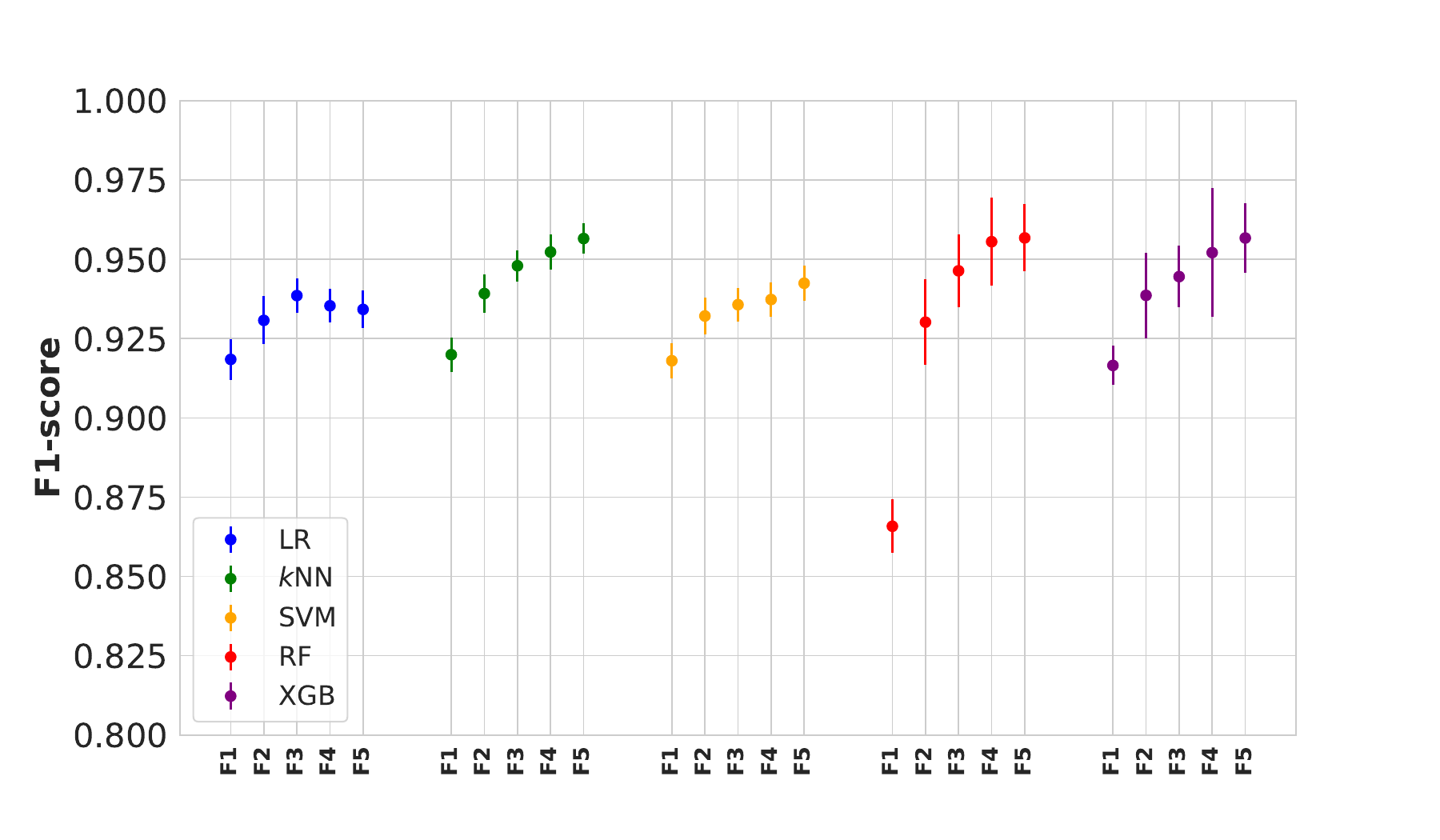} 
        \caption{Models trained on $40\%$ of the full ML dataset}\label{fig:4.8.3}
    \end{subfigure}\hfill
    \begin{subfigure}{0.46\textwidth}
        \includegraphics[width=1\linewidth]{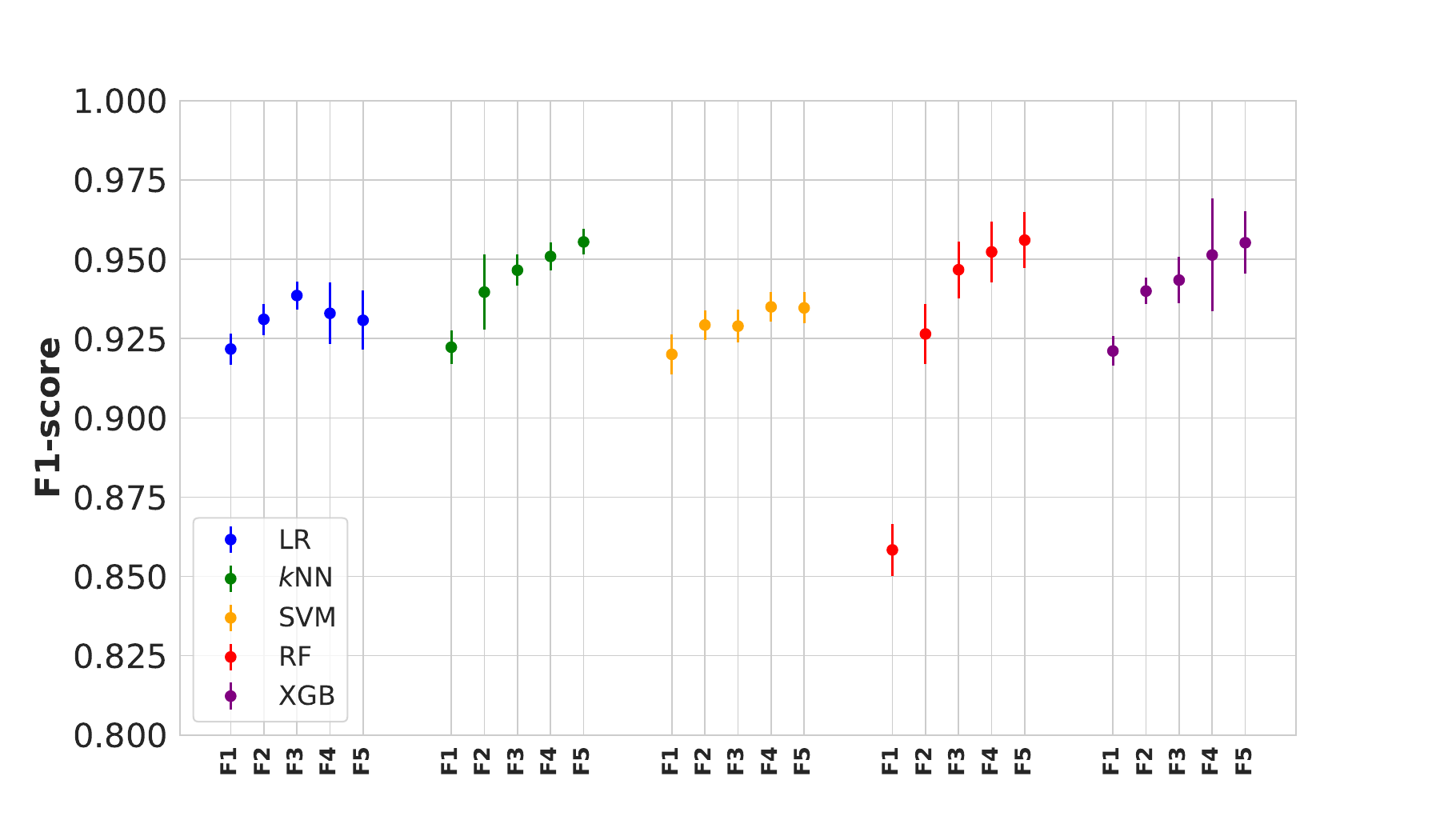} 
        \caption{Models trained on $20\%$ of the full ML dataset}\label{fig:4.8.4}
    \end{subfigure}
\caption{Evaluation of the performance of five supervised ML models, Logistic Regression (blue), $k$-Nearest Neighbour (green), Support Vector Machine (yellow), Random Forest (red) and XGBoost (purple), in classifying SFGs and AGN from the \textit{validation data}. The $F1$-score is used as the evaluation metric. Feature combinations F1 through F5, as outlined in Table \ref{tab:fea}, are used for training the ML models, respectively. Subplots (a), (b), (c), and (d) display the results of all five models trained on $80\%$, $60\%$, $40\%$, and $20\%$ of the complete ML dataset, respectively. Error bars represent the standard deviation derived from jackknife resampling.} 
\label{fig:4.8}
\end{figure}

Figure \ref{fig:4.8} shows the performance of LR, \textit{k}NN, SVM, RF, and XGB models trained on input feature combinations described in Table \ref{tab:fea} in distinguishing SFGs from AGN from the \textit{validation data}. Figures \ref{fig:4.8.1}, \ref{fig:4.8.2}, \ref{fig:4.8.3} and \ref{fig:4.8.4} show the results of models trained on $80\%$, $60\%$, $40\%$ and $20\%$ of the complete ML dataset respectively. All the classifiers are evaluated using the $F1$-score. The error bars shown in Figure \ref{fig:4.8} are the standard deviations of the $F1$ score calculated using jackknife resampling. 

\begin{table}
    \centering
    \caption{Five feature combinations used for training the ML models}
    \label{tab:fea}
    \ra{1.3}
    \begin{adjustbox}{width=0.49\textwidth}
        \begin{tabular}{@{}cl@{}}\toprule
        
       \textbf{Name of combination} &\textbf{ Features }\\\midrule
        F1 & $q_\mathrm{IR}$ \\
        F2 & $q_\mathrm{IR}$ and class$\_$star \\
        F3 & $q_\mathrm{IR}$, class$\_$star, and log$(M_{\rm star})$ \\
        F4 & $q_\mathrm{IR}$, class$\_$star, log$(M_{\rm star})$, and log($S_{8.0}/S_{4.5}$) \\
        F5 & $q_\mathrm{IR}$, class$\_$star, log$(M_{\rm star})$, log($S_{8.0}/S_{4.5}$), and log($S_{5.8}/S_{3.6}$) \\\bottomrule
        \end{tabular}
    \end{adjustbox}
\end{table}

\subsubsection{Results based on different feature combinations}
\label{different feature combinations results}

For each training data size, all five ML models demonstrate strong performance across various feature combinations, consistently achieving $F1$-scores$>90\%$, with the exception of the RF model trained using only the $q_\mathrm{IR}$ feature. The models exhibit slight variations in performance depending on the specific feature combinations. Notably, the LR model attains the highest $F1$-score when using the combination of $q_\mathrm{IR}$, class$\_$star, and log$(M_{\rm star})$. Introducing additional input features does not enhance the LR model’s performance and instead leads to a slight decline, indicating that the inclusion of the two IRAC colours, log($S_{8.0}$/$S_{4.5}$) and log($S_{5.8}$/$S_{3.6}$), may introduce redundancy or noise, thereby diminishing its discriminative effectiveness.

In contrast, the remaining four ML models benefit from the inclusion of the two IRAC colour indices, although the improvement for the boundary-based SVM classifier is generally marginal. For the other three models, excluding these IRAC colours leads to a noticeable decline in classification accuracy. This highlights the importance of these IRAC colour features for effective classification of SFGs and AGN in the MIGHTEE survey. Therefore, the absence of 5.8 and 8.0\,\mm\ observations will be a disadvantage for ML approaches in classifying radio-detected sources from future radio continuum surveys.

\begin{table*}
    \centering
    \caption{Recalls of X-ray-only and VLBI-only AGN}
    \label{tab:recall_xonly.tab}
    \ra{1.3}
    \begin{tabular}{@{}llcccccccccc@{}}\toprule
    ML models & & LR & $k$NN & SVM  & RF & XGB  \\\midrule
    trained on 80\% of the full ML dataset  
   & Recall of X-ray-only AGN & (3.3$\pm$0.1)\% & (16.7$\pm$0.2)\% & (26.7$\pm$0.3)\% & (20.0$\pm$0.3)\% & (13.3$\pm$0.2)\% \\
  &  Recall of VLBI-only AGN & 0 & (33.3$\pm$1.3)\% & (66.7$\pm$1.3)\% & (33.3$\pm$1.3)\% & (33.3$\pm$1.3)\% \\\midrule
  trained on 20\% of the full ML dataset  
   & Recall of X-ray-only AGN & (6.0$\pm$0.1)\% & (8.5$\pm$0.1)\% & (16.2$\pm$0.1)\% & (14.5$\pm$0.1)\% & (12.0$\pm$0.1)\% \\
  &  Recall of VLBI-only AGN & 0 & 0 & (16.7$\pm$0.3)\% & (16.7$\pm$0.3)\% & (8.3$\pm$0.3)\% \\\bottomrule
    \end{tabular}
\end{table*}

\subsubsection{Results based on different ML models and training sets}
\label{finalmlresults}
\begin{figure}
    \centering
    \includegraphics[width=0.92\linewidth]{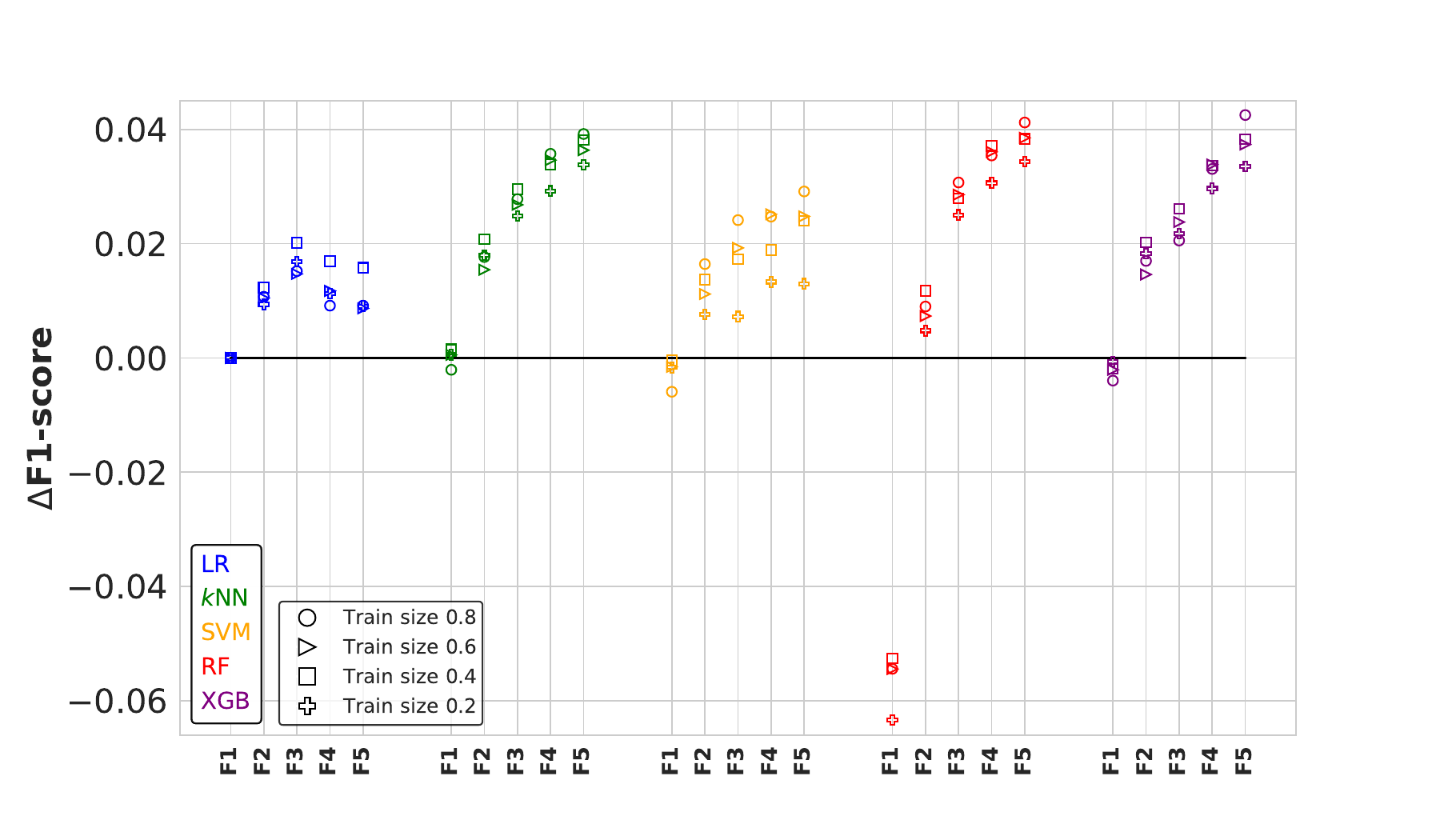}
    \caption{This figure mirrors Figure~\ref{fig:4.8}, but it uses the $F1$-score of the LR classifier trained on feature combination F1 ($q_\mathrm{IR}$) as a baseline, shown as a solid black line. Distinct symbols represent different training set sizes, providing a clear visualization of the impact of training set size on the performance of ML models in classifying SFGs and AGN from radio-detected sources.}
    \label{fig:4.110}
\end{figure}

We also evaluate the performance of all five ML models and compare their results across different training sets. For clarity, the $F1$-score of the LR classifier trained on the feature combination F1 ($q_\mathrm{IR}$) is used as \textit{baseline} (Figure~\ref{fig:4.110}). Figure~\ref{fig:4.8} and Figure~\ref{fig:4.110} show that, in classifying SFGs and AGN from the radio-detected sources, \textit{k}NN, RF, and XGB perform slightly better than the LR and SVM classifiers, particularly when trained on the feature combinations of F3, F4, and F5. However, the jackknife scatter for the two decision-tree-based models (RF and XGB) is notably higher. Therefore, among the five ML models considered, 
the \textit{k}NN classifier, which determines the membership of the class based on distance metrics (e.g. Euclidean distance) to identify the nearest neighbors, offers the most sustainable and interpretable approach. Its consistent performance and low variance make it a compelling choice for classifying SFGs and AGN in current and future radio continuum surveys.

Figure~\ref{fig:4.110} further demonstrates that as the size of the training dataset decreases, the performance of all ML classifiers experiences a slight decline. Nonetheless, all models achieve an $F1$-score\,$>90\%$ across various feature combinations, even when trained with only 20\% of the available data (with the exception of the RF model trained solely on the $q_\mathrm{IR}$ feature). This outcome underscores the robustness of ML approaches in classifying SFGs and AGN, even with limited training data.

Overall, these findings demonstrate the effectiveness of ML techniques in the classification of radio sources, thereby reinforcing their promise for application in forthcoming large-scale radio continuum surveys to be conducted with next-generation interferometric facilities, such as the SKA and the ngVLA.

\section{Discussions}\label{s:discussion}
In this study, we assess the performance of five widely used supervised ML algorithms in classifying SFGs and AGN from the MIGHTEE-COSMOS radio continuum survey. To construct training and test datasets, we use SFGs and AGN that have been conventionally classified from the MIGHTEE-COSMOS \citep{whittam2022mightee}. Additionally, we incorporate all available photometric data from the MIGHTEE-COSMOS multi-wavelength catalog \citep{Whittam24}, alongside conventional classification diagnostics, to inform the selection of ML input features. As expected, the five parameters used in conventional classification prove to be the most effective. Although the other two conventional classification features, X-ray luminosity and VLBI detection, are excluded from our ML analyses due to their limited completeness, the classification still achieves an $F1$-score\,$>90\%$. Consequently, the selection of input features is guided by an optimal balance between classification efficiency and completeness. 

In this section, we first examine the ability of the ML models to recover AGN that are identified exclusively by their X-ray luminosity or VLBI detection, despite the exclusion of these features from the input set. In Section \S\ref{s:discussion_if}, we further explore the characteristics of the selected input features for AGN versus SFG classification and evaluate the impact of incorporating additional features from the MIGHTEE-COSMOS multi-wavelength catalog on the performance of ML models. We then apply dimensionality reduction techniques and evaluate their influence on model performance (Section \S\ref{s:discussion_dr}), followed by an examination of the impact of data normalization (Section \S\ref{s:Data_normalization}). We also address the issue of class imbalance (Section $\S$\ref{s:Class Imbalance}), which may influence certain supervised ML algorithms, potentially causing them to neglect the minority class. Finally, we discuss the limitations of using ML approaches to classify SFGs and AGN from the extragalactic radio continuum survey.

\subsection{X-ray and VLBI classifications}
As presented in Section $\S$\ref{s:analyses}, the training set for our ML models is based on conventional classifications from \cite{whittam2022mightee}, which include X-ray and VLBI classifications. However, due to the limited completeness and unpredictability of these X-ray and VLBI classifications, they are not used as input features for the ML classification. In our full ML dataset, there are 146 AGN identified solely based on their X-ray luminosity, while 15 are classified exclusively through VLBI detection. Although the number of VLBI-only AGN is negligible, X-ray-only AGN constitute approximately 10\% of the total AGN sample. Table~\ref{tab:recall_xonly.tab} presents the recalls for these two AGN subpopulations as achieved by each ML model. Notably, when training with 20\% of the dataset, only about 10\% of the X-ray-only AGN are successfully recovered. This recovery fraction increases to approximately 20\% when 80\% of the dataset is used for training, with the exception of the LR and XGB models.

This result is significant because obtaining deep and wide X-ray data will remain a challenge for at least the next 15 years, until the launch of ESA’s Athena X-ray observatory\footnote{https://www.the-athena-x-ray-observatory.eu/en}. During the operational periods of MeerKAT and SKA1, the classification of radio continuum sources will, therefore, often proceed without X-ray data. Our ML approach indicates that incorporating even limited X-ray observations into model training can marginally improve classification recall.

\subsection{Input Features}\label{s:discussion_if}
As shown in Section $\S$\ref{s:analyses}, we select the five most effective features in classifying SFGs and AGN from the MIGHTEE-COSMOS survey. For the five selected features, our feature analyses consistently indicate that the $q_\mathrm{IR}$ parameter is the most effective feature in distinguishing between the two classes of radio sources. This is likely due to the fact that the majority (74\%) of AGN in the sample are radio-excess AGN, which are traditionally separated from SFGs using $q_\mathrm{IR}$. However, as evidenced by the ML cross-validation results (Section \S\ref{s:results}), all ML models, except the RF, achieve $F1$-scores exceeding 90\% when trained only with $q_\mathrm{IR}$.

The classification utility of $q_\mathrm{IR}$ arises from AGN-dominated sources exhibiting substantially more accelerated CR electrons than would be expected from star formation alone, producing an observable ‘excess’ in radio emission relative to infrared emission. Although this excess may vary with redshift, stellar mass, or radio spectral index \citep[e.g.,][]{delvecchio2021infrared, An21}, $q_\mathrm{IR}$ remains a robust and effective parameter to distinguish between SF- and AGN-dominated radio sources. 

The optical compactness parameter, class$\_$star, also ranks among the top three features across all feature selection methods, as discussed in Section $\S$\ref{s:analyses}. This parameter is particularly useful for identifying optical point-like AGN among radio sources, offering a straightforward approach to differentiation. While the IRAC colour index may not be the most individually effective feature for classification, our two-dimensional feature analyses underscore the significance of combining the two IRAC colours for improved separation of AGN from SFGs.

Overall, as shown in Figure~\ref{fig:4.8}, we observe improvements in ML model performance with additional features incorporated into the training dataset. However, beyond the selected five features, adding further optical or NIR photometric data or colours does not improve classification accuracy but slightly reduces dataset completeness, as shown in Appendix~\ref{A:feature-completeness}.

\subsection{Feature Space Dimensionality Reduction}\label{s:discussion_dr}

Our analysis reveals that $q_\mathrm{IR}$ serves as the most discriminative feature for distinguishing between SFGs and AGN in radio continuum surveys. While the inclusion of four additional features enhances the classification performance of ML models, it also introduces increased model variance in most cases (Figure~\ref{fig:4.8}). To address this, we implement a two-step dimensionality reduction approach: (1) feature selection to retain the most informative predictors, as described in Section $\S$\ref{s:f-analyses}; (2) non-linear compression of the feature space using either:
\begin{itemize}
    \item Autoencoder: A lightweight symmetric autoencoder trained over 10,000 epochs to minimize reconstruction error (MSE$=$0.32), compressing the selected feature set into a two-dimensional latent space; or 
    \item \textit{t}-SNE: As a comparative method, we also apply \textit{t}-distributed stochastic neighbor embedding to project the same feature set into two dimensions.
\end{itemize}

\begin{figure}
    \centering
    \includegraphics[width=1\linewidth]{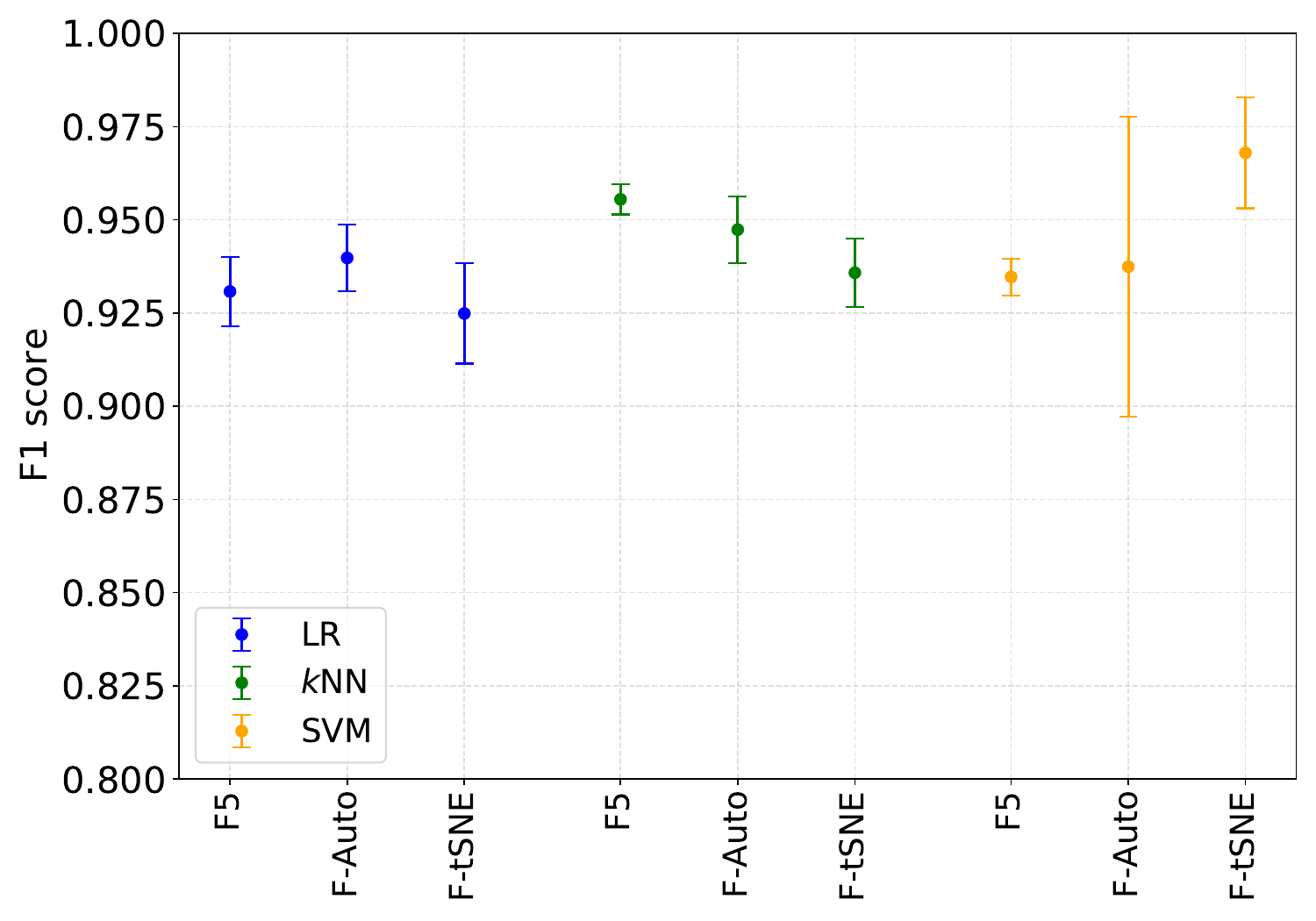}
    \caption{Comparison of $F1$-scores for feature combination F5, $q_\mathrm{IR}$ with autoencoder-compressed features (F-Auto), and $q_\mathrm{IR}$ and $t$-SNE-projected features (F-tSNE) across LR, \textit{k}NN, and SVM classifiers.} \label{fig:auto-tsne}
\end{figure}

Using the five most informative features, we train the LR, \textit{k}NN, and SVM classifiers on two compressed feature sets:
\begin{itemize}
    \item F-Auto: $q_\mathrm{IR}$ combined with two autoencoder-derived latent dimensions (Auto1, Auto2),
    \item F-tSNE: $q_\mathrm{IR}$ combined with $t$-SNE-projected dimensions (t-SNE1, t-SNE2).
\end{itemize}

Contrary to expectations, both dimensionality reduction methods increase model variance rather than stabilizing performance (Figure~\ref{fig:auto-tsne}). While the SVM classifier achieves a modest $\sim4$\% improvement with F-tSNE, the performance of \textit{k}NN and LR declined, suggesting that the two-dimensional projections may oversimplify complex non-linear relationships or that the limited input set ($N=5$) constrains the extraction of meaningful latent structure.

To further investigate, we conduct two additional tests. First, we increase the latent dimensionality (e.g., $n=3$), but observed similar performance degradation. Second, we expand the feature set to include the top nine features (excluding $q_\mathrm{IR}$), identified via ROC-based importance metrics: [class$\_$star, log$(M_{\rm star})$, log($S_{8.0}$/$S_{4.5}$), log($S_{5.8}$/$S_{3.6}$), log($i$/$z$), log($r$/$z$), log($g$/$z$), log($Y$/$H$), log($S_{4.5}$/$S_{3.6}$)], and repeated the compression experiments. In this case as well, both autoencoder and \textit{t}-SNE transformations generally reduced classifier performance across LR, \textit{k}NN, and SVM, with the only exception being a modest gain for SVM when combined with F-tSNE.

Taken together, these results indicate that dimensionality reduction is not effective in our case, possibly due to the relatively small sample size and the dominance of a single very prominent feature ($q_\mathrm{IR}$). We therefore conclude that retaining the original five-feature combination (F5), without additional dimensionality reduction, provides the most reliable classification performance for distinguishing SFGs and AGN in our MIGHTEE-COSMOS survey.

\subsection{Feature Scaling}\label{s:Data_normalization}
Data scaling is an important preprocessing step in data analysis and ML \citep{korobchynskyi2025systematic}. It involves transforming features into a consistent scale or format to enhance the efficiency and performance of computational models. This step is particularly essential when raw datasets contain variables with heterogeneous units, scales, or distributions, which may negatively impact model training and convergence \citep{ali2014data}. Common techniques include \citep{sujon2024use}: 
\begin{enumerate}
\item min-max scaling (normalization), which rescales values to a fixed range (typically [0,1]),
\item z-score standardization, which centers data to zero mean and unit variance, and
\item robust scaling, which uses medians and interquartile ranges to mitigate the influence of outliers.
\end{enumerate}

\begin{figure}
    \centering
    \includegraphics[width=1\linewidth]{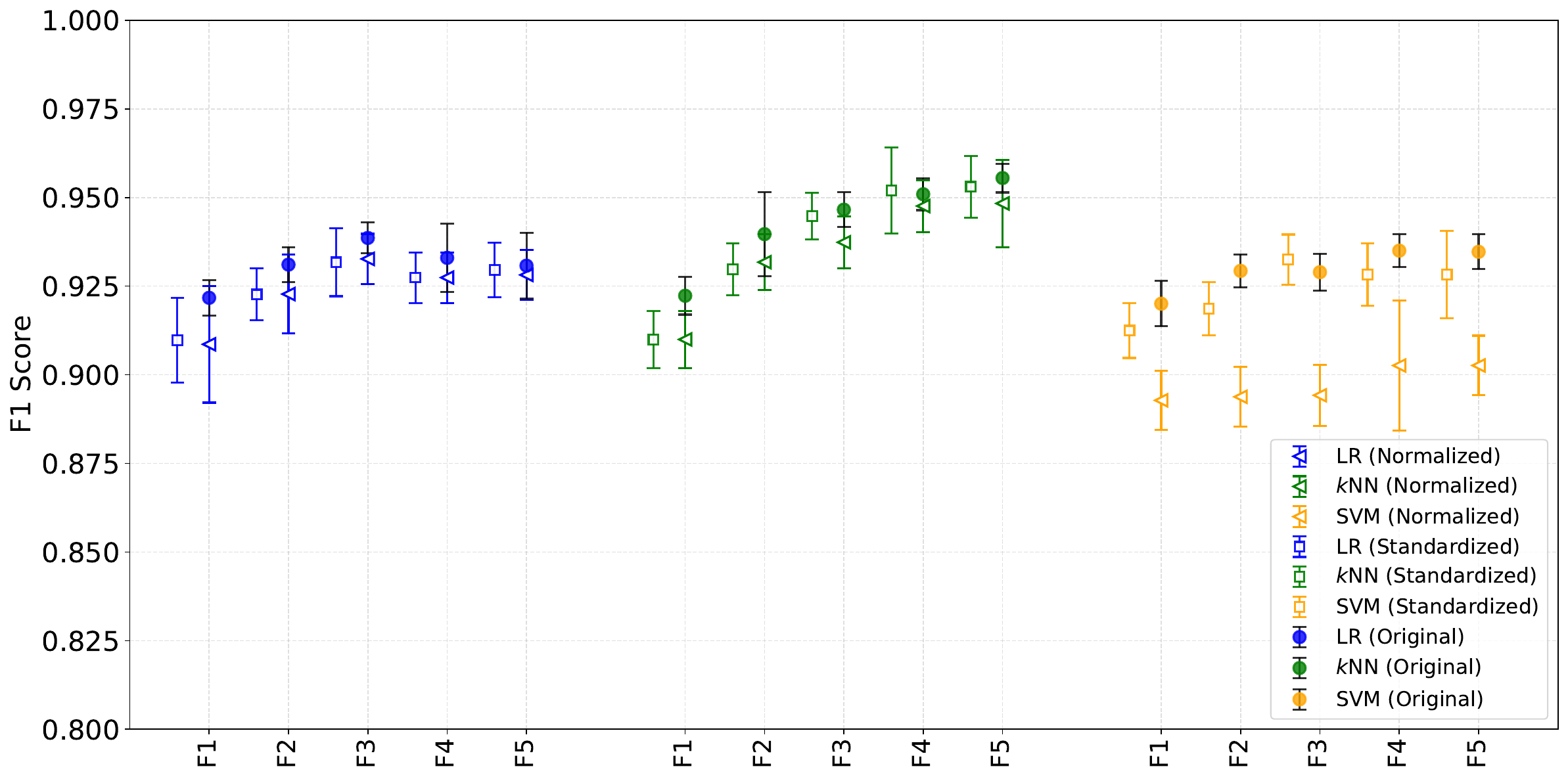}
    \caption{$F1$-score performance for the LR (blue), \textit{k}NN (green) and SVM (yellow) classifiers trained on the original (solid circles), min-max normalized (open triangles), and z-score standardized (open squares) datasets, across various feature combinations. The results indicate that feature scaling has a negligible impact on the performance of the LR and \textit{k}NN models. In contrast, the SVM demonstrates a statistically significant decrease in performance when trained on the normalized dataset.}\label{fig:normalised_data}
\end{figure}

In our study, we adopt min-max normalization and z-score standardization to all features to harmonize the feature space and assess its effect on model performance. We focus on LR, \textit{k}NN, and SVM models since they are known to be sensitive to feature scaling \citep{sujon2024use}. Both LR and SVM depend on the orientation of the decision boundary in the feature space, which can be skewed by unscaled inputs. Similarly, \textit{k}NN depends on raw distance metrics (e.g., Euclidean distance), making it susceptible to variations in feature scales. RF and XGB, on the other hand, are tree-based models and do not rely on distance calculations or gradient-based updates that depend on feature scale. These models split nodes based on feature thresholds, so they are largely invariant to monotonic transformations like normalization.

Figure \ref{fig:normalised_data} indicates that both implemented data scaling techniques have only a marginal effect on the performance of the \textit{k}NN and LR models. In contrast, the SVM exhibits a statistically significant degradation in performance when trained on the normalized dataset. Given this counterintuitive outcome, and in light of the established scale invariance of tree-based ensemble methods such as RF and XGB, we conclude that feature scaling provided no substantive benefit to our modeling framework. Consequently, we opt to employ the original, unscaled dataset in all our analyses to avoid introducing unnecessary preprocessing artifacts while maintaining the integrity of the underlying feature distributions.

\subsection{Class Imbalance}\label{s:Class Imbalance}

Imbalanced data refers to datasets with a pronounced skew in the distribution of class labels. This imbalance can affect many supervised ML algorithms, often causing them to overlook the minority class. This issue is particularly concerning, as predictions for the minority class are typically of the greatest importance \citep{das2022supervised}.

The typical approach to addressing data imbalance is to resample the training data randomly. The two standard methods are \textit{Undersampling} and \textit{Oversampling}. Undersampling reduces the number of sources in the majority class, while oversampling duplicates examples from the minority class. In this study, we used undersampling to balance training data by removing some SFGs, resulting in an equal number of AGN and SFGs. This approach is appropriate here, as AGN typically constitute the minority class in most radio continuum surveys.

We first assess the effect of class imbalance on our feature selection by comparing ROC-based AUC metrics derived from both the original and balanced datasets. As summarized in Table~\ref{tab:roc_balance}, the balanced dataset yields a feature importance ranking consistent with that of the original dataset (Table~\ref{tab:roc}), albeit with minor variations in absolute AUC values.

\begin{table}
    \centering
    \caption{ROC-based AUC values based on balanced dataset}
    \label{tab:roc_balance}
    \ra{1.3}
    \begin{tabular}{@{}ccc@{}}\toprule
     Feature ranking & Input features & AUC values\\\midrule
   1&  $q_\mathrm{IR}$ & 0.886 \\ 
   2&   class$\_$star & 0.630 \\
   3& log$(M_{\rm star})$ & 0.621 \\
   4& log($S_{8.0}$/$S_{4.5}$) & 0.574 \\
   5& log($S_{5.8}$/$S_{3.6}$) & 0.574 \\\bottomrule
    \end{tabular}
\end{table}

\begin{figure}
    \centering
    \includegraphics[width=1\linewidth]{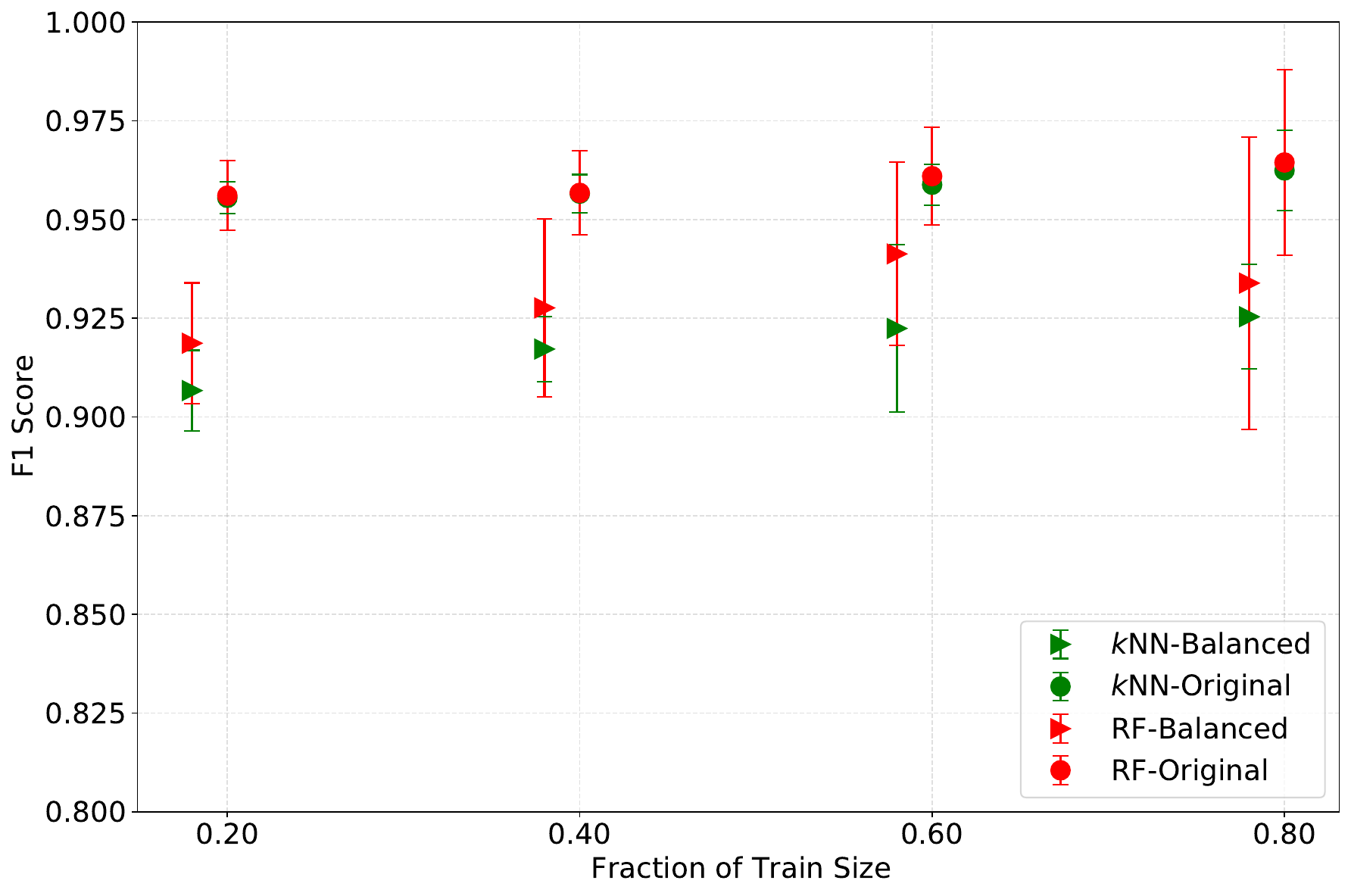}
    \caption{Comparison of $F1$-score performance for the \textit{k}NN and RF models trained on the original (circles) and class-balanced (triangles) datasets, shown as a function of the fraction of training data used. For visual clarity, data points corresponding to the balanced dataset are slightly offset leftward along the X-axis.} \label{fig:4.7}
\end{figure}

Secondly, we evaluate the performance of ML models in classifying SFGs and AGN from radio continuum survey data using both the original and balanced datasets. As shown in Figure \ref{fig:4.7}, we train the \textit{k}NN and RF models on varying fractions of the training data using the feature combination F5 ($q_\mathrm{IR}$, class$\_$star, log$(M_{\rm star})$, log($S_{8.0}/S_{4.5}$), log($S_{5.8}/S_{3.6}$)) and assess performance based on the $F1$-score. For both the original and balanced datasets, model performance remains consistently high, exceeding 90\% even when only 20\% of the training data is used. In the case of the balanced dataset, we further validate model robustness by using the remaining SFGs not included in the training set as an independent test sample. The resulting performance metrics remain high, with a recall of (96$\pm$0.01)\% and an $F1$-score of (97$\pm$0.01)\%, regardless of whether 20\% or 80\% of the dataset is used for training.

As also illustrated in Figure \ref{fig:4.7}, both ML models achieve slightly higher performance when trained on the imbalanced (original) dataset. This does not imply that imbalance is intrinsically beneficial, but rather reflects the fact that SFGs dominate the true underlying class distribution in deep radio continuum surveys. Although a more robust approach in ML classification is to train on balanced data to obtain well-calibrated probability models and subsequently incorporate prior information (such as the natural dominance of SFGs in this case), for simplicity and to remain consistent with the intrinsic survey distribution, we adopt the original dataset in our main analyses.

\subsection{Limitations}\label{s:limitation}
While supervised ML models deliver state-of-the-art classification results for MIGHTEE-COSMOS radio sources, several limitations merit attention, such as the quality of training data and challenges posed by missing or invalid measurements.

The ML algorithm’s mapping accuracy depends significantly on the quality of the labels in the training data. However, these outputs, derived using conventional methods, may be subject to biases and imperfections. For instance, our training set is based on the MIGHTEE-COSMOS multi-wavelength catalogue, where \cite{whittam2022mightee} employed five conventional techniques to label MIGHTEE-COSMOS radio sources. Although each diagnostic was applied independently, they are constrained by observational data quality, including data depth, coverage, and photometric accuracy.

Another challenge in ML-based classification of SFGs and AGN in radio continuum surveys is missing data or invalid measurements. Due to the unpredictability of astronomical properties, such as X-ray luminosity, VLBI detection, and optical or NIR photometry, we opted not to estimate these missing values using statistical imputation techniques \citep[e.g.,][]{Pelckmans05}. While XGB can manage missing data, it does so by inferring values based on the available measured features. Consequently, we restricted our ML analysis to samples with valid measurements across all five selected input features. This choice excluded approximately 7\% of radio sources with conventional labels in the MIGHTEE-COSMOS catalogue from the ML classification. Addressing such gaps will be a persistent challenge in applying ML classification to MIGHTEE and upcoming radio surveys.

\section{Conclusions}\label{s:conclusion}

In this study, we adopt and compare five supervised ML classification models, namely LR, SVM, \textit{k}NN, RF, and XGB, to classify star-formation-dominated or black-hole-accretion-dominated radio sources from the MIGHTEE-COSMOS survey. Using a sample of 4279 MIGHTEE-COSMOS radio sources labeled by \cite{whittam2022mightee} as either SFGs or AGN, along with their associated multi-wavelength measurements, we evaluate ML performance in classifying SFGs and AGN from radio continuum surveys. Our main conclusions are as follows:
\begin{enumerate}
    \item We analyze and select the most effective features for training and testing ML models. As expected, our one-dimensional, two-dimensional, ML-independent, ML-dependent, and ROC curve analyses indicate that the five parameters used in conventional classification prove to be the most effective. The IRRC parameter, $q_\mathrm{IR}$, is the most effective feature for distinguishing between SFGs and AGN. The optical compactness morphology parameter, class$\_$star, consistently ranks among the top three most effective features across all selection methods. While the IRAC colour may not be individually impactful, two-dimensional feature analyses reveal the importance of combining two IRAC colours for improved AGN-SFG separation. Therefore, the five features we selected to train ML models are $q_\mathrm{IR}$, class$\_$star, stellar mass, and two IRAC colours (log($S_{8.0}$/$S_{4.5}$) and log($S_{5.8}$/$S_{3.6}$)). The dataset completeness for sources with valid measurements across these five features is 93\%.
    \item We optimized the ML models using these selected features and evaluated classifiers with various feature combinations, guided by ROC-based AUC metric. Our results indicate that, for most models, ML performance generally improves as more feature combinations are included. Additionally, excluding the MIR colour features log($S_{8.0}$/$S_{4.5}$) and log($S_{5.8}$/$S_{3.6}$) leads to a noticeable performance drop for most ML models. This finding suggests that future radio surveys in regions lacking deep 5.8 and 8.0\,$\mu$m observations may experience a slight disadvantage in accurately classifying radio sources as either SFGs or AGN.
    \item We assess ML classification performance dependency on training data size by using $20\%$, $40\%$, $60\%$ and $80\%$ of the full dataset. All models achieve $F1$-scores greater than $90\%$ with any training size, except for the RF model when trained with the single feature $q_\mathrm{IR}$ and a training set size of $20\%$.
    \item Due to the limited completeness and unpredictability of X-ray and VLBI classifications, we do not include them as input features for training the ML models. However, our ML approach indicates that incorporating even limited X-ray observations into model training can marginally improve classification recall.
    \item We assess the impact of dimensionality reduction strategies and feature scaling and find that neither provides substantive benefits to our modeling framework. We also examine the effect of class imbalance in the MIGHTEE-COSMOS data and find that class imbalance does not impact ML model performance in our case. We therefore conclude that the unscaled dataset, combined with the original five-feature set (F5) and without additional dimensionality reduction, yields the most robust and reliable classification of SFGs and AGN in the MIGHTEE-COSMOS survey.
    \item Overall, our results demonstrate that all ML models perform well in classifying SFGs and AGN from radio sources, achieving $F1$-score $>90\%$ even with a small fraction (20\%) of the training data and a few key input features. Among the models assessed, the distance-based $k$NN classifier consistently emerges as the most accurate and stable, making it a compelling choice for the classification of SFGs and AGN in future large-scale radio continuum surveys, such as those by next-generation radio interferometric facilities.
\end{enumerate}

\section*{Acknowledgements}
We are grateful to the anonymous referee for a detailed report and valuable comments that improved the quality of this work. FXA acknowledges the support from the National Natural Science Foundation of China (12303016) and the Natural Science Foundation of Jiangsu Province (BK20242115). WS is grateful for support from the South African National Research Foundation (NRF) and National Astrophysics and Space Science Programme (NASSP). WS and MV acknowledge financial support from the Inter-University Institute for Data Intensive Astronomy (IDIA - a partnership between the University of Cape Town, the University of Pretoria and the University of the Western Cape).
MV acknowledges financial support from the South African Department of Science and Innovation's National Research Foundation under the ISARP RADIOMAP Joint Research Scheme (DSI-NRF Grant Number 150551) and the CPRR HIPPO Project (DSI-NRF Grant Number SRUG22031677). EH expresses gratitude for the valuable discussions with Prof. Chris Thron. FXA and WS sincerely thank Prof. Ian Smail and Prof. Seb Oliver for their insightful suggestions. We acknowledge the use of the \href{https://github.com/Hack4Dev/rooibosTea_classification}{rooibosTea\_classification code} described by \cite{hussein2022comparison}
as a guideline in our work.

The MeerKAT telescope is operated by the South African Radio Astronomy Observatory, which is a facility of the National Research Foundation, an agency of the Department of Science and Innovation. We acknowledge the use of the ilifu cloud computing facility -\href{https://www.ilifu.ac.za/}{www.ilifu.ac.za}, a partnership between the University of Cape Town, the University of the Western Cape, Stellenbosch University, Sol Plaatje University, the Cape Peninsula University of Technology and the South African Radio Astronomy Observatory. The ilifu facility is supported by contributions from IDIA, the Computational Biology division at UCT and the Data Intensive Research Initiative of South Africa (DIRISA).

\section*{Data Availability}
The MIGHTEE Early Science continuum data used in this work is extensively detailed in \citet{heywood2022mightee}. The MIGHTEE-COSMOS conventional classification catalogue and the cross-matched multi-wavelength catalogue were released with \citet{whittam2022mightee} and \citet{Whittam24}, respectively. All codes used in our analyses are publicly available on GitHub: \href{https://github.com/pfunzowalter/mightee-class-pub}{https://github.com/pfunzowalter/mightee-class-pub}.



\bibliographystyle{mnras}
\bibliography{references}




\appendix
\section{Additional features}
\label{A:feature-completeness}
To efficiently classify SFGs and AGN from the radio continuum survey, we derive colour indices using flux densities in the MIR, NIR, and optical wavelengths from the MIGHTEE-COSMOS multi-wavelength catalogue \citep{Whittam24}. Details on these multi-wavelength data are provided in Section $\S$\ref{sec: catalogue}. Briefly, we use twelve photometric bands, including HSC $griz$-band, IRAC 3.6, 4.5, 5.8, and 8.0\,$\micro$m data, along with UltraVISTA $YJHK_{\rm s}$-band photometries. In addition, other measurements available in the MIGHTEE-COSMOS catalogue, such as $q_\mathrm{IR}$, class$\_$star, and stellar mass are incorporated. 

From these data, we select the most effective input features for ML analyses from a total of 18 parameters: $q_\mathrm{IR}$, class$\_$star, log(M$_{\rm star}$), three MIR colours (log(S$_{8.0}$/S$_{4.5}$), log(S$_{5.8}$/S$_{3.6}$), log(S$_{4.5}$/S$_{3.6}$)), and 12 NIR and optical colours (log($g/r$), log($r/i$), log($i/z$), log($g/i$), log($g/z$), log($r/z$), log($Y/J$), log($J/H$), log($H/K_{\rm s}$), log($Y/H$), log($Y/K_{\rm s}$), log($J/K_{\rm s}$)). 

\begin{figure}
    \centering
    \includegraphics[width=1\linewidth]{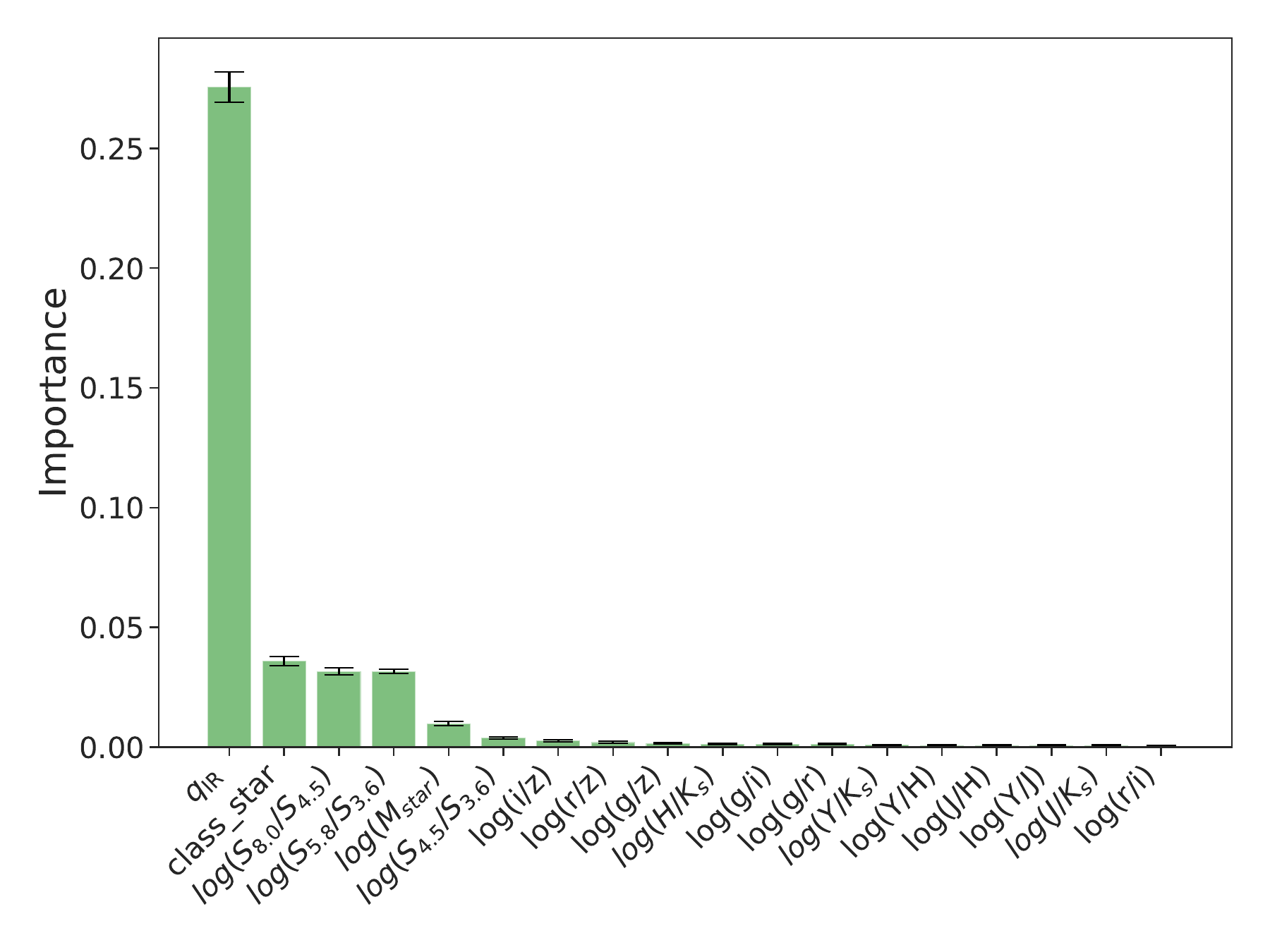}
    \caption{Permutation feature importance of 18 measurements.} 
    \label{fig:allperm}
\end{figure}

Figure~\ref{fig:allperm} illustrates the permutation importance of these features, with our selected five features demonstrating the highest effectiveness in classifying SFGs and AGN among radio-detected sources.

\begin{figure}
    \centering
    \includegraphics[width=1\linewidth]{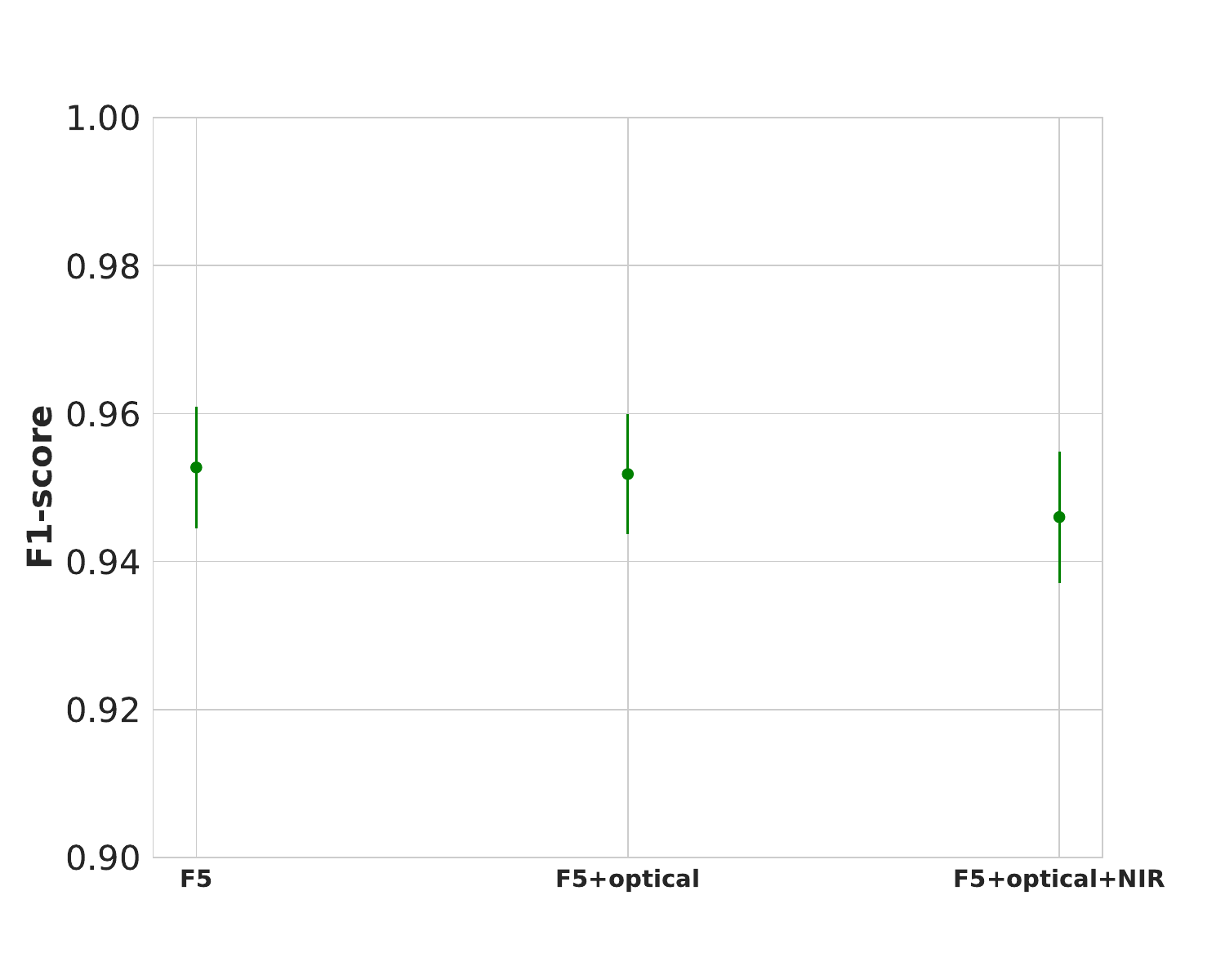}
    \caption{The results of applying the \textit{k}NN classifier to distinguish between AGN and SFGs trained using three different feature combinations, namely, F5, F5+optical and F5+optical+NIR. 
    The evaluation metric is the $F1$-score, with error bars representing the standard deviation obtained through jackknife resampling.}
    \label{fig:color_ml_results}
\end{figure}

In addition, we incorporate these optical and NIR colours as input features to assess the performance of ML models. Using the \textit{k}NN model as an example, we present results in Figure~\ref{fig:color_ml_results}. Three feature combinations are used to train the \textit{k}NN model: 1)F5, which includes class$\_$star, $q_\mathrm{IR}$, log(S$_{\rm 8.0}$/S$_{\rm 4.5}$), log(S$_{\rm 5.8}$/S$_{\rm 3.6}$); 2)F5 + optical colours (log($g/r$), log($r/i$), log($i/z$), log($g/i$), log($g/z$), log($r/z$)); and (3) F5 + optical + NIR colours (log($Y/J$), log($J/H$), log($H/K_{\rm s}$), log($Y/H$), log($Y/K_{\rm s}$), log($J/K_{\rm s}$)). The data are randomly split into 80\% for training and 20\% for testing, with model performance evaluated using the $F1$-score as the classification metric. As shown in Figure~\ref{fig:color_ml_results}, adding these features does not improve or even slightly decrease the performance of the \textit{k}NN classifier. This outcome is likely due to the additional features introducing confusion, which hampers the model’s ability to effectively distinguish between SFGs and AGN. Furthermore, the completeness of the ML dataset is marginally reduced if all optical and NIR photometric measurements are required.

\section{Feature correlation plots}\label{B:correlation}
\begin{figure}
\centering
    \begin{subfigure}{0.15\textwidth}
        \includegraphics[width=\linewidth]{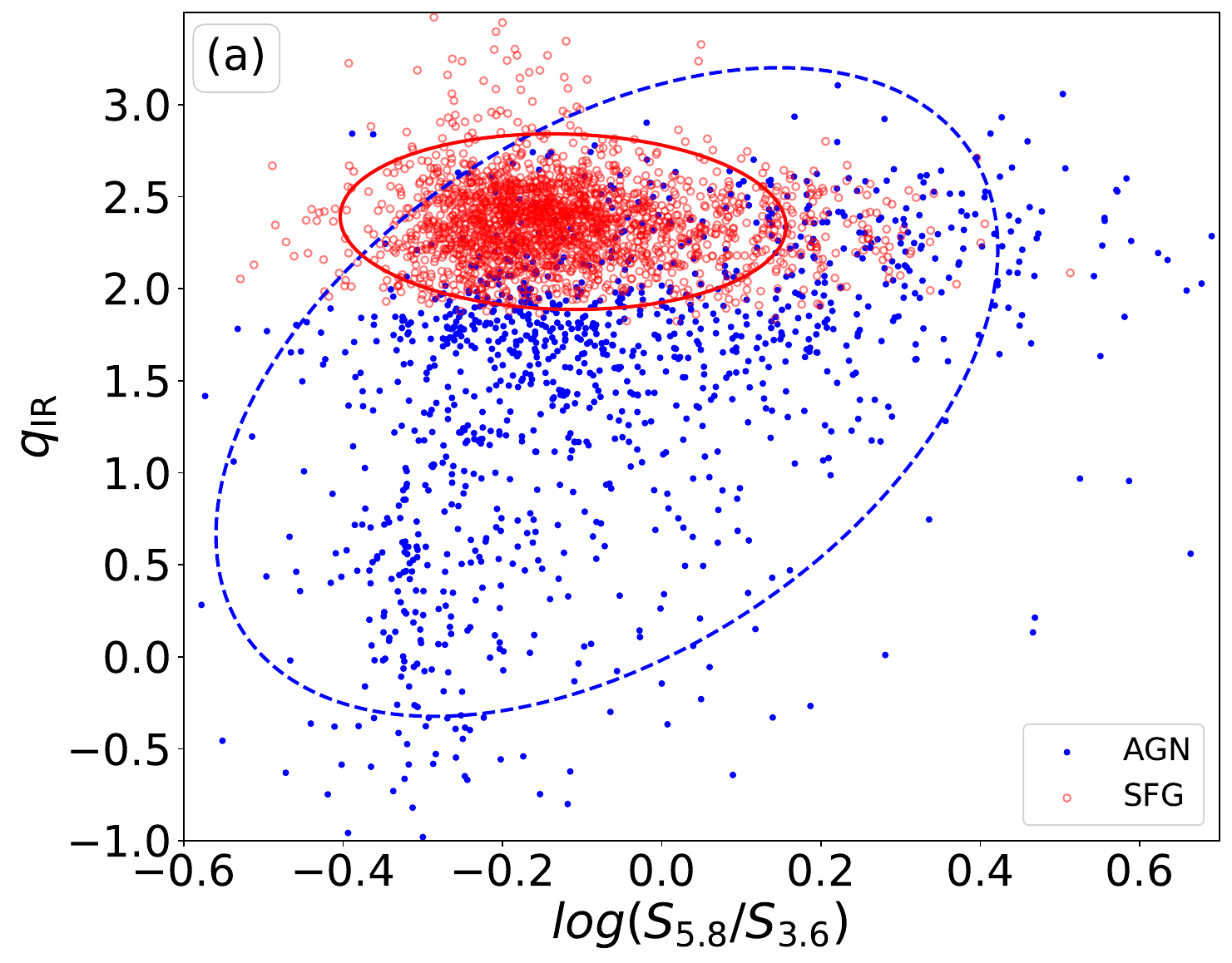}
        \phantomcaption\label{fig:Appc}
    \end{subfigure}
    \begin{subfigure}{0.15\textwidth}
        \includegraphics[width=\linewidth]{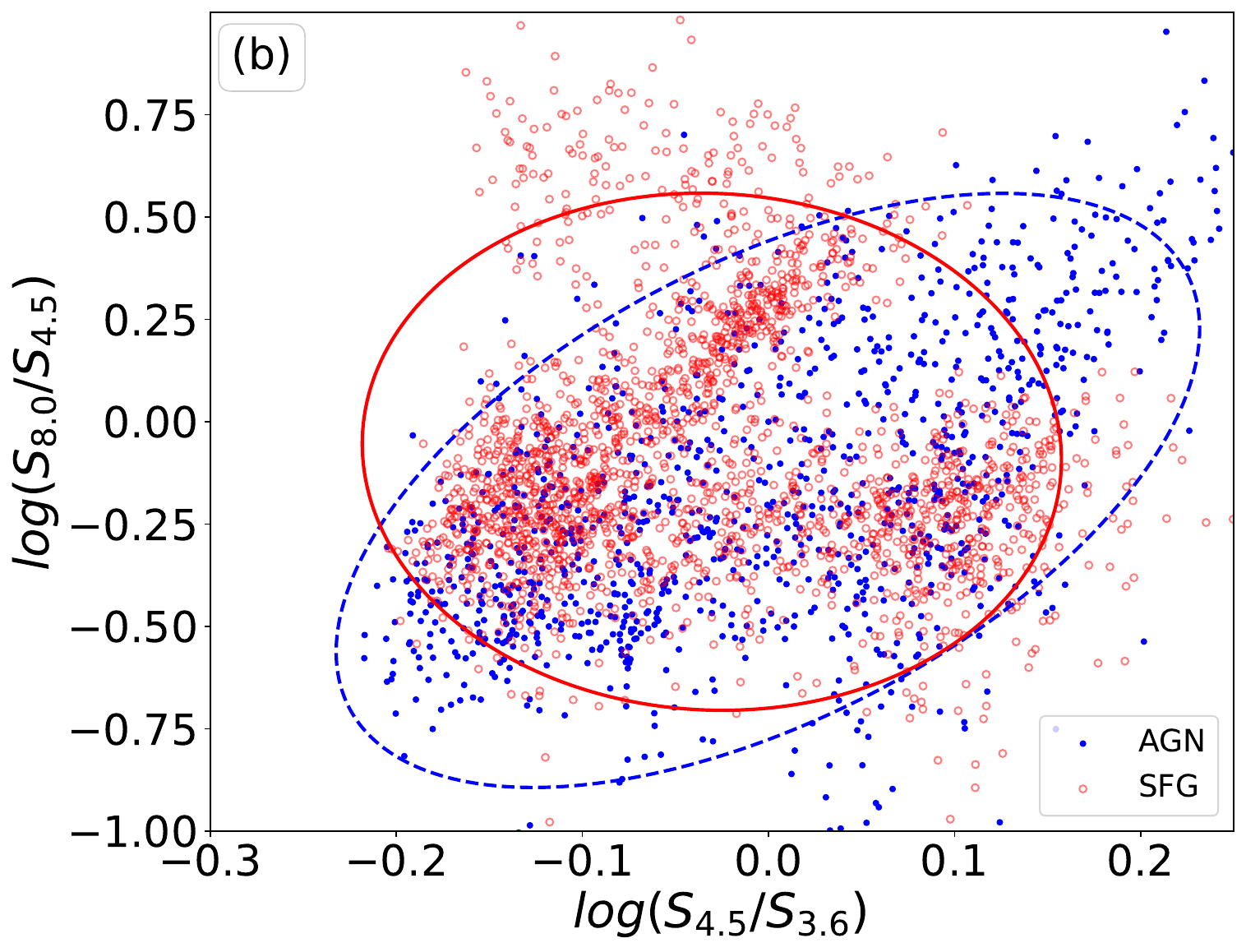}
        \phantomcaption\label{fig:Appd}
    \end{subfigure}\hfill
    \begin{subfigure}{0.15\textwidth}
        \includegraphics[width=\linewidth]{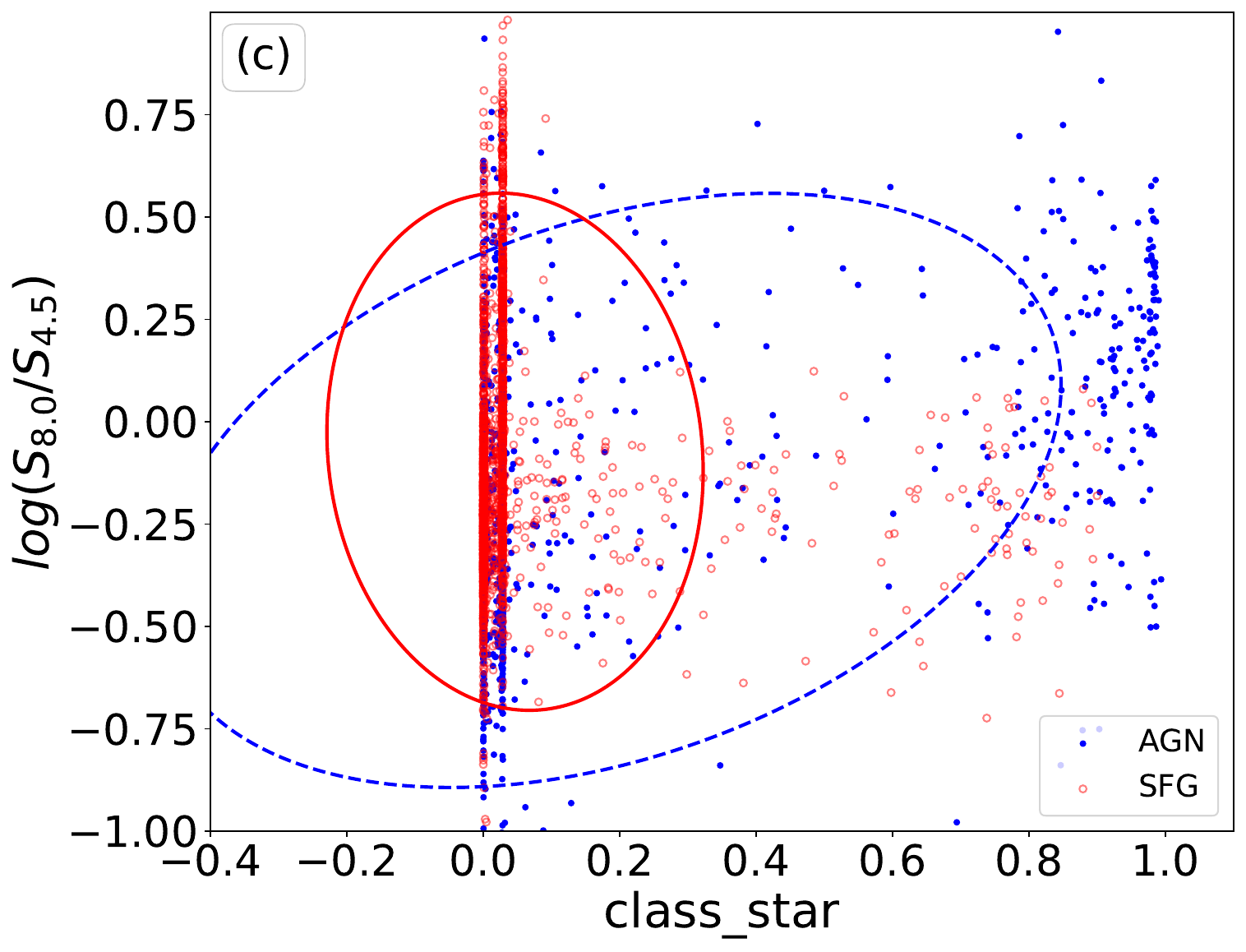}
        \phantomcaption\label{fig:Appe}
    \end{subfigure}\hfill
    \begin{subfigure}{0.15\textwidth}
        \includegraphics[width=\linewidth]{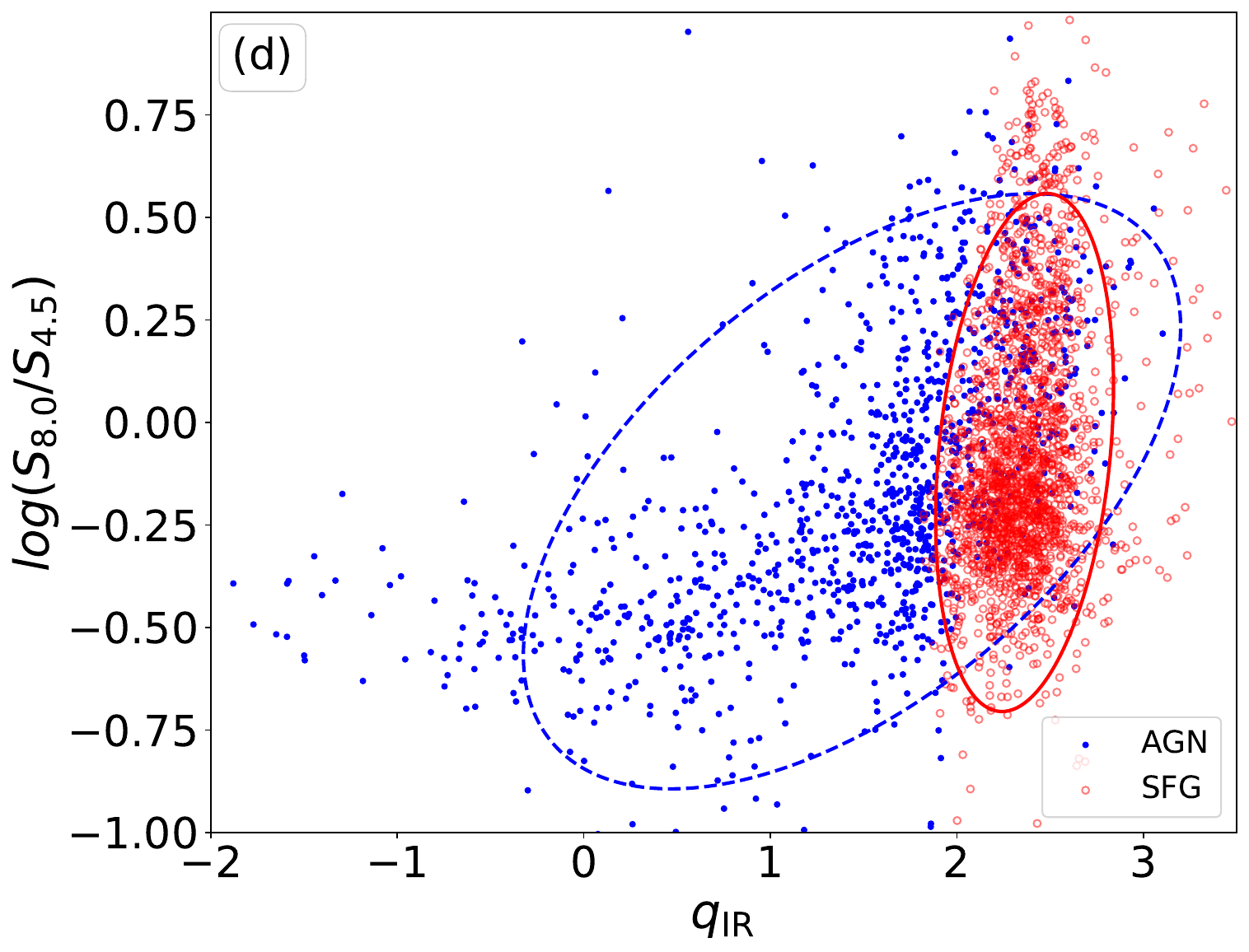}
        \phantomcaption\label{fig:Appf}
    \end{subfigure}
    \begin{subfigure}{0.15\textwidth}
        \includegraphics[width=\linewidth]{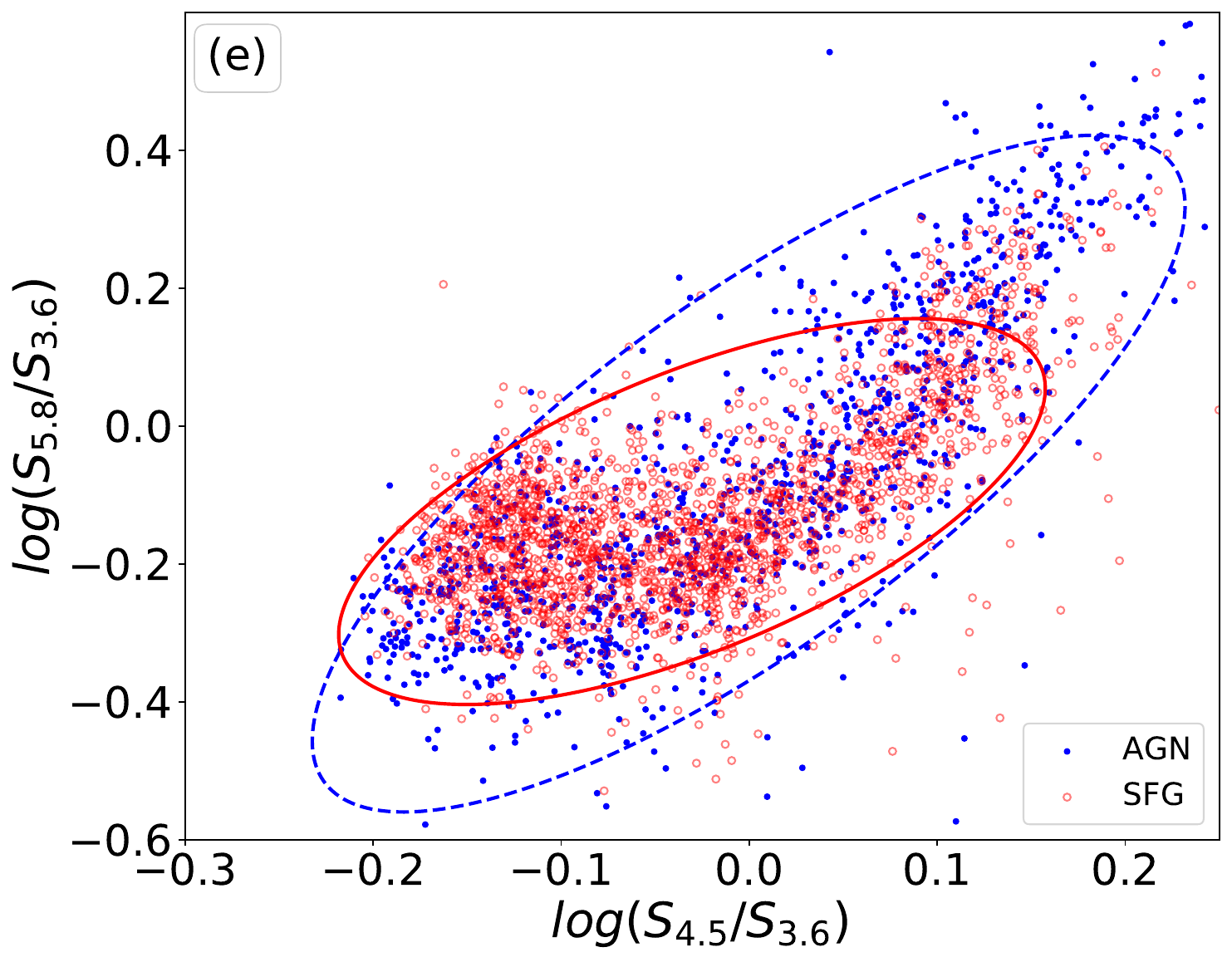}
        \phantomcaption\label{fig:Appg}
    \end{subfigure}\hfill
    \begin{subfigure}{0.15\textwidth}
        \includegraphics[width=\linewidth]{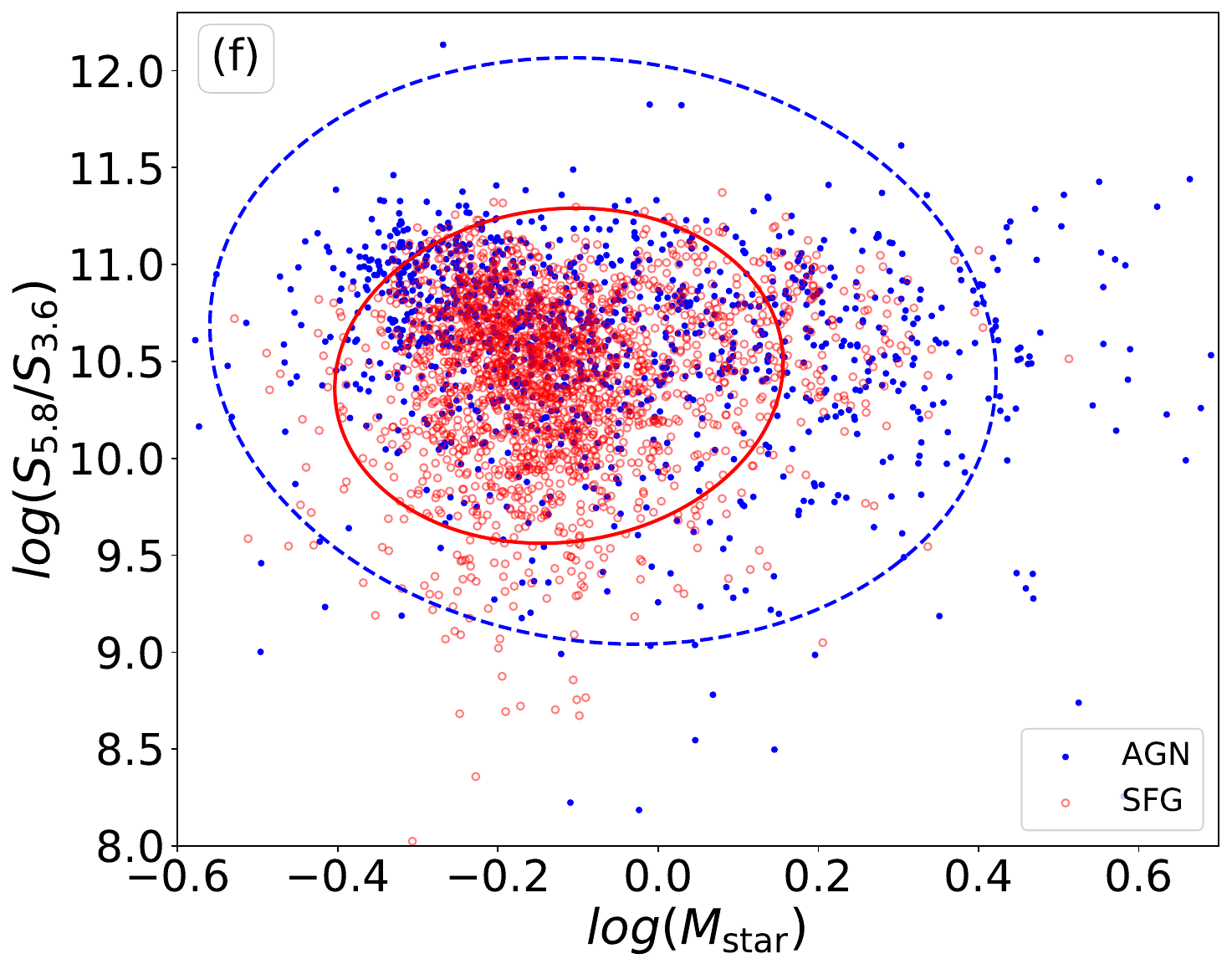}
        \phantomcaption\label{fig:Appi}
    \end{subfigure}
    \begin{subfigure}{0.15\textwidth}
        \includegraphics[width=\linewidth]{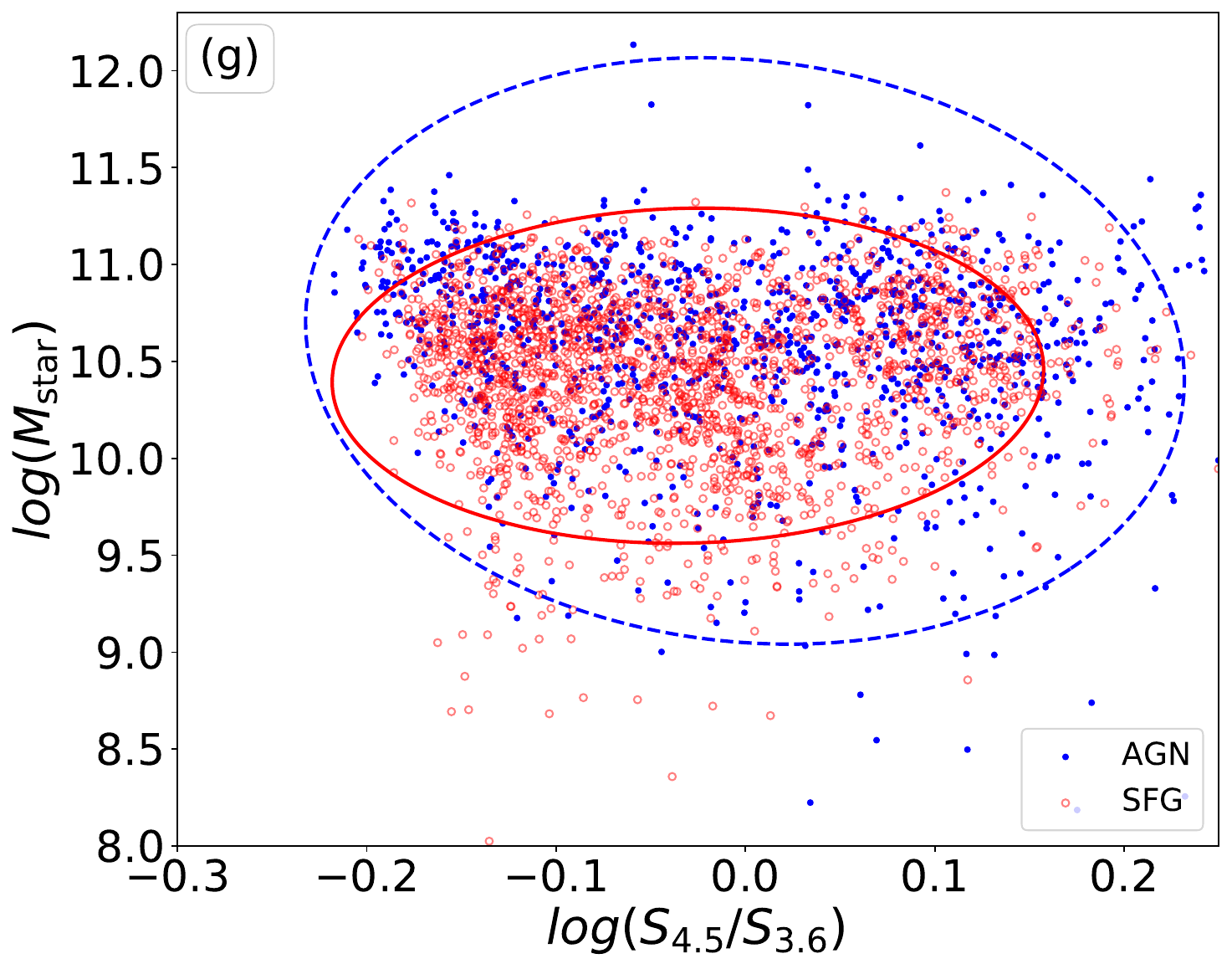}
        \phantomcaption\label{fig:Appj}
    \end{subfigure}\hfill
    \begin{subfigure}{0.15\textwidth}
        \includegraphics[width=\linewidth]{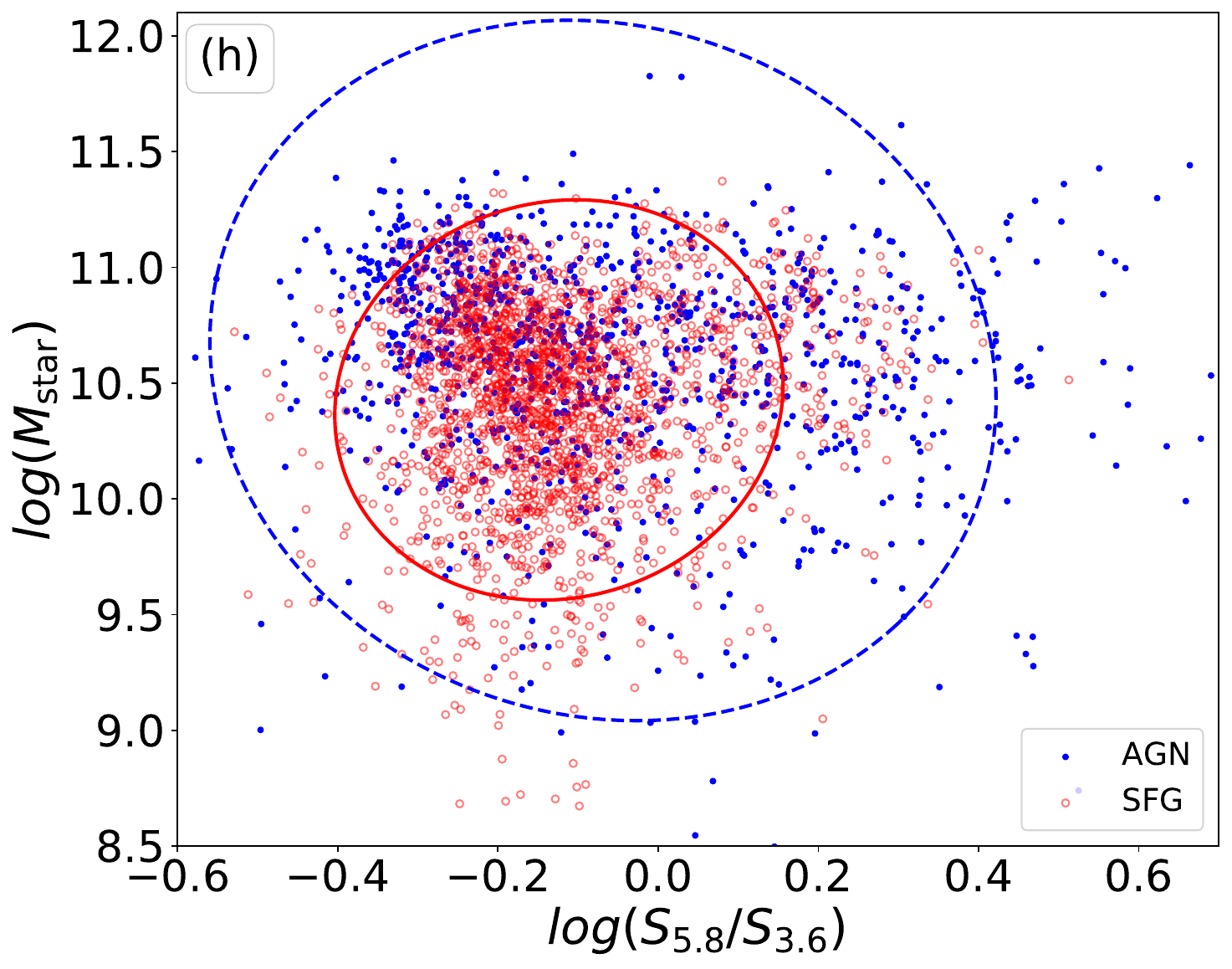}
        \phantomcaption\label{fig:Appk}
    \end{subfigure}\hfill
    \begin{subfigure}{0.15\textwidth}
        \includegraphics[width=\linewidth]{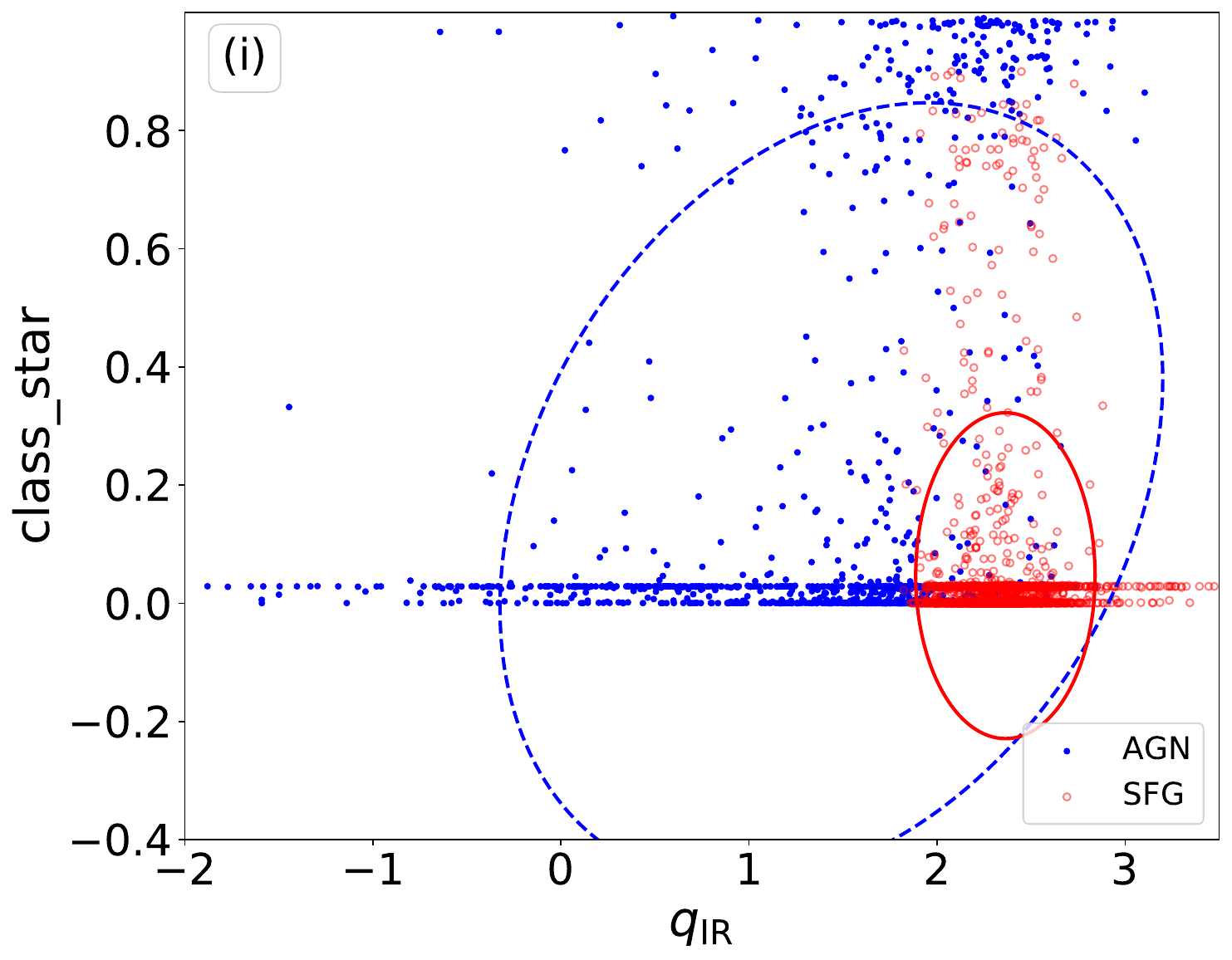}
        \phantomcaption\label{fig:Appl}
    \end{subfigure}
    \begin{subfigure}{0.15\textwidth}
        \includegraphics[width=\linewidth]{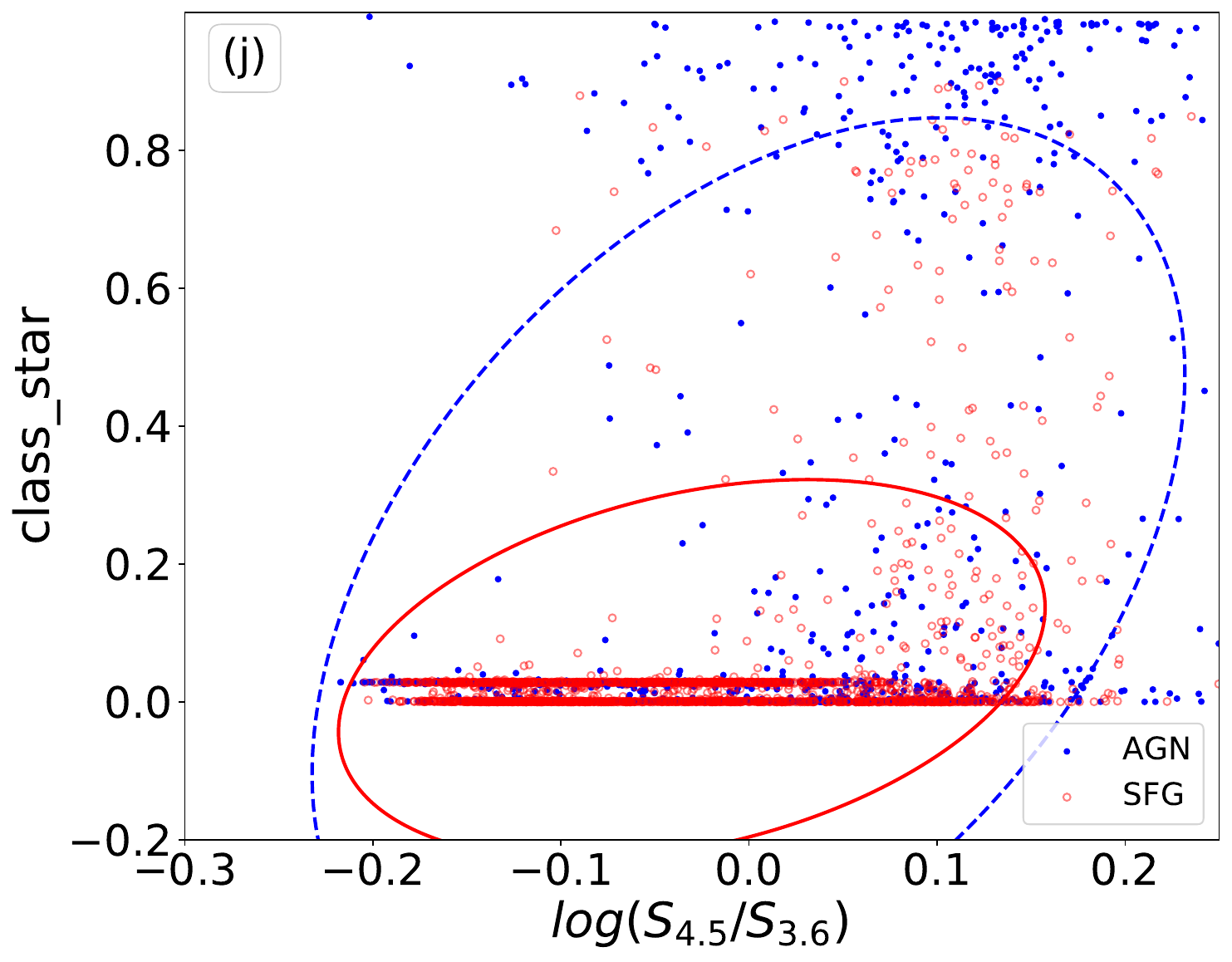}
        \phantomcaption\label{fig:Appm}
    \end{subfigure}\hfill
    \begin{subfigure}{0.15\textwidth}
        \includegraphics[width=\linewidth]{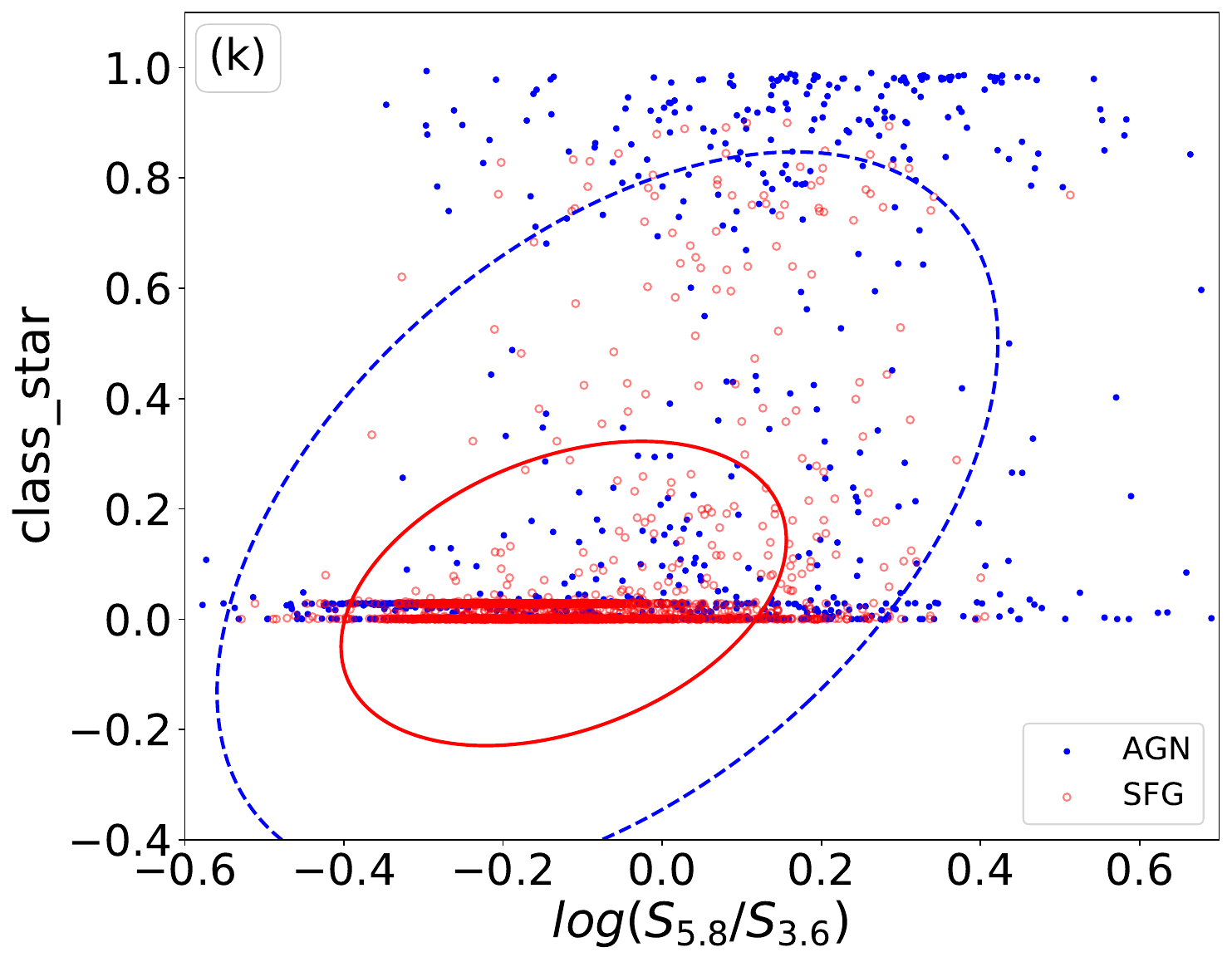}
        \phantomcaption\label{fig:Appn}
    \end{subfigure}\hfill
    \begin{subfigure}{0.15\textwidth}
        \includegraphics[width=\linewidth]{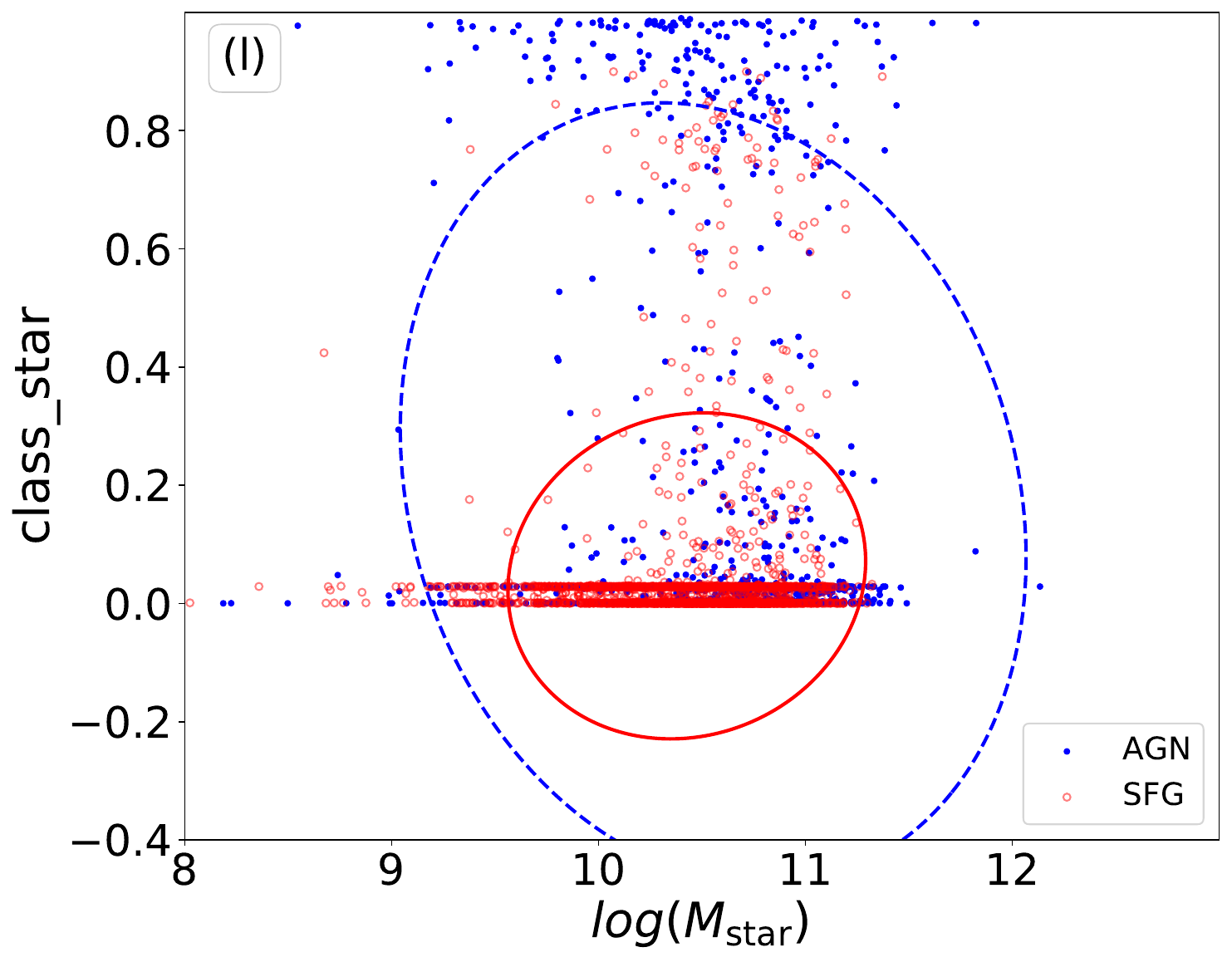}
        \phantomcaption\label{fig:Appo}
    \end{subfigure}
    \caption{As with Figure~\ref{fig:4.3 corelation}, the feature correlation plots for the remaining 12 feature pairs are shown.} 
    \label{fig:4.3App-corr}
\end{figure}

As described in Section $\S$\ref{s:3.2.2}, we examine the correlations among the six features shown in Figure~\ref{fig:4.2}, resulting in 15 correlation plots. Three of these plots are presented in Figure~\ref{fig:4.3 corelation}, with the remainder shown in this section (Figure~\ref{fig:4.3App-corr}). As further discussed in Section $\S$\ref{s:3.2.2}, combining certain feature pairs can, in some cases (for instance, the two IRAC colours), improve the performance of ML models for classifying SFGs and AGN from the radio continuum surveys, despite significant overlap between the confidence ellipses of these two populations.

\section{Hyperparameters}
\label{app:hyperparameter}
Hyperparameter optimization is a key step in ML classification. 
Section $\S$\ref{subs:hyperparameter} details the methods employed for hyperparameter tuning in this study. Here, we provide examples illustrating the optimization of hyperparameters across various ML models.

\begin{figure}
    \begin{subfigure}{0.24\textwidth}
        \includegraphics[width=\linewidth]{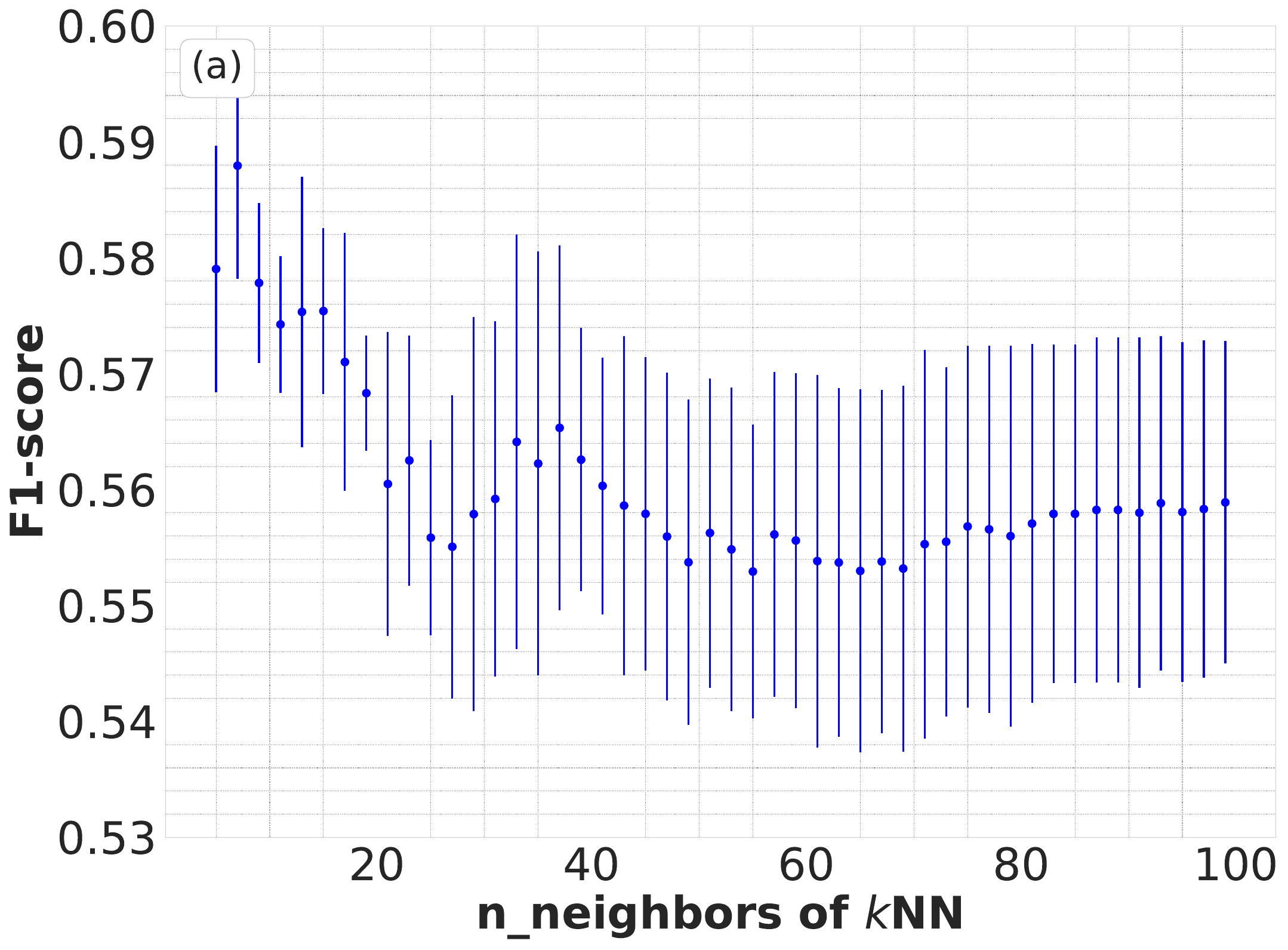}
        \phantomcaption\label{fig:hyp-a}
    \end{subfigure}\hfill
    \begin{subfigure}{0.24\textwidth}
        \includegraphics[width=\linewidth]{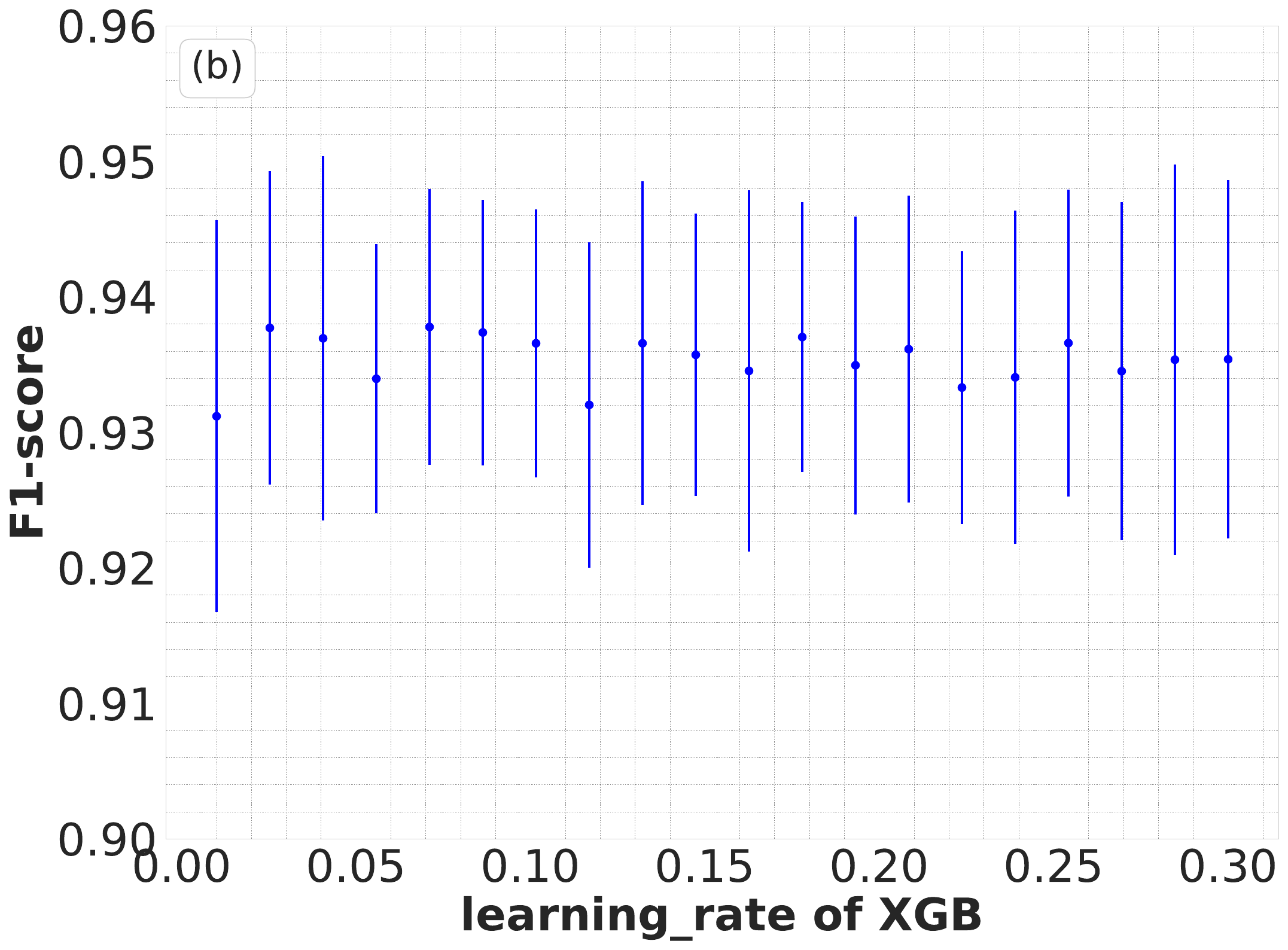}
        \phantomcaption\label{fig:hyp-b}
    \end{subfigure}
    \medskip
    \begin{subfigure}{0.24\textwidth}
        \includegraphics[width=\linewidth]{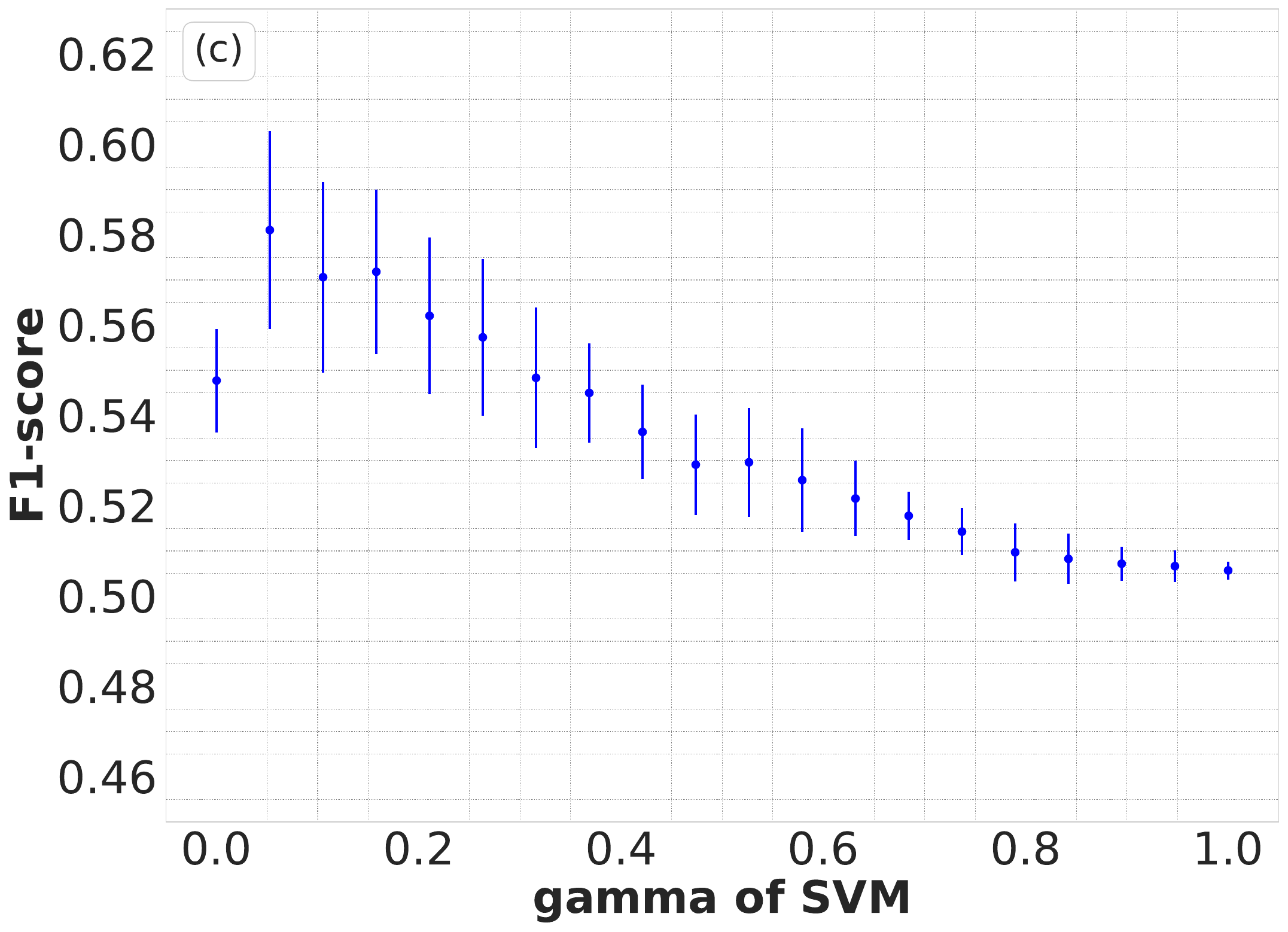}
        \phantomcaption\label{fig:hyp-c}
    \end{subfigure}\hfill
    \begin{subfigure}{0.24\textwidth}
        \includegraphics[width=\linewidth]{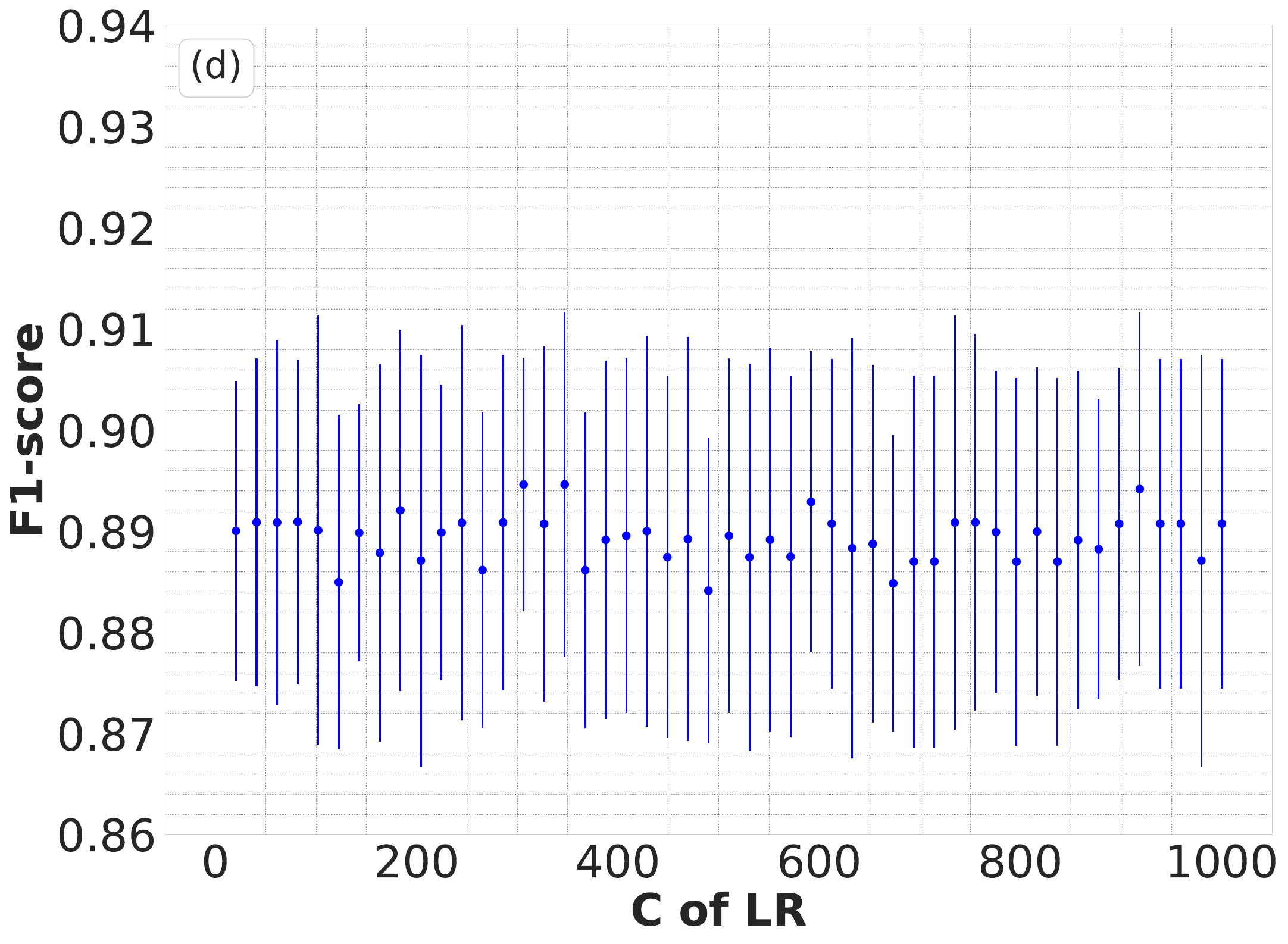}
        \phantomcaption\label{fig:hyp-d}
    \end{subfigure}
    \medskip
    \begin{subfigure}{0.24\textwidth}
        \includegraphics[width=\linewidth]{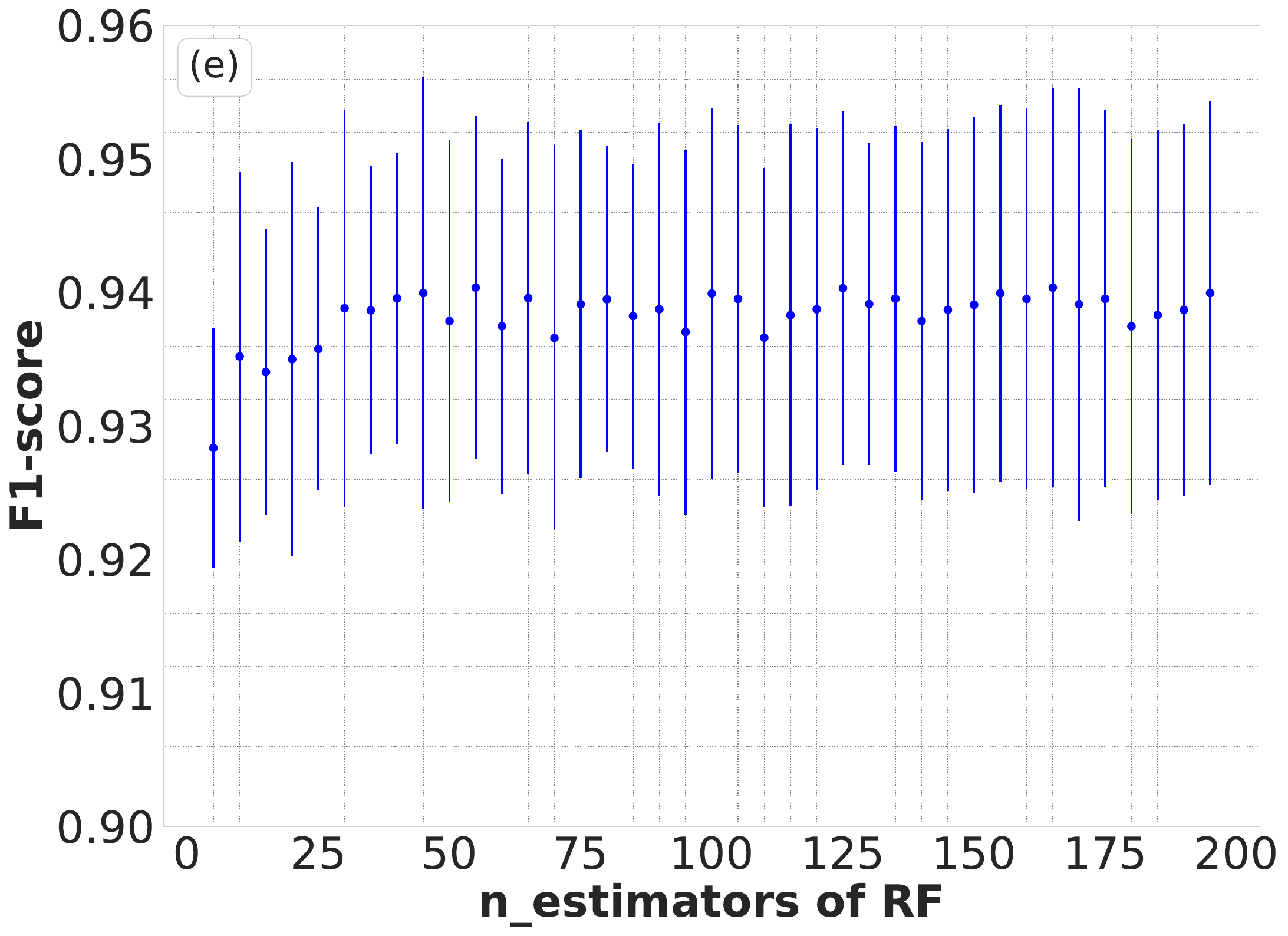}
        \phantomcaption\label{fig:hyp-e}
    \end{subfigure}\hfill
    \begin{subfigure}{0.24\textwidth}
        \includegraphics[width=\linewidth]{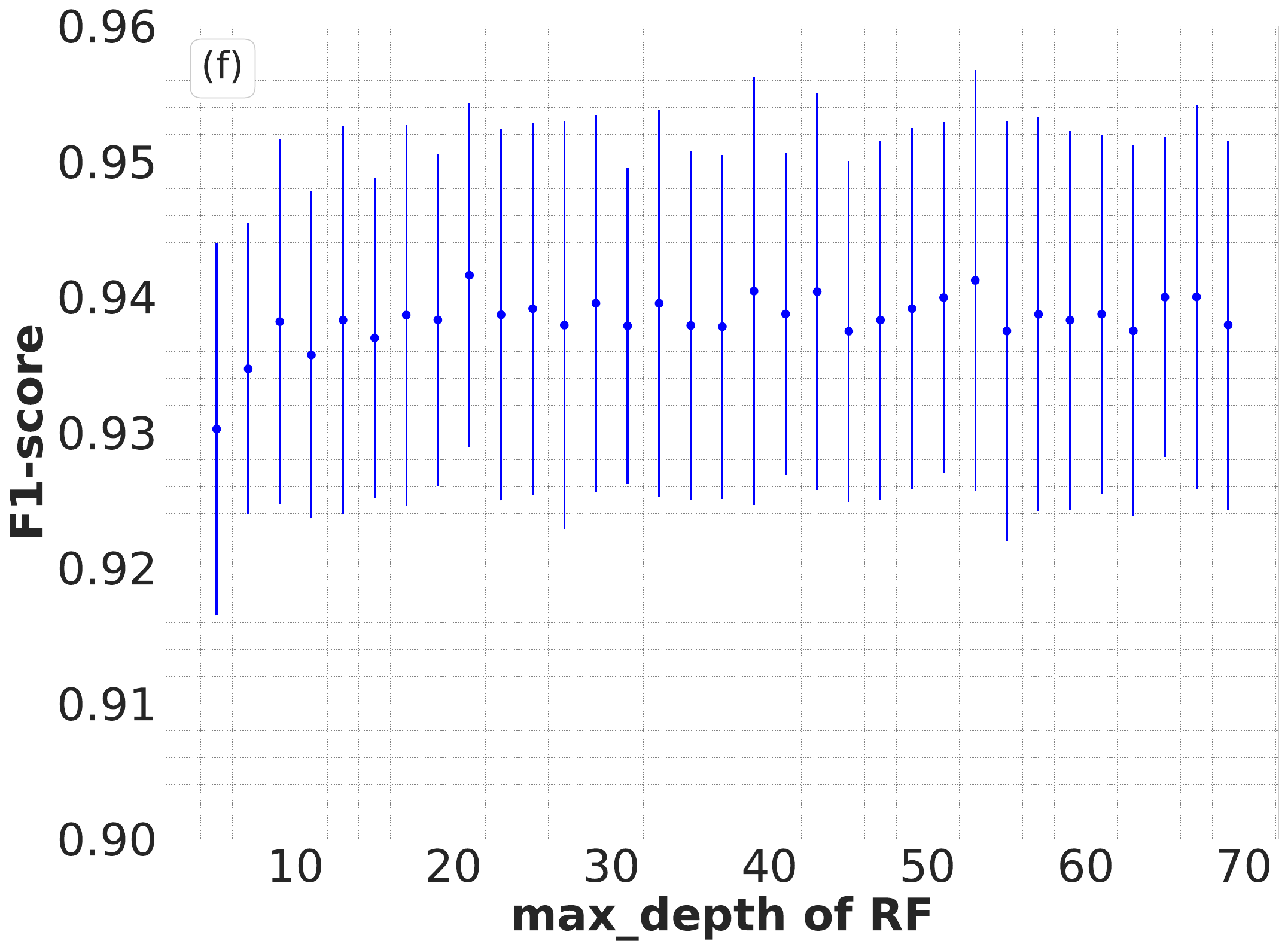}
        \phantomcaption\label{fig:hyp-f}
    \end{subfigure}

    \caption{Six examples illustrate how the performance of ML models varies with specific hyperparameters, evaluated using the $F1$-score. For each model, the selected hyperparameter is varied while others remain at their default values. Error bars represent the $F1$-score standard deviation, calculated via jackknife resampling. This analysis provides insights into the sensitivity of model performance to parameter tuning, highlighting optimal configurations and trade-offs for each model type.}
    \label{fig:hyper}
\end{figure}

As outlined in Section $\S$\ref{subs:hyperparameter}, we perform a three-fold split of the sample and apply a grid search technique to identify the optimal hyperparameters for each ML model. This process involves adjusting one hyperparameter at a time, while holding the others at their default values, and evaluating performance changes based on the $F1$-score. Figure ~\ref{fig:hyper} illustrates examples of how ML model performance varies with specific hyperparameters. For instance, in the \textit{k}NN classifier, performance decreases as the \textit{Number of Neighbors} increases, leading us to select a value of $<15$ for this hyperparameter. Figure~\ref{fig:hyp-c} demonstrates that SVM performance improves with an increase in $\gamma$, a parameter used in the Radial Basis Function (RBF) kernel and other nonlinear kernels. Higher $\gamma$ values increase the flexibility of the decision boundary, allowing it to adapt more closely to individual data points.  

Not all hyperparameters, however, exhibit a monotonic relationship with model performance. Figure~\ref{fig:hyp-b} shows fluctuations in XGB performance in response to the learning rate parameter, while Figure~\ref{fig:hyp-c} suggests that LR classification performance remains largely unaffected by variations in the $C$-value, which controls regularization strength by balancing the trade-off between model fit and weight minimization to prevent overfitting. Additionally, Figures~\ref{fig:hyp-e} and~\ref{fig:hyp-f} demonstrate that RF classifier performance reaches its maximum when the hyperparameters $n\_estimators$ and $max\_depth$ exceed values of 40 and 10, respectively. 
We point out that this section provides only selected examples of hyperparameter optimization for ML models. For a comprehensive list of hyperparameters for each model, we refer readers to the \href{https://scikit-learn.org/stable/}{scikit-learn} and the \href{https://xgboost.readthedocs.io/en/stable/}{\textit{XGBoost}}.

\bsp	
\label{lastpage}
\end{document}